\documentclass[12pt]{article} 
\usepackage[hyperfootnotes=false]{hyperref}
     \usepackage{amsmath}
      \usepackage{amssymb}
  \usepackage{graphicx}
  \usepackage{tikz}
\usepackage{tikz,pgfplots}
  \usepackage{xcolor}
  \usepackage{framed}
  \newlength{\abstractwidth}
 \usetikzlibrary{decorations.pathmorphing}
\tikzset{snake it/.style={decorate, decoration=snake}}
\usetikzlibrary{shapes}
\usepackage{empheq}

  \usepackage{subfigure}
  \usepackage{caption}
 \usepackage{mciteplus}

  \newcommand{\be}{\begin{equation}}
  \newcommand{\bea}{\begin{eqnarray}}
  \newcommand{\eea}{\end{eqnarray}}
  \newcommand{\beq}{\begin{equation}}
  \newcommand{\ee}{\end{equation}}
  \newcommand{\eeq}{\end{equation}}

  \newcommand{\half}{{1\over 2}}

\def\la{\label}

\def\32{{3 \over 2 } }

  \def\ba{\begin{eqnarray}}
  \def\ea{\end{eqnarray}}

 \def\simleq{\; \raise0.3ex\hbox{$<$\kern-0.75em
      \raise-1.1ex\hbox{$\sim$}}\; }
 \def\simgeq{\; \raise0.3ex\hbox{$>$\kern-0.75em
      \raise-1.1ex\hbox{$\sim$}}\; }

\def\nref#1{(\ref{#1})}

\begin{document}

\begin{titlepage}
  \bigskip

  \bigskip\bigskip

  \bigskip

\begin{center}
{\Large \bf { Two dimensional Nearly de Sitter gravity }}
 \bigskip
{\Large \bf { }} 
    \bigskip
\bigskip
\end{center}

  \begin{center}

 \bf {Juan Maldacena$^1$, Gustavo J. Turiaci$^2$  and Zhenbin Yang$^3$   }
  \bigskip \rm
\bigskip
 
   $^1$Institute for Advanced Study,  Princeton, NJ 08540, USA  \\
   \bigskip
     $^2$Physics Department, University of California, Santa Barbara, CA 93106, USA  \\
\rm
 \bigskip
 $^3$Jadwin Hall, Princeton University,  Princeton, NJ 08540, USA

  \bigskip \rm
\bigskip
 
\rm

\bigskip
\bigskip

  \end{center}

 \bigskip\bigskip
  \begin{abstract}

 We study some aspects of the de Sitter version of Jackiw-Teitelboim gravity. Though we do not have propagating gravitons, we have a boundary mode when we compute observables with a fixed dilaton and metric at the boundary. We compute the no-boundary wavefunctions and probability measures to all orders in perturbation theory. We also discuss contributions from different topologies, borrowing recent results by Saad, Shenker and Stanford. We discuss how the boundary mode leads to gravitational corrections to cosmological observables when we add matter. Finally, starting from a four dimensional gravity theory with a positive cosmological constant, we consider a nearly extremal black hole and argue that some observables are dominated by the two dimensional nearly de Sitter gravity dynamics.   

 \medskip
  \noindent
  \end{abstract}
\bigskip \bigskip \bigskip

  \end{titlepage}

   \tableofcontents


\section{Introduction }

Recently, there has been  intensive study of one of the simplest two dimensional gravity theories; the JT gravity theory 
\cite{Jackiw:1984je,Teitelboim:1983ux}, see also \cite{Henneaux:1985nw,LouisMartinez:1993eh}.  It has been mostly studied with negative curvature, see eg \cite{Jensen:2016pah,Maldacena:2016upp,Engelsoy:2016xyb,Grumiller:2016dbn,Almheiri:2016fws,Cvetic:2016eiv,Sarosi:2017ykf,Mertens:2017mtv,Lam:2018pvp,Harlow:2018tqv,Larsen:2018iou,Saad:2018bqo,Goel:2018ubv, Lin:2018xkj,Kitaev:2018wpr,Yang:2018gdb,Brown:2018bms,Alishahiha:2018swh,Blommaert:2018iqz,Blommaert:2019hjr,BF, SSS}\footnote{The zero curvature case was recently discussed in \cite{Dubovsky:2017cnj,Dubovsky:2018bmo}.}. 
This theory arises as the universal description of gravitational effects for near extremal black holes \cite{Almheiri:2014cka,Nayak:2018qej,Li:2018omr}.

 Here we consider the JT theory with positive curvature (see also \cite{Anninos:2017hhn, Anninos:2018svg} and \cite{Dong:2018cuv, Gorbenko:2018oov}).  
 This describes nearly-de-Sitter two dimensional gravity. We first study the pure gravity theory and argue that the physics is very similar to the nearly Anti-de-Sitter case. We explain that the computation of the no-boundary wavefunctions is very similar to the computation of anti-de-Sitter partition functions. These are probability amplitudes for finding a universe with a 1-geometry which is a circle of a given size and with a given value of the dilaton. We explain how perturbative quantum effects can be incorporated by the dynamics of boundary modes similar to the ones appearing in the nearly-$AdS_2$ case. In addition, we mention that one can also consider non-perturbative effects that correspond to summing over geometries that are essentially the same as the ones considered recently by Saad, Shenker and Stanford \cite{SSS}. Our discussion can be viewed as a reinterpretation of those results into a de-Sitter context. 
 This is perhaps not too surprising given the simplicity of this two dimensional gravity theory. 

 These two dimensional gravity results can be viewed as giving four dimensional gravity results in a limit where the dynamics is dominated by a two dimensional nearly-$dS_2$ region. This arises as follows. We can consider black holes in four dimensional de-Sitter space. Such black holes have a maximum mass (or maximum horizon area). As we approach this maximum the geometry develops a region with a  nearly $dS_2 \times S^2$  geometry. In this case, the leading effects that break the $dS_2$ isometries are due to gravitational modes captured by the nearly-$dS_2$ gravity described above. 
    We discuss some aspects of the computation of the no boundary proposal when the boundary is $S^1\times S^2$ (see \cite{Laflamme:1986bc} and more recently \cite{Anninos:2012ft,Banerjee:2013mca}), and argue that when the size of the $S^1$ becomes very large, the computation reduces to the one in two dimensional gravity. In this regime, there is a long period of two dimensional expansion that transitions to a four dimensional expansion. 
    We will also mention that the four dimensional correlators, in the same long distance limit, reduce to the two dimensional correlators, together with their quantum corrections.

 We then discuss matter correlators in exactly $dS_2$ with no gravity.  The simplicity of nearly-$dS_2$ gravity allows us to compute quantum gravity corrections to these correlators. The gravitational corrections to the wavefunctions reduce to an integral over the boundary modes (as in the nearly-$AdS_2$ case). On the other hand, in de Sitter it is natural to compute expectation values. These can be computed by performing an integral over the boundary values of the fields in the square of the wavefunction. This leads to expressions for quantum corrections of expectation values which are relatively simple. Note that now we have two boundary modes, one for the bra and the other for the ket of the wavefunctions.

   This paper is organized as follows. In section \ref{sec:nearlyds2g}, we discuss various aspects of two dimensional nearly-$dS_2$ gravity, its action, boundary modes, no boundary wavefunctions, sums over topologies, etc. 
   In section \ref{4dSec}, we describe how this two dimensional theory arises from a four dimensional gravity theory. 
In section \ref{PureMatter}, we recall the computation of matter correlators in $dS_2$ with no gravity. In section \ref{sec:corrgb}, we compute the gravitational correction to matter correlators by integrating over the boundary modes. We finish with some conclusions and a discussion. 
  
   As a side comment, in appendix \ref{AdSPartition},   we show how the nearly-$AdS_2$ partition function is related to the Klein Gordon inner product of two wavefunctions, where the slice independence of the Klein Gordon inner product is interpreted as the RG flow of the boundary theory.    
  
\section{Nearly $dS_2$ gravity } \la{sec:nearlyds2g}
 
\subsection{Action } 

The action for nearly-$dS_2$ gravity can be written as 
\be \la{ActdS}
i S = i \phi_0 \left[ \int d^2 x \sqrt{g} R  - 2 \int_{Bdy} K \right] + 
i \int d^2 x \sqrt{g} \phi ( R - 2)  - i2 \int_{Bdy} \phi_b K 
\ee
The first term, proportional to $\phi_0$,  is purely topological and captures the de Sitter entropy. 
The second term is the de-Sitter version of the Jackiw-Teitelboim theory \cite{Jackiw:1984je,Teitelboim:1983ux}. 
We have indicated also the boundary terms.  

\begin{figure*}[h!]
  \centering
  \subfigure[Poincare]{%
    \includegraphics[width=0.4\textwidth]{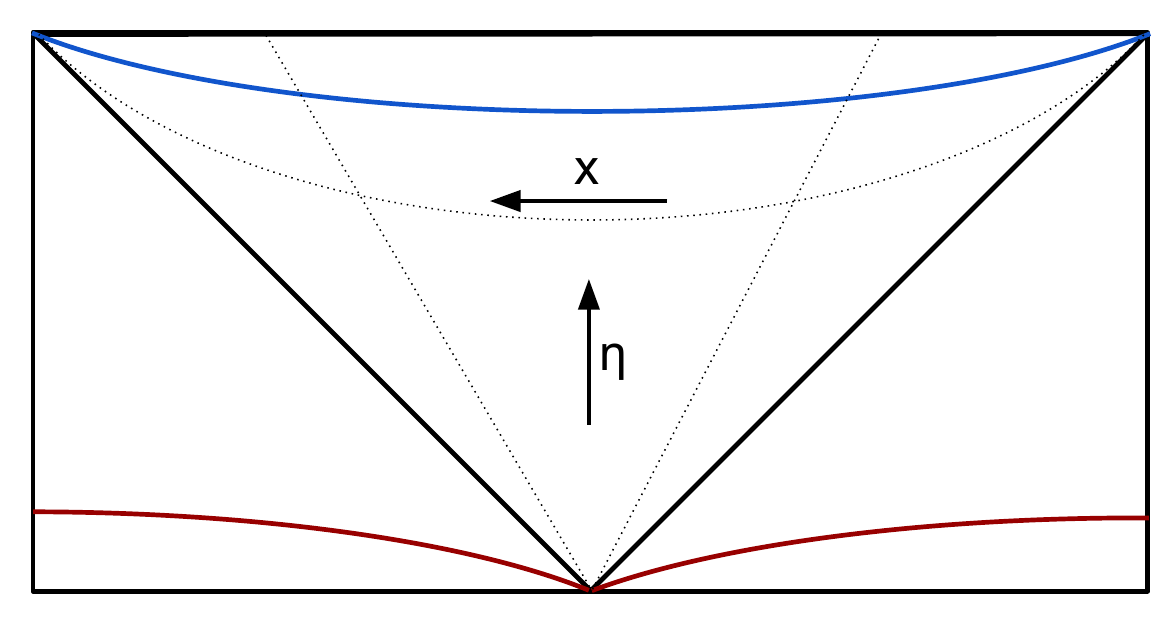}%
    \label{fig:Poincare}%
    }\hspace{2cm}
    \subfigure[Global]{%
    \includegraphics[width=0.4\textwidth]{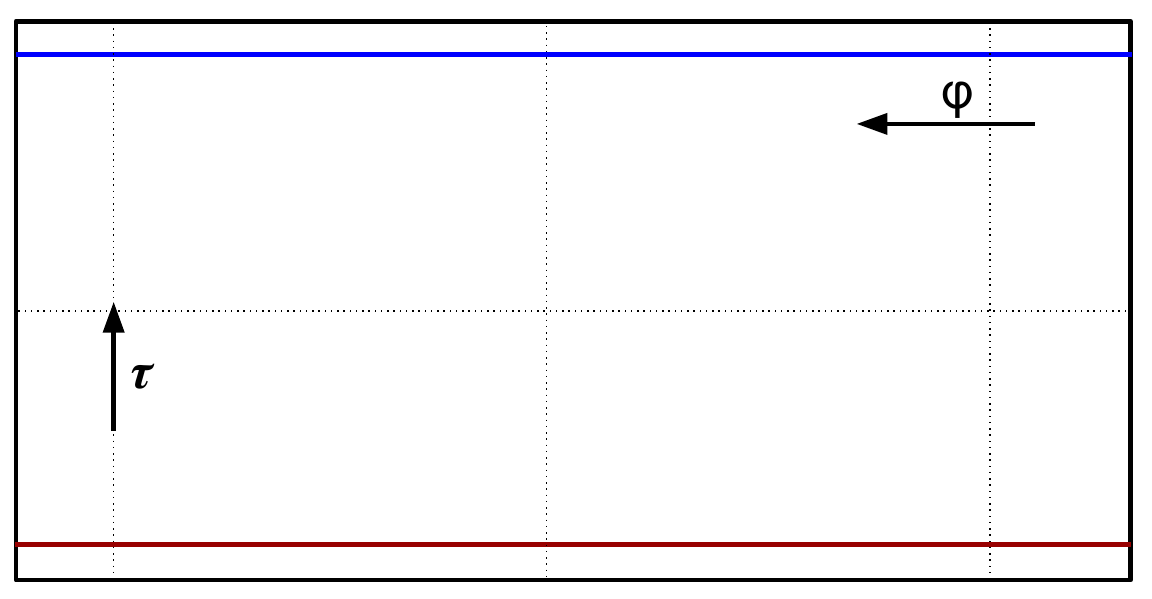}%
    \label{fig:Global}%
  }\hspace{2cm}%
  \subfigure[Milne]{%
    \includegraphics[width=0.5\textwidth]{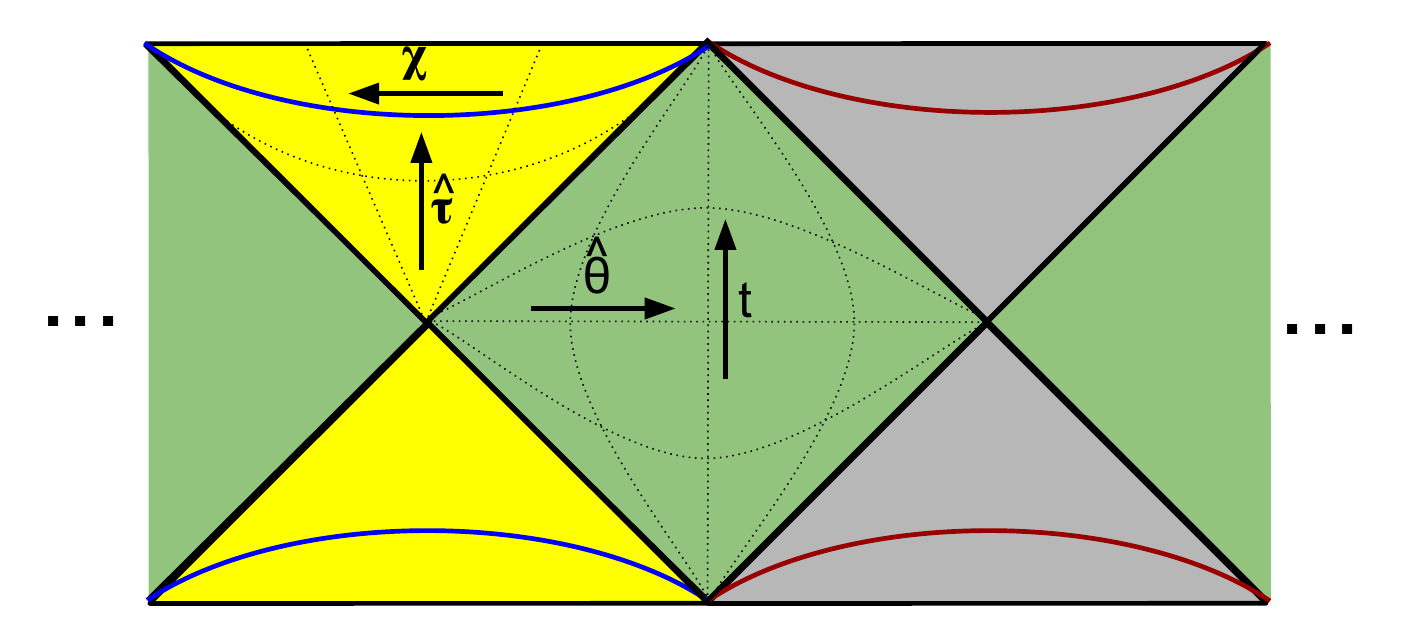}%
    \label{fig:Milne}%
  }%
  \caption{The Carter-Penrose diagram of $dS_2$ is a horizontal strip. On top of this diagram we 
display the regions where $\phi $ goes to $+\infty$ in blue and to  $-\infty$ in red, the latter representing 
some kind of ``singularity''. In (a) we display \nref{Poincare} in (b) \nref{Global} in (c) we see \nref{Milne}. 
The region covered by the Milne type coordinates is yellow and the region covered by the static coordinates in green. In addition we have a region that corresponds to the interior of a black hole shaded in grey.  }
  \label{Penrose}
\end{figure*}

The equations of motion for $\phi$ set the metric to be $dS_2$. The equations of motion for the metric
determine $\phi$ up to a few constants. 
In fact, 
as in any two dimensional dilaton gravity theory, all solutions have a Killing vector, given by 
$\zeta^\mu = \epsilon^{\mu \nu } \partial_\nu \phi$ \cite{Mann:1992yv}. 
 Therefore we can use it to simplify the form of the solution, which differs depending on the location of the fixed point of this Killing vector. 
We find three different types of solutions
\bea \la{Poincare}
 & ~&~~~~ ds^2_{\rm Poincare} = { - d \eta^2 + dx^2 \over \eta^2 } ~,~~~~~~~~~~~~ \phi =  { \phi_p \over \eta } ~
\\
& ~&~~~~ ds^2_{\rm Global}  =  - d \tau^2 + \cosh^2 \tau d\varphi^2 ~,~~~~~~ \phi = { \phi_g  \sinh \tau }  
\la{Global}
\\
&~& \left\{ \begin{array}{rl} ds^2_{\rm Milne}  &=    - d\hat  \tau^2 + \sinh^2 \hat \tau d\chi^2 ~,~~~~~~ \phi = \phi_m \cosh \hat \tau 
\cr
ds^2_{\rm Static} & =   d\hat \theta^2 - \sin^2 \hat \theta dt^2 ~,~~~~~~~~~~~ \phi = \phi_m \cos \hat \theta \la{Milne} \end{array} \right.
\eea
The last two are simply different coordinate patches of the same solution. The two patches are connected
via $\hat \tau =0$ which is just a coordinate singularity. 
The coordinates are related by 
 $\hat \tau \to i \hat \theta $ and $\chi \to i t $ , which produces a metric with a time translation symmetry. The static patch has two horizons, one of which can be viewed as a black hole horizon and
the other as a cosmological horizon. See figure \ref{Penrose}(c) for the regions of the Penrose diagram covered by these coordinates. We can view \nref{Milne} as describing a nearly-$dS_2$ expanding universe with two black holes at its two ``ends'' and $\chi = \pm \infty$.
 The ``static'' coordinates in \nref{Milne} cover the region in the vicinity of the black hole horizon while the ``Milne'' patch in \nref{Milne} covers the expanding region. 

As we will discuss below, when nearly-$dS_2$  arises from a higher dimensional theory, we find that 
$\phi_0 + \phi$ is typically the area of a transverse sphere. So, the regions where $\phi$ becomes very negative
will lead to a singularity and the regions where $\phi$ increases tend to match into expanding higher dimensional spaces. 
Notice also that, independently of any higher dimensional consideration, $\phi_0 + \phi$ governs the cost to produce a topology change in the geometry, so when this becomes small we would need to include other topologies. In this paper we will mostly imagine that $\phi_0 + \phi$ is large, so that we do not
have to worry about topology change\footnote{The only exception is subsection \ref{OtherTopologies}.}.

\subsection{The asymptotic nearly-symmetries and the boundary modes }\label{sec:NdSschwarzian}

We are interested in computing observables of the cosmological type, so we will be looking at the expanding regions, and also the regions where $\phi$ is growing. 
In principle, the wavefunction of the universe depends on the metric of the spatial slice and the value of
$\phi$ along that slice. However,  the constraints (coming from the reparametrization gauge symmetry) 
determine much of this dependence. 
So it is convenient to pick a gauge where $\phi$ takes a constant value along the spatial slice. This is equivalent to a choice of time. This choice of time is well defined in the region that $\phi$ is becoming very large. We will also pick a gauge where the spatial coordinate is chosen so as to represent proper distance, $u$. 
These two choices are essentially the same as the ones we do when we analyze the Nearly-$AdS_2$ problem \cite{Maldacena:2016upp}.

\begin{figure*}
	\centering
	  \subfigure{%
    \includegraphics[width=0.3\textwidth]{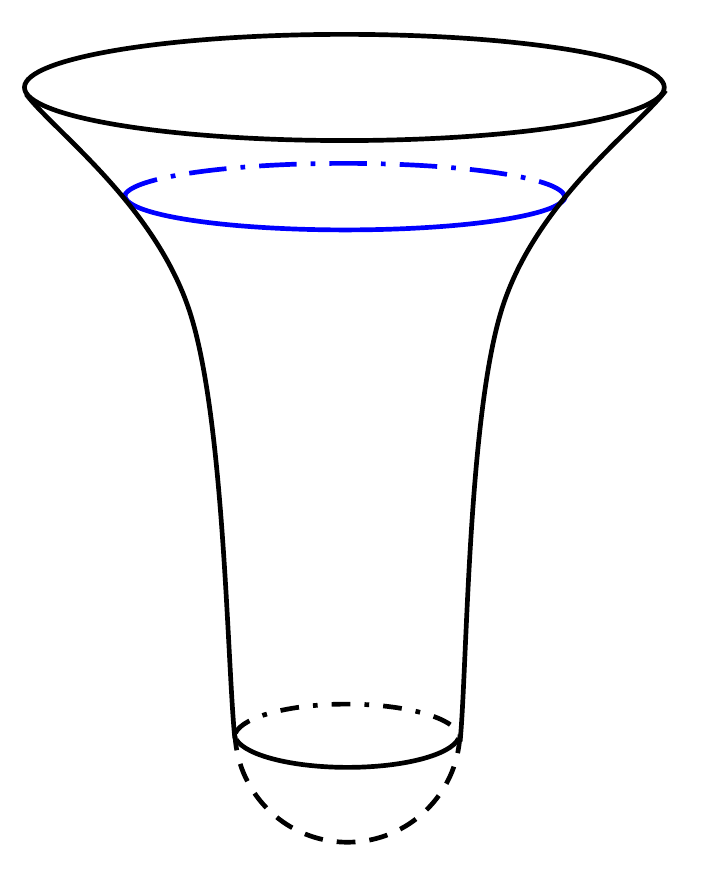}
    }\hspace{2cm}
    \subfigure{%
    \includegraphics[width=0.3\textwidth]{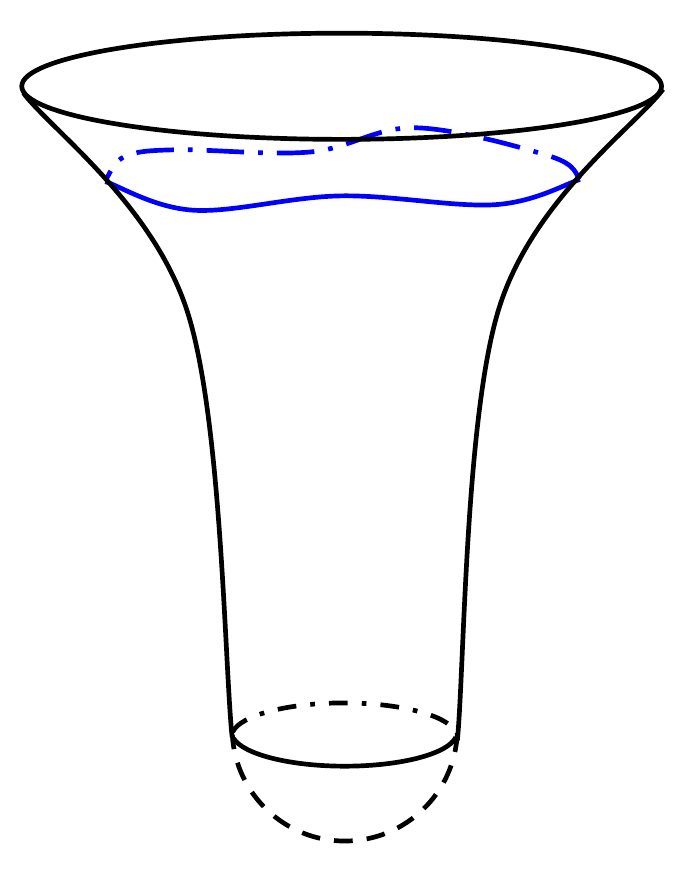}%
  }\hspace{2cm}%
\caption{Different configurations of $dS_2$ related by an asymptotic symmetry.  }
\label{fig:dSconfiguration}
\end{figure*}

As in $AdS_2$, 
$dS_2$ has an asymptotic symmetry group corresponding to reparametrizations, 
\be \la{AsympSym}
x \to  \tilde x( x) ,~~~~~~~\eta  \to \tilde \eta =  \eta  { d \tilde x \over d x }   
\ee
where $x$ and $\eta$ are the coordinates in \nref{Poincare}.
The growing dilaton breaks this symmetry and leads to change in action for two configurations related by this transformation. This change can be computed by doing the path integral over $\phi$ in the bulk, which sets the metric to be exactly $dS_2$. We also fix the value of $\phi$ at the boundary. Then the transformations \nref{AsympSym} act on the shape of the boundary in $dS_2$ (see figure \ref{fig:dSconfiguration}). 
More explicitly, we can think of the boundary curve as parametrized by a function $x(u)$ where 
$x$ is the spatial coordinate  in \nref{Poincare} and $u$ is proper time. Then $\eta$ is fixed by the condition
\be
 du^2 = { dx^2 \over \eta^2} \longrightarrow  \eta = - { d x \over d u } 
 \ee
 for very small $\eta$. In addition we are assuming that we will look at large proper distances $u \propto 
 1/|\eta|$ corresponding to fixed comoving coordinate intervals in $x$.  
Then the extrinsic curvature term gives 
\be \la{actw}
 i S  =  - i 2 \phi_b \int K  = - i 2 \phi_b \int du  +  i 2 \phi_b \int du \{x(u) , u \} 
\ee
for small $\eta$. 
The first term gives a result proportional to the total length. This term says that the universe is expanding, since the momentum conjugate to $\phi_b$ is $- i \partial_{\phi_b} = \pi_{\phi_b}     \propto  -   \dot \rho$ where $\rho$ is the conformal 
factor of the metric, see \nref{Momentum}. By acting with $ - i \partial_{\phi_b}$ on \nref{actw} we see that we get 
an expanding solution. 
 The second term in \nref{actw}  involves the Schwarzian 
derivative, 
\be
 \{x , u \} = {  x''' \over x'}  - {3\over 2 }  { {x''}^2 \over {x'}^2}
 \ee
  and it depends on the shape of the boundary curve. It is 
   the first term violating the asympotic symmetry
\nref{AsympSym}  of exact $dS_2$.

\subsection{The wavefunction of the universe }  \label{sec:HHWavefunction}

Here we consider the computation of the  wavefunction of the universe, 
\be \la{WdW}
\Psi_+ = \sum_{\rm Geometries} e^{ i S } 
\ee
where the sum is over geometries that end on the particular spatial slice and have a large 
value of the dilaton on that slice. The plus means that we will be focusing on the part of the wavefunction that contains the expanding universe. We can call 
this the ``positive frequency" part of the wavefunction, as we will discuss in more
detail below. 
The  sum in \nref{WdW} also 
 contains a path integral over the boundary modes,  $x(u)$,  with the action \nref{actw}. This is the same as what
we had in the case of nearly-$AdS_2$  \cite{Jensen:2016pah,Maldacena:2016upp,Engelsoy:2016xyb}. 
 
In addition, the computation of the wavefunction of the universe should include some prescription for
how to deal with the geometries at early time. One such prescription is the Hartle-Hawking no boundary prescription \cite{Hartle:1983ai}. 
This consists of summing over geometries that have been continued into the  ``positive imaginary time" direction and have no boundaries at early times. 
It is possible for the geometry to have no boundary because of the continuation to Euclidean space. What we mean by  ``positive imaginary time" direction is that it selects the vacuum for the high frequency modes of the fields. It is the extension of Feynman's $i\epsilon$ prescription\footnote{ Other possibilities for the wavefunction of the universe have been proposed, for example, the tunneling proposal 
\cite{Vilenkin:1986cy,Vilenkin:1994rn,Vilenkin:2018dch}. It would be interesting to explore it in this two dimensional context.}. 

In our perspective, as opposed to  \cite{Hartle:1983ai}, we focus only on the positive frequency part and we will use the Klein-Gordon inner product. In addition, we will only apply \nref{WdW} for a 
boundary in the asymptotically expanding region. As we discuss in more detail in 
appendix \ref{AdSPartition},  this is analogous to the way we treat the $AdS$ problem (for another perspective see also \cite{Freidel:2008sh}). 

\begin{figure}[h]
\begin{center}
\includegraphics[scale=.45]{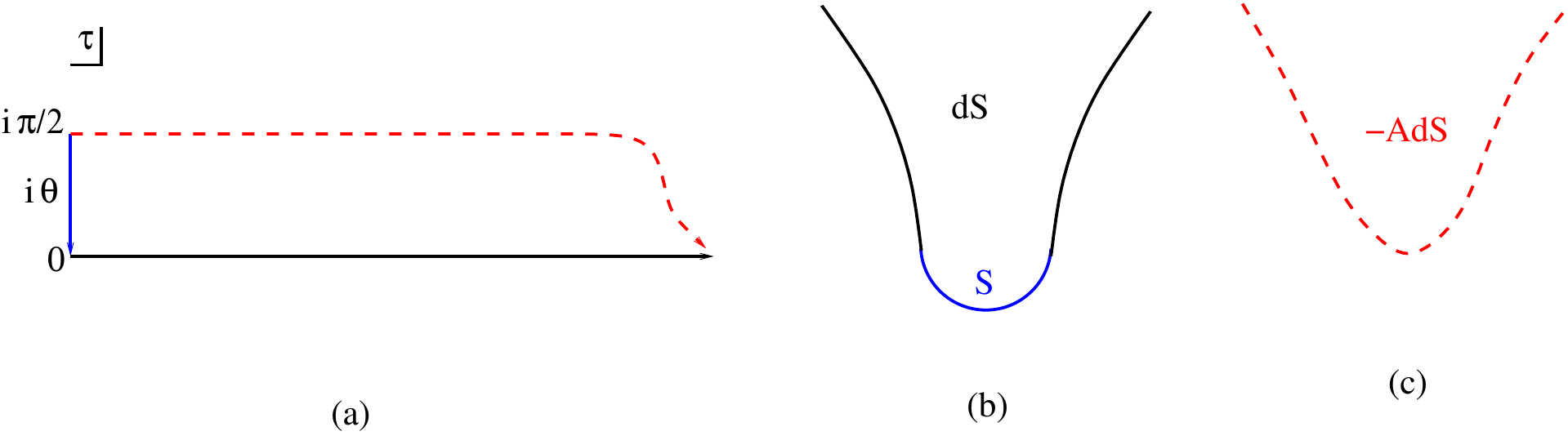}
\caption{ In (a) we see the integration contour in the complex $\tau$ plane, where $\tau$ is the time coordinate in 
 \nref{Global}. The contour starts at $\tau = i \pi/2$. The usual choice is to run it down along $\tau = i \theta$, for real $\theta$, to 
 $\tau =0$ and then continue along the real and positive $\tau $ axis.    (b) Picture for the geometry along the traditional contour. First we get a half sphere and then it is joined to half of global de-Sitter. 
 An alternative choice of contour is displayed by the red dashed line 
 where $\tau = i { \pi \over 2} + \tilde \tau$ for real $\tau$. (c) For real $\tilde \tau$  the metric is minus the metric in hyperbolic space. }
\label{HHContour}
\end{center}
\end{figure}

A simple example of such a geometry is the following. 
Imagine that we fix the future slice to be a circle of a size $\ell$. 
Then we can consider the lorentzian geometry \nref{Global} from the large times $\tau$ down to 
$\tau =0$. Then we can join this to its euclidean continuation 
\be \la{HHusual}
ds^2 = d\theta^2 + \cos^2 \theta ~d\varphi^2,~~~~~\phi = i \phi_g  \sin \theta ~,~~~~~~ \tau = i \theta  ~,~~~~
 0 \leq \theta \leq { \pi \over 2 } ~,~~~~\varphi \sim \varphi + 2 \pi
\ee
Then the geometry shrinks smoothly at $\theta = \pi/2$, and has no boundary. See figure \ref{HHContour}(b). 
An important point to note is that $\phi_g$ is real if we want the future geometry to end on a real value of 
$\phi_b$. However, we see that in Euclidean space we have an imaginary value for $\phi$. When 
$dS_2$ comes from a higher dimensional theory, this means that the higher dimensional geometry is complex. 
This is not a problem and it is consistent with the no boundary prescription as  discussed in e.g. 
\cite{Halliwell:1984eu} and more recently in \cite{Hertog:2011ky}.

In conclusion, in the classical approximation, we can use this analytically continued geometry to get
\be \la{ClaWdW}
\Psi_{+ \rm Classical}  \sim  \exp\left[  - i 2 \phi_b \ell  +  4 \pi \phi_0  + i  { 4 \pi^2  \phi_b\over \ell} \right]  ~,~~~~~~ S_{0,dS} = 8 \pi \phi_0 
\ee
The term involving the de-Sitter entropy comes from the first term in \nref{ActdS}, and from the Euclidean part of the geometry. The term going like $1/\ell$ comes for the classical solution for the Schwarzian action, after inserting $x(u) = \tan( { \pi u \over \ell }) $ in \nref{actw}.  Equivalently, we could have
had the Schwarzian discussion in the global coordinates \nref{Global} which would have given us 
\nref{actw} with $x(u) = \tan{ \varphi(u) \over 2}$. Then we can set $\varphi = 2\pi u/\ell$. 

The simplicity of two dimensional gravity allows us to compute perturbative gravity corrections
to this result \nref{ClaWdW}. The computation is essentially the same as for Nearly-$AdS_2$ \cite{Stanford:2017thb}. The quantum corrections arise by doing the path integral over the Schwarzian degrees of freedom. To be more precise, this is the $\ell$ dependent part of the quantum corrections. Other quantum corrections, that could arise from other fields moving the bulk, give an answer which is $SL(2)$ invariant and contributes only to $S_{0,dS}$. 

Thus, all perturbative quantum corrections give
\be \la{FinWf}
\Psi_+ \propto    \left( { \phi_b  \over  \ell} \right) ^{3/2}  \exp\left[  - i 2 \phi_b \ell  + { 1 \over 2} S_{0,dS}  + i  { 4 \pi^2  \phi_b\over \ell} \right] 
\ee
where the $1/\ell^{3/2}$ arises from a one loop effect and there are no further corrections \cite{Stanford:2017thb}. As in \cite{Stanford:2017thb}, we can express the answer as an integral of the form 
\bea \la{Dens}
\Psi_+ &\propto&   \int_0^\infty  dE \rho(E) e^{ i \ell E} , ~~~~~~~~~ \rho(E) = e^{ S_{0,dS} \over 2 } \sinh\left( 4 \pi  \sqrt{   \phi_b E } \right)
\\ \la{EntDens}
\Psi_+& \propto & e^{ S_{0,dS} \over 2 } \int_{-\infty}^{\infty}  d s s \, e^s \exp\left( i { \ell s^2 \over 16 \pi^2 \phi_b } \right)
\eea
 where in the second equation we defined $s$ as the argument of the $\sinh$ in \nref{Dens} to obtain an integral of the form 
   $\int_0^\infty ds s \sinh s e^{ i c s^2} $. We then extended the range of the integral and used the symmetry of the integrand under
 $s \to -s$.  
 
\subsection{Superspace analysis} 

It is interesting to compare \nref{FinWf} to what we can obtain from a mini-superspace\footnote{We follow the usual tradition of naming the space of all metric ``superspace" (maybe a better name would be  ``megaspace"). 
 Do not confuse this with the superspace that arises in supersymmetric theories.} analysis. Aspects of canonical quantization were discussed in 
 \cite{Henneaux:1985nw,LouisMartinez:1993eh}. Since there are no propagating modes, 
 the minisuperspace analysis is  the same as all of superspace 
since we can fix the gauge where the only variables are a constant value of $\phi$ and the proper length
$\ell$ of the circle (see appendix \ref{App: MomentumConstraint}). This gauge is good when we are in the asymptotic future and we ignore other topologies, etc. We expect it to fail when the size of the circle shrinks to zero, for example.  We write 
 the spacetime metric as $ ds^2 = e^{ 2 \rho} ( -d\eta^2 + d\varphi^2)$, with $\varphi \sim \varphi + 2 \pi$.  
After integrating by parts in \nref{ActdS} we find an action of the form 
 \be \la{Momentum} 
 iS = -  4 \pi  i   \int d \eta [ \dot \phi \dot \rho  + \phi e^{ 2 \rho} ]     ~,~~~~  \longrightarrow ~~~\pi_{\phi} = -
 4 \pi   \dot \rho
\ee
restricting ourself to $\partial_\varphi \phi=0$  and $\partial_\varphi \rho =0$ while keeping the physical degree of freedom.  We still need to impose the Hamiltonian constraint. 
The action implies that the momentum conjugate for $\phi$ is indeed proportional to  $ -  \dot \rho $ as we remarked earlier. 
The only variables on the wavefunction are a constant value of $\phi$ and 
$\rho$, $\Psi[ \phi, \rho]$.  
The Hamiltonian constraint contains a term that looks like the wave 
equation 
\be \la{WdWE}
 \partial_{ \phi} \partial_\rho \Psi + 16 \pi^2   \phi  e^{ 2 \rho } \Psi =0 ~,~~~~ {\rm or} ~~~ [4 \partial_u \partial_v + 4 ] \Psi =0~,~~~~u \equiv  \phi^2 ~,~~~~~v \equiv  (2\pi)^2 e^{ 2 \rho} 
 \ee
 After using that our previous definition of $\ell$ was such that $\ell = 2 \pi e^\rho$,
  we see that we get  a wave equation with positive square mass in Minkowski space  
 \bea ds^2 & =& -du dv = -d T^2 + T^2 d \sigma^2 , ~~~~  \la{SuperSpace}
 \\
  & ~&{\rm  with }~~~ T = \sqrt{uv} = \phi 2\pi e^{ \rho} = \phi \ell ~,~~~~~~~e^{\sigma} = \sqrt{ v \over u} = {  2 \pi e^\rho \over \phi  }   = { \ell \over \phi }
 \eea
 In other words, the superspace is flat Minkowski space. More precisely, the region where we trust this superspace description is the far future of the its   
  Milne wedge, $ T\gg 1$. 
 We are interested in the asymptotic form of the solutions for large $T$. Writing the wavefunction as 
 \be 
 \Psi \sim  e^{ i m \sigma } F_m(T)
 \ee
  we find that \nref{WdWE} 
  becomes a Bessel function for $F$. Then, after expanding for large $T$ we find that the  general solution 
  for large $T$ has the form 
  \be  
  \Psi = { 1 \over \sqrt{T} } e^{ - 2 i T } f(\sigma) + { 1 \over \sqrt{T}} e^{ + 2 i T} \tilde f(\sigma)  ~,~~~~~~~T\gg 1 \la{AsFor}
  \ee
  where $f$ and $\tilde  f$ are arbitrary functions which are not fixed by looking purely at the large $T$ region. 
  In principle, they should be found by looking at the solutions in the full superspace with the appropriate boundary conditions. Notice that at $v=0$ the circle shrinks to zero size and we will need special boundary conditions there. 
  The expanding, or ``positive frequency" solutions, correspond to the first term in \nref{AsFor}, and we will focus purely on that term from now on. 
  We see that the result in \nref{FinWf} corresponds to a particular 
  function $f \propto  e^{-\sigma } \exp[ i  4 \pi^2 e^ {-\sigma } ]$ which then gives \footnote{Note added in v4: Recently \cite{Cotler:2024xzz} pointed out there should be an extra ${1\over \sqrt{\ell}}$ factor in front of the wavefunction. We discuss this correction in appendix \ref{App:corr}.}
  \be \la{WdWtot}
  \Psi_+ = { 1 \over \phi_b } \left( {\phi_b \over \ell } \right)^{3\over 2 }  \exp\left[  - i 2 \phi_b \ell  + 4\pi \phi_0  + i  { 4 \pi^2  \phi_b\over \ell} \right]  ~,~~~~~\phi_b \ell \gg 1 
  \ee
  where we have indicated that this is the asymptotic form of the solution in the 
  large $T$ region. Note that we determined this solution from \nref{WdW}, rather than by solving the wave equation \nref{WdWE} in the full space. 
 
\begin{figure}
\begin{center}
\includegraphics[scale=0.35]{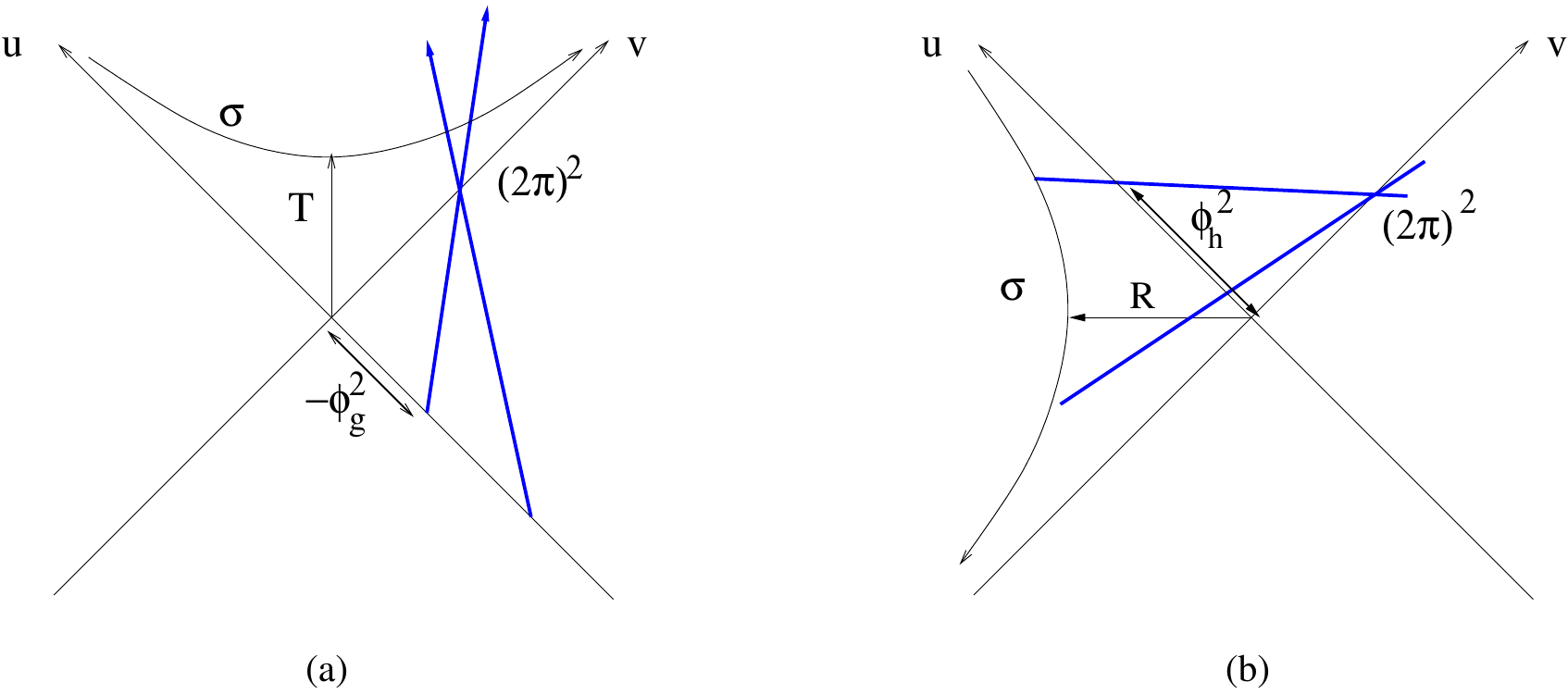}
\end{center}
\caption{ (a) The superspace for the JT gravity theory after a partial gauge fixing is just the Minkowski space parametrized by $u$ and $v$. The late time region 
corresponds to $T\gg 1$. Classical solutions correspond to the usual straight lines for classical particles in Minkowski space. The condition that space shrinks smoothly corresponds to a boundary condition at $v =0$ which sets the momentum to be such that all classical trajectories go through the point $u=0$, $v= (2\pi)^2$. 
The region $u< 0$ corresponds to imaginary $\phi$ and arises after euclidean continuation of the solution. We display two possible classical trajectories. (b) We show the same space for the Euclidean AdS JT gravity theory analyzed in appendix \ref{AdSPartition}. The asymptotic region corresponds now to the left Rindler wedge.  We also display the trajectories of the classical solutions.}
\label{SuperSpaceFig}
\end{figure}

It is interesting to note that the classical solutions \nref{Global} become
simply   
 straight lines in Minkowski space 
  \be \la{Slope}
  e^\gamma u - e^{-\gamma} v = {\rm const}  \longrightarrow e^{\gamma } = { 2 \pi \over \phi_g } = { \ell \over \phi_b} 
  \ee
  where $\gamma$ is the rapidity or ``boost angle'' of the classical particle trajectory, and we used the classical solution \nref{Global}. 
  Notice that the region $u> 0$ corresponds to the Lorentzian region, while we move towards $u< 0$ with \nref{HHusual}. The no boundary proposal says that we always have $v> 0$ and we have no interpretation for the region $v< 0$. 
  The boundary condition at $v=0$ is that the solution shrinks smoothly, which implies a condition on the (euclidean) conformal time derivative of the scale factor, $\partial_\theta \rho =-1$. This in turn fixes the momentum conjugate to $\phi$, and then implies that $\pi_u = -{ 2\pi\over  \phi_g}$ where $\phi_g$ is related to the
  value of $\phi$ at the tip of the sphere,  \nref{HHusual}.  This, together with \nref{WdWE}, sets the slope \nref{Slope} and implies that 
  the trajectory crosses $u=0$ at $v = (2\pi)^2$. So, all classical trajectories pass through this point, see  figure \ref{SuperSpaceFig}(a).
   This is a consequence of the no-boundary proposal. 
   Of course, the wavefunction \nref{WdWtot} gives us a particular superposition of trajectories or more precisely a wavefunction for a relativistic particle on Minkowski space. 
   
   We could wonder whether the wavefunction \nref{WdWtot} can be reproduced by a simple computation for a massive field in Minkowski space. Since all classical trajectories pass through the point $(u,v) =( 0, (2\pi)^2 )$, it is natural to conjecture that the wavefunction could be given by the insertion of a field operator at this point. Just by looking at the answer,  \nref{WdWtot}, we find that it matches what 
   we would have if we had inserted an operator $ \partial_{v} \Phi(u,v)|_{u=0,v=(2\pi)^2}$, where
   $\Phi$ is the field operator for the massive field in Minkowski space\footnote{The exact expression is $\Psi_+ =i\sqrt{u\over v-(2\pi)^2-i\epsilon}K_1(2i\sqrt{u(v-(2\pi)^2-i\epsilon)})$. It is interesting that the Feynman $i\epsilon$ prescription gives us both growing wavefunction in the forbidden region and an expanding universe.}. 
   We do not know the significance of this fact. In particular, note that the boundary conditions for the no-boundary wavefunction were imposed at the line $v=0$, and not this particular point! 
   Note also that once we include other topologies, as we will discuss in section \ref{OtherTopologies}, we will have other contributions to the wavefunction. 
   
  Note that \nref{WdWtot}, compared to \nref{FinWf},  has an additional factor of $1/\phi_b$, which is 
  $\ell $ independent. 
    In general, the two dimensional Klein Gordon inner product
takes the form 
\be
\langle \bar \Psi , \Psi \rangle_{KG} =  i \int * ( \bar  \Psi d \Psi - \Psi d \bar \Psi ) 
 \ee
 where the integral is over a codimension one slice. As usual, to get a non-zero and positive answer we should consider just the  ``positive" frequency components, which in terms of the asymptotic form of the solutions in \nref{AsFor} corresponds to considering only the first term. This is natural if we view these wavefunctions in the context of a ``third quantization" where they are multiplying universe creation and annihilation operators. In any case, we will only be applying these formulas for very large $T$ where the notion of positive and negative frequencies is well defined and it is related to whether we consider an expanding or contracting universe.  
 In our case we can take slices with constant $\phi_b$ and the integral becomes \cite{Strobl:1993yn} 
 \be  
 \langle \bar \Psi , \Psi \rangle_{KG} =   i\int d\ell    \left(  \bar \Psi \partial_{\ell}  \Psi - \Psi \partial_{\ell} \bar  \Psi \right) \la{KGProd}
  \ee
  The derivative with respect to $\ell$ gets its leading contribution from the term $e^{- 2i \phi_b \ell}$ and
  produces a factor of $\phi_b$. As usual, we see that the Klein Gordon norm gives a positive value when we 
  consider a wavefunction with a purely ``positive'' frequency part. If we had a wavefunction with both 
  positive and negative frequencies we would need to project on to the positive frequency part before 
  computing the Klein Gordon norm.  
In summary, the final expression involves an integral of the form 
\be \la{FinalProb} 
\langle \bar \Psi , \Psi \rangle_{KG} =  e^{ S_{0,dS} }  \int { d\ell \over 
\phi_b }  { \phi_b^3 \over \ell^3 } = e^{ S_{0,dS} } \int { d\hat \ell \over { \hat 
\ell}^3 } ~,~~~~~~~~~~~ \hat \ell = { \ell \over \phi_b  } 
  \ee
  So, we derived  the measure of integration
for $\hat \ell$, up to an overall numerical constant that we have absorbed in $S_{0,dS}$. 

This integral is divergent at small $\hat \ell$. This suggests that the sphere partition function diverges in the JT theory. However, at small $\ell$ we are going 
away from the Schwarzian limit and more information about the exact wavefunction is needed.   In principle, one could attempt a computation of the sphere partition function by performing the functional integral. We have not managed to get a concrete answer due to the presence of zero modes (both bosonic and fermionic, from
ghosts), whose precise treatment we leave to the future.  


The conclusion is that \nref{FinalProb} represents the final probability measure for different sizes for the 
universe. Notice that it depends on $\phi$ and $\ell$ only through their ratio, $\hat \ell$, which is a quantity that is independent of time for late times, while each of them individually is growing in time. 
This computation involves the simplest topology but is valid to all orders in gravity perturbation theory. 

One might be surprised by the factor of $\phi_b$ in \nref{WdWtot} since we did not have such a factor in the computation of the partition function of nearly-$AdS_2$. In appendix \ref{AdSPartition} we explain the connection between Wheeler de Witt wavefunctions and the partition function of the dual field theory.

\subsection{Sum over topologies } 
\la{OtherTopologies} 

In principle, the sum \nref{WdW} includes also a sum over topologies. 
For pure JT gravity with negative cosmological constant this sum was done in \cite{SSS}. 
An important element was that the $\phi$ integral sets the geometry to have negative curvature. Then there are many possible topologies with one boundary. The genus suppression factor was $e^{- 2 S_0}$. 

In our case,  we have positive curvature, so naively we cannot use the same idea. 
Before giving up completely, let us look again at the computation we discussed near \nref{HHusual} 
and figure \ref{HHContour}(a). The usual presentation of this computation corresponds to picking a contour in the 
$\tau$ plane describing the metric in  \nref{Global}. We chose a contour that starts at $\tau = i \pi/2$ and 
runs along the imaginary axis up to $\tau=0$ and then along the real $\tau $ axis, see figure \ref{HHContour}. 
However, we could as well choose a contour that also starts at $\tau= i\pi/2$, but we keep the imaginary 
part fixed and increase the real part. In other words, we set 
\be
\tau = i { \pi \over 2} + \tilde \tau 
\ee
with $\tilde \tau$ real. In these coordinates the metric \nref{Global} takes the form 
\be
ds^2 = - ( d\tilde \tau^2 + \sinh^2 \tilde \tau d\varphi^2 )  \la{MinusMe}
\ee
So the topology looks like a disk, where the circle shrinks smoothly at $\tilde \tau =0$. The overall sign of the metric is negative and this makes this spacetime a solution of the JT action with {\it positive} cosmological constant (see also
\cite{Maldacena:2002vr,Harlow:2011ke}).
\begin{figure}
\begin{center}
 \begin{tikzpicture}[scale=1]
 \node[inner sep=0pt] (russell) at (0,0)
    {\includegraphics[width=.7\textwidth]{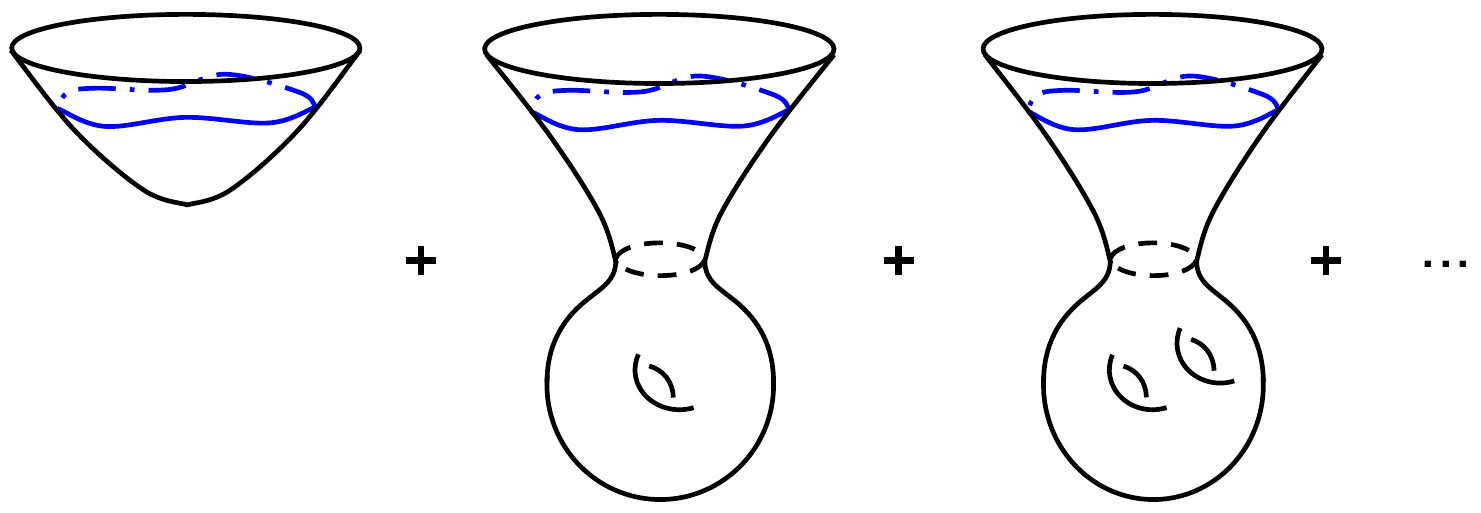}};
    \draw (-4.39,-0.1) node {\small $-AdS$};
    \draw (0.1,0) node {\small $b$};
     \end{tikzpicture}
    \end{center}
    \vspace{-0.2cm}
\caption{Sketch of the sum over ``negative" trumpet geometries \eqref{CoshGeo} \cite{SSS}. To the left we have $-AdS$ as in figure \ref{HHContour}(c) and we add contributions with different topologies, also integrated over the size $b$ of the throat. }
\label{fig:trumpetSSS}
\end{figure}

This inspires us to consider instead geometries that have the form 
\be \la{CoshGeo}
ds^2 = - ( d\tilde \tau^2 + \cosh^2 \tilde \tau d x^2 ) 
\ee
where we set $x\sim x + b$, see figure \ref{fig:trumpetSSS}. This is a trumpet like geometry, as the ones considered in \cite{SSS}. We can cut this geometry at $\tilde \tau=0$ and join a genus $g$ constant negative curvature Riemann surface with a single boundary with proper length $b$. The metric is just minus the metric considered in \cite{SSS}. When we integrate over the moduli space of Riemann surfaces we do not expect this overall minus sign to give any difference. The only difference will be in the action of the trumpet and the integral over the Schwarzian modes, which will have a difference due to the extra $i$, as we had for the disk.  
The geometry \nref{CoshGeo} is singular when continued to the original coordinate $\tau$, giving $ds^2 = -d\tau^2 + \sinh^2\tau dx^2 $, with a
singularity at $\tau=0$. The geometries we discussed above are non-singular and represent an analytic continuation used for the purpose of defining the no boundary wavefunction for these geometries.

So the final conclusion is that the sum over all single boundary topologies gives us 
\be \la{PartFu}
\Psi_+ =  Z_{SSS}\Big( { \beta \over \phi_b } \to - i { \ell \over \phi_b } ;  S_0 \to \half S_{0,dS} \Big)  \sim \langle Tr[ e^{ i \hat \ell H } ] \rangle 
\ee
where $ Z_{SSS}( { \beta \over \phi_b }   ;  S_0  ) $ is the sum obtained in \cite{SSS}.  Even though higher genus contributions are suppressed by factors like $e^{ - g S_{0,dS}}$, the sum over genera is divergent.  In \cite{SSS} a possible completion 
was chosen. In this completion, the result can be viewed in terms of a random 
hermitian matrix theory, where the matrix is the Hamiltonian of the $AdS_2$ problem. 
 Here we have a similar result, except that \nref{PartFu} can be viewed as 
 $Tr[ e^{ i \hat \ell H } ] $ instead of $Tr[ e^{ -\beta H} ]$, where $H$ is the random hermitian matrix. 
What was the Hamiltonian in the Nearly-$AdS_2$ case becomes an operator performing space translations
along the spatial boundary in the $dS_2$ problem. If we view Milne space \nref{Milne} as describing a black hole, then $H$ seems to be related to (minus) its energy. 
 
 When we consider $\langle \bar \Psi_+ , \Psi_+ \rangle $ we get two copies of
 the system, and we can now get geometries that connect the two sides (the
 dS analog of \cite{Saad:2018bqo}). These are interesting new geometries in the de-Sitter context that we plan to explore in more detail in the future. 
 But the sum over all such connected geometries corresponds to computing 
 $\langle Tr[e^{ i \hat \ell  H} ] Tr[e^{ - i \hat \ell H }] \rangle$. 
 The idea of averaging over Hamiltonians looks somewhat natural in de-Sitter. 
 It could correspond to the following. Assuming that some form of dS/CFT is true, we find that that the wavefunctions would be dual to some field theory. 
 That correspondence would be precise if we wait to the infinite future. However, if we put a finite (but large) time cutoff, then we do not know what will happen in the future, and this could give rise to an average over the precise couplings of the boundary  ``CFT". It is conceivable that in the case of pure
 JT gravity this boils down to an average of Hamiltonians as was found in 
 \cite{SSS}.

\section{Nearly $dS_2$ from four dimensional gravity} 
\la{4dSec}

 In this section we will discuss a particular four dimensional theory where the 
 two dimensional 
 gravity that we discussed arises in some limit. The general $D$ dimensional case is very similar
 and discussed in appendix \ref{GeneralD}. 
 
 \subsection{Four dimensional gravity and the Schwarzschild de-Sitter solution}
 
We start with the standard four dimensional gravity action with a positive cosmological constant 
$\Lambda = 3/R_{dS}^2$  
\be
S_L =   { R_{dS}^2  \over 16 \pi G_N}  \left[ \int \sqrt{g} (R- 6)  - 2 \int K \right],
\ee
 where we rescaled the metric by an overall factor of $R_{dS}^2$ so that now the simplest  classical solution is de Sitter space with unit radius. 
 
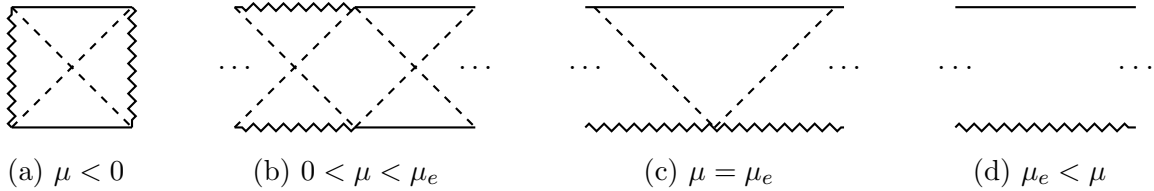
\begin{figure} 
\begin{center}
\begin{tikzpicture}[scale=0.4]
\draw[thick,dashed] (-2,2) -- (2,-2);
\draw[thick,dashed] (-2,-2) -- (2,2);
\draw[thick,decoration = {zigzag,segment length = 2mm, amplitude = 0.5mm},decorate] (-2,-2) -- (-2,2);
\draw[thick,decoration = {zigzag,segment length = 2mm, amplitude = 0.5mm},decorate]  (2,2) --  (2,-2);
\draw[thick] (-2,2)--(2,2);
\draw[thick] (-2,-2)--(2,-2);
\draw (-0.2,-3.5) node {\small (a) $\mu<0$};
\end{tikzpicture}\hspace{0.8cm}
\begin{tikzpicture}[scale=0.4]
\draw[thick,dashed] (-2,2) -- (2,-2);
\draw[thick,dashed] (-2,-2) -- (2,2);
\draw[thick,dashed] (-6,2) -- (-2,-2);
\draw[thick,dashed] (-6,-2) -- (-2,2);
\draw[thick,decoration = {zigzag,segment length = 2mm, amplitude = 0.5mm},decorate] (-2,2) -- (-6,2);
\draw[thick,decoration = {zigzag,segment length = 2mm, amplitude = 0.5mm},decorate]  (-2,-2) --  (-6,-2);
\draw[thick] (-2,2)--(2,2);
\draw[thick] (-2,-2)--(2,-2);
\draw (2,0) node {\small$\ldots$};
\draw (-6,0) node {\small$\ldots$};
\draw (-2.3,-3.5) node {\small (b) $0<\mu<\mu_e$};
\end{tikzpicture}\hspace{0.6cm}
\begin{tikzpicture}[scale=0.4]
\draw[thick,dashed] (-4,2) -- (0,-2);
\draw[thick,dashed]  (0,-2) --  (4,2);
\draw[thick] (-4.3,2)--(4.3,2);
\draw[thick,decoration = {zigzag,segment length = 2mm, amplitude = 0.5mm},decorate] (-4.3,-2)--(4.3,-2);
\draw (4.3,0) node {\small$\ldots$};
\draw (-4.3,0) node {\small$\ldots$};
\draw (-0.2,-3.5) node {\small (c) $\mu=\mu_e$};
\end{tikzpicture}\hspace{0.6cm}
\begin{tikzpicture}[scale=0.4]
\draw[thick] (-3,2)--(3,2);
\draw[thick,decoration = {zigzag,segment length = 2mm, amplitude = 0.5mm},decorate] (-3,-2)--(3,-2);
\draw (3,0) node {$\ldots$};
\draw (-3,0) node {$\ldots$};
\draw (-0.2,-3.5) node {\small (d) $\mu_e<\mu$};
\end{tikzpicture}
\end{center}
\vspace{-0.5cm}
\caption{Penrose diagrams for a black hole in dS$_4$ as a function of the mass parameter $\mu$. The dots on the sides indicate the possibility of continuing the diagram or identifying the endpoints. }
\label{Penrosefd}
\end{figure}

Let us recall the Schwarzschild de-Sitter solution 
 \be \la{SchdS4}
 ds^2 = - f d\tau_r^2 + { d\rho^2 \over f} + \rho^2 d\Omega_2^2 ~,~~~~~~f = 1 -\rho^2 - { \mu \over \rho } 
 ~, ~~~ \mu = 2 M G_N/R_{dS}^2 
 \ee
  Penrose diagrams   for various ranges of $\mu$ are shown in 
  figure   \ref{Penrosefd}. For the case in figure \ref{Penrosefd}(b) we have a 
  static patch region, with $f>0$, bounded by two horizons, 
   where $f(\rho)=0$;  the black hole horizon $\rho_+$ and the cosmological horizon $\rho_c$, 
 with $0 < \rho_+ < \rho_c < 1$.   For $\rho_c < \rho $,  
 $\tau_r$ is spacelike and we are in  an expanding region.  For this
 range of $\mu$, the solution can be viewed as a pair of black holes in four dimensional de-Sitter. 
 
  If we go to Euclidean time, so that
 $\tau_r$ is purely imaginary, and set its period to $\beta$, $\tau_r \sim \tau_r - i \beta$, with 
 \be \la{betva}
 \beta = { 4 \pi \rho_+ \over1 - 3  \rho_+^2  }  ~,~~~~~~~1- \rho_+^2 - \mu/\rho_+=0 ~,~~~
 \ee
then the Euclidean circle shrinks smoothly as $\rho \to \rho_+$. 
Just for later purposes let us record here the values of the mass and entropy of the black hole
in terms of $\rho_+$ 
\be
 M =  \half { R_{dS}^2 \over G_N}  ( \rho_+ - \rho_+^3 ) 
   ~,~~~~~~~~S = {\pi  R_{dS}^2   \over   G_N} \rho_+^2  \la{MaEnt}
   \ee
   The entropy here is only the entropy coming from the black hole horizon, excluding any contribution from the cosmological horizon. 
   
If we demanded that the euclidean solution is smooth at the cosmological horizon we would have obtained 
the same equation as \nref{betva} but with $\beta \to -\beta$. ~\footnote{ It is not possible to remove the singularity at both horizons  with the same $\beta$.}

We see that the temperature goes to zero, or  $\beta \to \infty $, when 
$ \rho_+ \to \rho_e $, with 
\be \la{ExtVa}
\rho_e = { 1 \over \sqrt{3} }  ~,~~~~~~~~ \mu\to \mu_e = { 2 \over 3 \sqrt{3} } ~,~~~~~~f \to - { 3 } (\rho - \rho_e)^2 ~,~~~~{\rm for} ~~ \rho \sim \rho_e
\ee
At this point,  the black hole horizon and the cosmological horizon have the same area, 
$ \rho_+ = \rho_c = \rho_e$. However, they are separated by a non-zero proper distance because $f$ is also going to zero near $\rho\sim \rho_e$.
 In fact, in this region,  the metric becomes $dS_2 \times S^2$. 
This solution is sometimes called an 
``extremal'' black hole or ``Nariai'' solution \cite{Nariai}\cite{Ginsparg:1982rs} (see also \cite{Bousso:1995cc,Bousso:1996au} for applications in inflation). This is the maximal mass (and entropy) black hole that we
can have in de Sitter space.

If the value of $\mu$ is slightly below $\mu_e$ then we get a long region where the metric is 
approximately $dS_2 \times S^2$ which then transitions to a full four dimensional de-Sitter solution. 
In the nearly $dS_2$ region the metric is described by 
\be  \la{AppMe}
d s^2 = \rho_e^2 ds^2_{dS_2} +  \rho_e^2(1 + \delta )^2 d\Omega_2^2  ~,~~~~~ \delta = { \rho \over \rho_e } -1 
\ee 
The metric describing the transition to the $dS_4$ region can be  approximated by  \nref{SchdS4} with $\mu \to \mu_e$.
The number of e-folds during the nearly-$dS_2$  region is given by 
\be
{\cal N}^{dS_2}_{\rm e-folds} \sim \log \beta \sim - { 1 \over 2 } \log |\mu_e -\mu |
\ee
up to additive constants.

  We conclude that, for very large $\beta$, there is a region of the geometry where we can perform a 
  Kaluza Klein reduction to two dimensional nearly-$dS_2$ gravity. 
   The effective two dimensional 
gravity action becomes \nref{ActdS} with  
\be \la{ParamJT}
\phi_0 = { R_{dS_4}^2 \rho_e^2   \over 4 G_N } ~,~~~~~~~ \phi = 2 \phi_0  \delta 
\ee
where $\delta$  is  defined in 
\nref{AppMe}.

 We can also consider solutions where $\mu> \mu_e$, see figure \ref{Penrosefd}(d). 
 These solutions have a singularity in the past.   Again, when $\mu$ is very close to $\mu_e$,  the solutions have a long period where 
 the metric looks like a nearly $dS_2$ gravity configuration, of the form in  \nref{Global}. 
 
 Finally, for $\mu< 0$ we get solutions as in figure \ref{Penrosefd}(a) which 
 contain a naked singularity in the static patch region. 
 
 Below, we will be performing various analytic continuations of this basic solution. 

  \subsection{The classical no-boundary wavefunction for $S^1 \times S^2$ }
 
 We start by reviewing 
 the computation for these wavefunctions that was 
 done in section 5.2 of \cite{Anninos:2012ft} (this was studied originally in \cite{Laflamme:1986bc}, see also \cite{Castro:2012gc}, \cite{Banerjee:2013mca}, \cite{Das:2013mfa} and \cite{Conti:2014uda}). First we will describe
 the geometries that give rise to saddle points. There are four saddle points. 
 We will argue that two of them do not satisfy the no-boundary conditions because they go into the wrong half of the complex plane. Whether the other two contribute or not depends on the contour of integration. We will discuss various
 plausible choices below. 
  
 In this section we will discuss the computation of the no-boundary wavefunction for  three geometries 
 that asymptote to $S^1 \times S^2$, or 
 \be \la{AsyMe}
 ds^2 \sim     - { d \rho^2  \over \rho^2} + \rho^2 ( d \tau_r^2 + d\Omega_2^2 )  ~,~~~~~~~\tau_r  \sim  \tau_r + \lambda ~,~~~~\rho \to \infty
 \ee
 The metric \nref{SchdS4} has this asymptotic form and    $\tau$ is a spacelike   for large $\rho$. 
 Here we are also compactifying   $\tau_r \sim \tau_r + \lambda$  so that we have a circle at late times. We see that $\lambda$ has the interpretation of the proper length of the circle at late times in units of the radius of the sphere. Both the size of the sphere and the size of the circle are expanding, but their ratio, $\lambda$, is staying fixed at late times. 
 
 This identification   introduces a singularity at the cosmological horizon of the static patch when $\mu < \mu_e$. 
 For $\mu > \mu_e$ the metric was singular in the past, and compactifying the $\tau_r$ direction does 
 not introduce an additional singularity. 
 So, all the real metrics are singular. Nevertheless we can still apply the no boundary proposal 
 because this proposal instructs us to continue $\rho $ in the positive imaginary direction and find a 
 complex geometry with no boundary. 
 A simple example of such geometries is the following. We set $\rho = i r$ and write the metric 
 \be \la{mSchAdS4}
 ds^2 = - [   f d\tau_r^2 + { dr^2 \over  f } + r^2 d\Omega_2^2 ] ~, ~~~~f =  1 + r^2 - { \tilde \mu \over r } ~,~~~~ \tilde \mu  = - i \mu 
 \ee
 Up to an overall sign this looks like the metric for the Euclidean 
 $AdS$ Schwarzschild black hole (when $\tilde \mu$ is real). In this case,  the period,  $\lambda$, of the $\tau_r$ circle is
 related to $r_+$ via 
 \bea
 \lambda &=&  { 4 \pi r_+ \over 1+  3 r_+^2   } ~,~~~~~~\tilde  \mu = r_+ (1 + r_+^2)  ~,~~~~~{\rm or } ~~ \la{LeAdS}
\\
 i \lambda &=& { 4 \pi \rho_+ \over 1 - 3 \rho_+^2  } ~,~~~~~~  \mu = \rho_+   (1 - \rho_+^2) \la{LedS}
 \eea
and $\lambda$ has the interpretation of the inverse temperature of the $AdS$ black hole. We get these
geometries by thinking of the no boundary proposal along a contour like the one in dashed lines in 
figure \ref{HHContour}(a). Notice that due to the overall minus sign in \nref{mSchAdS4} this is a 
solution of the Einstein equations with {\it positive} cosmological constant. Normally we require that
physical solutions have positive metrics. However, in computing the no boundary proposal,  we are 
doing an analytic continuation,  and along the dashed path in figure \ref{HHContour}(a)  we actually  
have a negative metric. 

Evaluating the de-Sitter action on this geometry,  and being careful both with the boundary terms (see appendix \ref{Cones}), we find (as in \cite{Hawking:1982dh,Witten:1998zw})
\bea\la{OffSha}
iS &=& iS_ {\rm Large} + iS_{\rm Finite} 
\\
iS_{\rm Large} &=& - i  { R^2_{dS}  \over   G_N}\lambda  ( \rho_b^3 - { \rho_b \over 2 } ) = - i 
{ R^2_{dS} \over 4 \pi G_N } \left[  \int \sqrt{g^{(3)} } \left( 1 - { R^{(3)} \over 4 } \right) \right] 
\\ 
& ~& 
\cr
iS_{\rm Finite}& =& { \pi R_{dS}^2 \over   G_N }   { r_+^2 (1-r_+^2 ) \over 1 + 3 r_+^2 }  =-
{ \pi R_{dS}^2 \over  G_N }   { \rho_+^2 (\rho_+^2 +1) \over 1 - 3 \rho_+^2 }   ~,~~~ \rho_+ = i r_+  \la{rhopl}
\eea
The large terms give a rapidly oscillating  phase 
 in the wavefunction of the universe and  are  the ones responsible for giving the expansion of 
the universe. Here $\rho_b$ is the value of $\rho$ at the future slice where we are evaluating the wavefunction of the universe. 
 We have also expressed the large terms 
 as a function of the local metric and curvature of the three-metric of that slice. 
  Here we will be interested in the finite terms\footnote{  
  $iS_{\rm Finite} $ is minus the final value that we would have had for $\log Z$ if we were doing the   $AdS$ problem.}.

\begin{figure}[h]
\begin{center}
 \includegraphics[scale=.45]{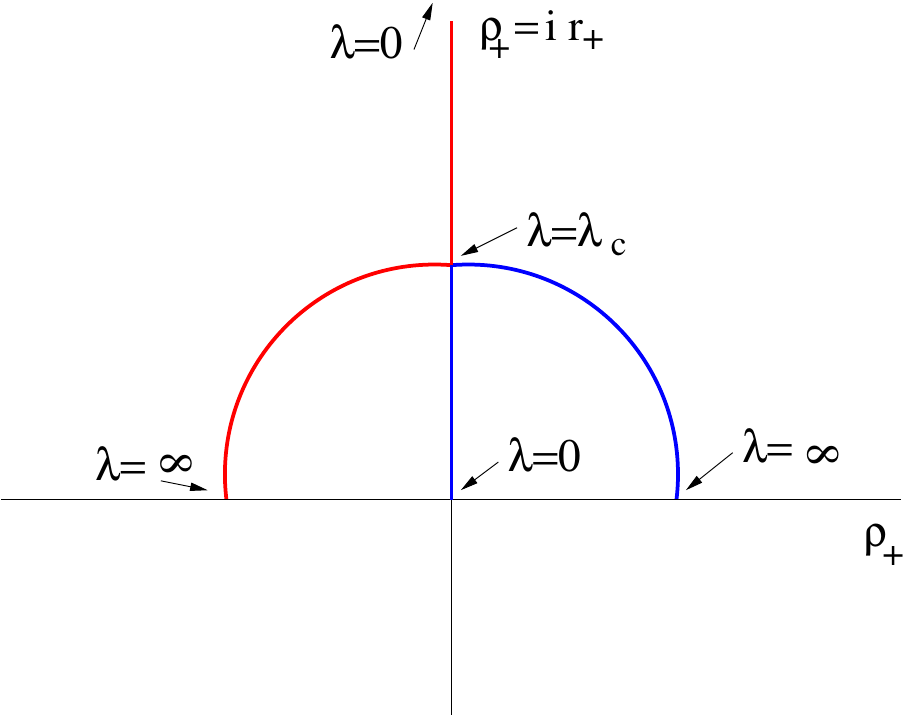}
\caption{ We plot the   values of $\rho_+$ for the allowed saddle points as we change $\lambda$. 
For each $\lambda$ there are two solutions one on the red line (the left and top line) and one on the
blue line (the right and bottom line). We suspect that the blue ones are the physically relevant solutions. 
 For $\lambda> \lambda_c$ we move along the circle, reaching the real axis at $\lambda = \infty$. For $\lambda < \lambda_c$ we move along the purely imaginary axis towards
 the origin, which we reach as $\lambda\to 0$.   }
\label{Saddles}
\end{center}
\end{figure}

Now, for each value of $\lambda$ in \nref{LeAdS} there are two values of $r_+$ or equivalently two 
values of $\rho_+$. For $\lambda < \lambda_c = 2 \pi/\sqrt{3} $, these two values of $r_+$ are purely real, or 
$\rho_+$ purely imaginary. While for $\lambda > \lambda_c$ they are both complex. 
Now, in all these cases the value of $\mu$ of the corresponding de-Sitter solution is complex. This means that these geometries cannot be interpreted as real geometries but rather as complex solutions that are only used to compute the wavefunction of the universe according to the no-boundary proposal. 
It is convenient to solve the equation for $\rho_+$  
\bea \la{RhoPE}
{ \rho_+ \over \rho_e} & = & { i \over y }  \pm  \sqrt{1- { 1 \over y^2}  }  ~,~~~~~~y = { \lambda  \over \lambda_c } ~,~~~~~\lambda_c = { 2 \pi \over \sqrt{3}}
\eea
and $\rho_e$ is the extremal value in \nref{ExtVa}.
In figure \ref{Saddles} we plot the values of $\rho_+/\rho_e$ for various values of $\lambda$. Note that 
as $\lambda \to \infty$, then $\rho_+ \to \rho_e$ and the solution develops the nearly $dS_2$ region. For both of these saddles we can evaluate the total 
amount of imaginary proper  time evolution and we get 
\be \la{ImTime}
\Delta {\rm Im} (t_{\rm proper}) = {\rm Im } \left[ \int_{\rho_+}^\infty { d \rho \over \sqrt{ - f(\rho) } }  \right] < 0 
\ee
The fact that it is less than zero implies that the saddles are consistent with
the no-boundary proposal, i.e. we are going into the right direction of the complex plane, so as to prepare a reasonable state for the matter fields.  Just for reference, the total imaginary proper time  evolution for the configurations in section \ref{sec:HHWavefunction} is $-\pi/2$  (from \nref{HHusual}).

We had mentioned that there exist other possible classical solutions which are obtained by 
analytically continuing the euclidean black hole solution which is regular at the cosmological horizon. These
are obtained from the previous ones, \nref{RhoPE}, 
 by changing $\lambda \to - \lambda$.
These solutions have the opposite sign for imaginary proper time, \nref{ImTime}, basically because they 
are the complex conjugates of the solutions we discussed above. For this reason, they 
do {\it not} obey the no-boundary condition. They are going into the wrong side of the complex plane. So we will not include these other solutions\footnote{
This   removes the solution that had similar behavior to the proposed 
dual of Vasiliev's theory, see near equation (5.17) of  \cite{Anninos:2012ft}.}. 


Now, out of the solutions that are allowed, \nref{RhoPE}, we need to select the ones that actually do contribute. 
  This is something that 
requires us to know the integration contour. This is a contour on the whole functional integral and 
it is  beyond the scope of this paper to determine it.  We will make some remarks on the choice of contour in section \ref{sec:ContChoices} and appendix \ref{Contours}. For now we will just mention that we think that the following saddles contribute (these are in blue in figure \nref{Saddles}.). 
\bea
{ \rho_+ \over \rho_e } &=& { i \over y } +  \sqrt{ 1 - { 1 \over y^2} } ~, ~~~~~~~~~{\rm for} ~~ y= { \lambda \over \lambda_c } > 1 \la{SaddWant}
\\
{ \rho_+ \over \rho_e } &=&i \left[  { 1 \over y } - \sqrt{  { 1 \over y^2} -1} \right] ~, ~~~~{\rm for} ~~ y= { \lambda \over \lambda_c } < 1  \la{SaddWant2}
\eea
We see that as $\lambda \to \infty$ we get  $\rho_+ \to \rho_e$. 
 It also includes the one that looks similar to the small $AdS$ black holes for  $\lambda < \lambda_c$.

 In the next section we will connect the saddle \nref{SaddWant} to the 
 nearly $dS_2$ computation of the wavefunction. Therefore, we are assuming this saddle does indeed contribute.  
 
  This analysis can be repeated for an asymptotic $S^1 \times H^{d-1}$ geometry. The bulk is a hyperbolic black hole in Euclidean AdS \cite{Birmingham:1998nr}.   But these geometries do not   develop  a nearly dS$_2$ region, so we will not discuss them further.

\subsection{Relation between the four dimensional and the two dimensional Hartle-Hawking wavefunctions   } \la{Sec:4d-2d HHwavefunction}

In the previous section we described bulk complex geometries that   prepare 
a no boundary  state in the asymptotic future on a $S^1 \times S^{2}$ slice. In this section,  we will focus on the   limit of a  large $S^1$, namely $\lambda \to \infty$. 

To begin with, we will compute the classical no boundary wavefunction in this limit. For this we will use the saddle \nref{SaddWant}. Expanding \eqref{OffSha} for large $\lambda$ gives, up to a divergent phase, 
\be \la{PsiExpan}
\Psi_{cl}(\lambda)   \sim \exp \left(-i \lambda M_e  +  S_e + i \frac{  C_1}{\lambda}- \frac{C_2}{\lambda^2} +\ldots \right)
\ee
where the dots indicate subleading terms \footnote{Curiously the subleading terms indicated by $+\ldots$ in \eqref{PsiExpan} are all pure phases (for $\lambda > \lambda_c$) and there are no further corrections to the absolute value $|\Psi_{cl}(\lambda)|$.} and 
\bea
\la{ExtDef} 
M_e &=& {R_{dS}^2 \over G_N } { \mu_e \over 2 } ~,~~~~~~~~S_e ={ R_{dS}^2 \over G_N} \pi  \rho_e^2 \\
  C_1 &=&  {R_{dS}^2 \over G_N} 2 \pi^2 \rho_e^3 ~,~~~~~~~~ C_2 =  { R_{dS}^2 \over G_N}  { 8 \pi^3  \over 27 }  
\eea
where $\rho_e = 1/\sqrt{3}$ and $\mu_e = 2 \rho_e^{3} $. 
Note that $S_e$ is the extremal black hole entropy and it is given by the area of its horizon,
 or the area of the cosmological horizon, but {\it not} the sum of the two. 
 
The second and third terms in \nref{PsiExpan} can be related to the ones that 
appear in the JT gravity theory for parameters \nref{ParamJT}. The second term, related to the entropy, is straightforward. 
The third term requires us to identify $\ell/\phi_b$ with $\lambda$. These parameters are defined in different regions of the geometry, $\ell/\phi_b$ is 
defined in the region still well approximated by JT theory, while $\lambda$ is defined in the asymptotic four dimensional region. To relate these two parameters
it is necessary to discuss the metric connecting the nearly $dS_2$ region with the four dimensional geometry. This metric can be approximated 
by the extremal metric \nref{SchdS4} with $\mu = \mu_e$ given by 
\be 
ds^2 = (-f) d\tau_r^2  - { d\rho^2 \over (-f) } + \rho^2 d\Omega_2^2 ~,~~~~~~  -f =    ( \rho -\rho_e)^2( 1 + { 2 \rho_e \over \rho} ) \la{ExtGeo}
 \ee  
  In practice, $\lambda$ is  the period of the $\tau_r$ coordinate. Expanding 
 this metric near $\rho \sim \rho_e$ and using the definition of the dilaton in \nref{ParamJT}  we find 
 \be
\ell \sim 3 \lambda (\rho -\rho_e) ~,~~~~\phi_b = 2 \phi_0  (\rho - \rho_e)/\rho_e ~,~~~~~~  { \phi_b \over \ell  } ={ 2 \phi_0 \rho_e \over \lambda }  
\ee
After this identification, third term of \nref{ClaWdW} agrees with the  third of \nref{PsiExpan}. 
The first term in \nref{ClaWdW} does not agree with the first term in  \nref{PsiExpan}. This is not a problem, these are just phases that depend on the late time regularization, and we are implicitly using different late time regularizations when we look at   the wavefunction from the two or four dimensional perspective.  

Finally, the last term in \nref{PsiExpan} is a new term, not captured by the JT gravity theory. Of course, it becomes very small when $\lambda \to \infty$, which is the regime of validity of the JT gravity. However, this term renders the wavefunction normalizable, eliminating the small $\ell$ divergence of the integral over the probability measure in
\nref{FinalProb}.
 So the important point about the last term in \nref{PsiExpan} is that it is real and it provides a suppression at small $\lambda$ in the classical answer  
\be
\left|\Psi_{cl}(\lambda)\right|^2 =  e^{2 S_e} e^{- 2  C_2/\lambda^2}   
\ee 

In conclusion, for very large $\lambda$ we can approximate the computation of the full Hartle-Hawking wavefunction in terms of the computation in the nearly-$dS_2$ theory. This computation correctly gives the 
contribution from the saddle point geometry and all its perturbative corrections in the large $\lambda$ region. 
It includes the corrections from the determinants of the four dimensional fields, etc. The point is that those
determinants are consistent with the isometries of $dS_2$ and only contribute to the extremal entropy $S_e$. 
The only $\lambda$ dependent contribution comes from the two dimensional gravitational modes and gives a 
factor of $1/\lambda^3$ in the measure. So the total contribution to the square of the wavefunction is \footnote{To be precise, the quantum correction is not only given from the JT gravity mode but also from an $SO(3)$ gauge field arising from dimensional reduction. The final answer (which can be extracted from section 3.4 of \cite{Iliesiu:2020qvm}) gives $\left|\Psi_s(\lambda)\right|^2 =e^{2 S_e- 2 C_2/\lambda^2}\lambda^{-3}$ (for $\lambda \gtrsim R_{dS}^2/G_N$) and $\left|\Psi_s(\lambda)\right|^2 =e^{2 S_e- 2 C_2/\lambda^2} \lambda^{-6}$ (for $\lambda \lesssim R_{dS}^2/G_N$). This does not affect the conclusion that the wavefunction is normalizable and peaked around a size of order $\lambda \sim R_{dS}/\sqrt{G_N}$. We thank Henry Lin for pointing this out.}
\be
\left|\Psi_s(\lambda)\right|^2 =e^{2 S_e} e^{- 2 C_2/\lambda^2} \frac{1}{\lambda^3} \la{ConvExp}
\ee
Since $C_2 \propto  {R^2_{dS} \over G_N} $ we see the wavefunction is peaked at a large $\lambda \sim R_{dS}/\sqrt{G_N} \gg 1$.   This means that the approximations we made are correct at that maximum. Note that the peak value 
of $\lambda$ arises from a balance between classical and quantum effects. The reason that the final answer is trustworthy is that the classical part is very small 
for large $\lambda$.

\subsection{Some remarks on contour choices and contributing saddle points}
\la{sec:ContChoices}

In the Euclidean $AdS_4$ problem with a $S^1_\beta \times S^2$ boundary  we face a similar issue when we have to decide which 
  saddle points contribute. There, a reasonable way to proceed would be to 
  do the functional integral over the metrics with  fixed black hole area.
  This introduces a conical singularity at the black hole horizon. Then integrating over this last variable selects the particular black hole  areas (or masses) which make the solution smooth at the horizon. The black hole solutions can
  be viewed as saddle points for this last integral. In this case it is reasonable to assume that the contour integral is over positive masses, or positive values of $r_+$. This selects which saddles contribute. See appendix \ref{AdSContour}. 
 Note that we are talking about whether a saddle contributes or not, which is independent of
  the question of whether it is the {\it largest} contribution. This question arises after we decide which ones contribute. 
    
   We can mimic this in the $dS_4$ context.  We can start with 
   the real solution \nref{SchdS4} and compactify the $\tau_r$ circle to $\tau_r \sim \tau_r + \lambda$. This produces a conical singularity at $\rho_+$, where
   $f(\rho_+)=0$. We 
   then fill the geometry near this singularity with an off shell rounded complex cone which obeys the no boundary condition. This allows us to evaluate the action, see appendix  \ref{Cones},    
\be
i S_{\rm Finite}^{\rm off}  =   { R^2_{dS}   \over G_N } \left[  - i  { \lambda \over 2 }   (\rho_+ - \rho_+^3)  + 
  \pi  \rho_+^2  \right] = - i \lambda M + S    \la{OffSh}
\ee
where $M$ is formally the analytic continuation of the 
 mass of the de-Sitter black hole and $S$ is its entropy \nref{MaEnt}. Notice that the final form 
 is the same as that of the free energy if we set $\beta = i \lambda$.
 Note that we did not assume the last equality in \nref{OffSh}, we
  derived it from the action of the off shell configuration.
   This derivation made most sense for $\rho_+ > \rho_e$ so that $\rho_+$ is the first 
 conical singularity that we encounter. But it can be analytically continued to any value of $\rho_+$ without any change in the answer. When it is analytically continued we might encounter other singularities, then
  we go around them again the positive imaginary direction and find  no extra   contribution, 
 see appendix \ref{Cones}. 
 Of course, the saddle points for this action, \nref{OffSh},  reproduce \nref{RhoPE}. 
 
 In order to seek some guidance on the choice of contour, we will start by analyzing \nref{OffSh} 
  near $\rho_+\sim \rho_e$. This is the regime where we can trust the two dimensional gravity answer.  In this regime the action looks like 
 \be 
 i S_{Finite}^{off} \sim  -i \lambda M_e + S_e + { R_{dS}^2 \over G_N} \left( i { \lambda  \over 2 \rho_e}   (\rho -\rho_e)^2 + 2 \pi \rho_e (\rho - \rho_e) \right)  
 \ee
 with $M_e$, $S_e$ given in \nref{ExtDef}. 
 This looks similar to the exponent in expression in \nref{EntDens}, where  
  $\rho-\rho_e$ is proportional to $s$ in \nref{EntDens}.
 The term quadratic in $\rho-\rho_e$ is  related  to the energy, $E$,  in 
 \nref{EntDens}, \nref{Dens},  and it goes as   $E \propto - (\mu -\mu_e) \propto  (\rho-\rho_e)^2$.  
 So the fact that we have positive energies in \nref{Dens} is related to the fact that black hole masses should be less than the extremal mass. On the other hand, 
 for a given mass we have two entropies, $S_e\pm |s|$, which correspond
 to the black hole entropy or to the cosmological horizon entropy. 
 Note that this is {\it not} equal to the naive entropy of the static patch, which would have been the sum of the two entropies. 
 
 From the fact that in \nref{EntDens} we have an integral over 
 $s$ we conclude that the integral over $\rho_+$ should extend to both sides of $\rho_e$. And also, from \nref{Dens}, we conclude that the integral is restricted to $E\geq0$ or $\mu \leq \mu_e$. 
 
 Now, this analysis only tells us what the contour in $\rho_+$ looks like for the region near $\rho_+\sim \rho_e$. As we go away from this region we do not have the luxury of the perturbatively exact results of nearly $dS_2$ gravity. 
  
  So one guess for the contour is that $\rho_+$ starts at $\rho_+=0$ and ends at $\rho_+=1$. This then covers all physical black holes $0 \leq \mu \leq \mu_e$. 
  Another possibility is to continue to $\rho_+ \to \infty$. This is reasonable because in this case the geometries with a conical defect angle seem well defined. However, for $\rho_+>1$, we have $\mu<0$ and the spacetime contains a naked singularity, visible from the far future, which does not look too reasonable.  
  
  With both of these possibilities we find that the saddles that contribute are the ones in \nref{SaddWant} \nref{SaddWant2}. 
  In addition we have an endpoint contribution at $\rho_+=0$, which is subleading and, in the case that we end the contour
  at $\rho_+=1$, we have an endpoint contribution that is actually 
  dominant over the saddles \nref{SaddWant} \nref{SaddWant2}. This endpoint contribution corresponds to $dS_4$ spacetime in Milne type coordinates. The action is equal to the entropy of $dS_4$. 
  The details of this analysis are presented in appendix \ref{Contours} \footnote{ See \cite{Conti:2014uda} for another perspective on this issue.}. It would be interesting to understand this point further.

\subsection{Four dimensional cosmological correlators} 

Here we connect  four dimensional correlators to two dimensional ones. In sections \ref{PureMatter} and \ref{sec:corrgb} we will compute the two dimensional correlators and their quantum corrections. Here we are just motivating that computation from a four dimensional point of view. 

\begin{figure}
\begin{center}
\includegraphics[scale=0.5]{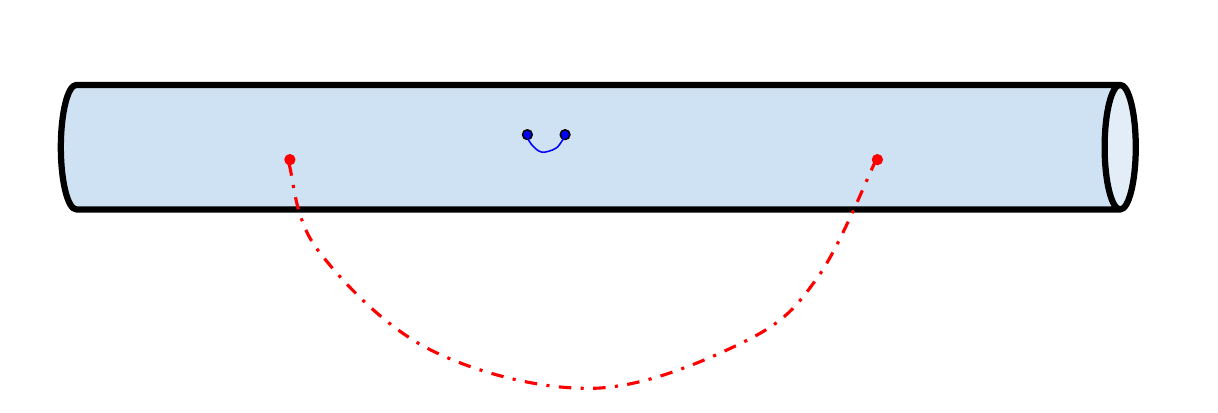}
\caption{Correlators on $R\times S_2$, the blue solid line represents short distance correlation functions and the red dashed line represents long distance correlation functions (distance larger than the size of sphere).}    \label{CylinderCorr}
\end{center} 
\end{figure}

\begin{figure}
\begin{center}
\includegraphics[scale=0.5]{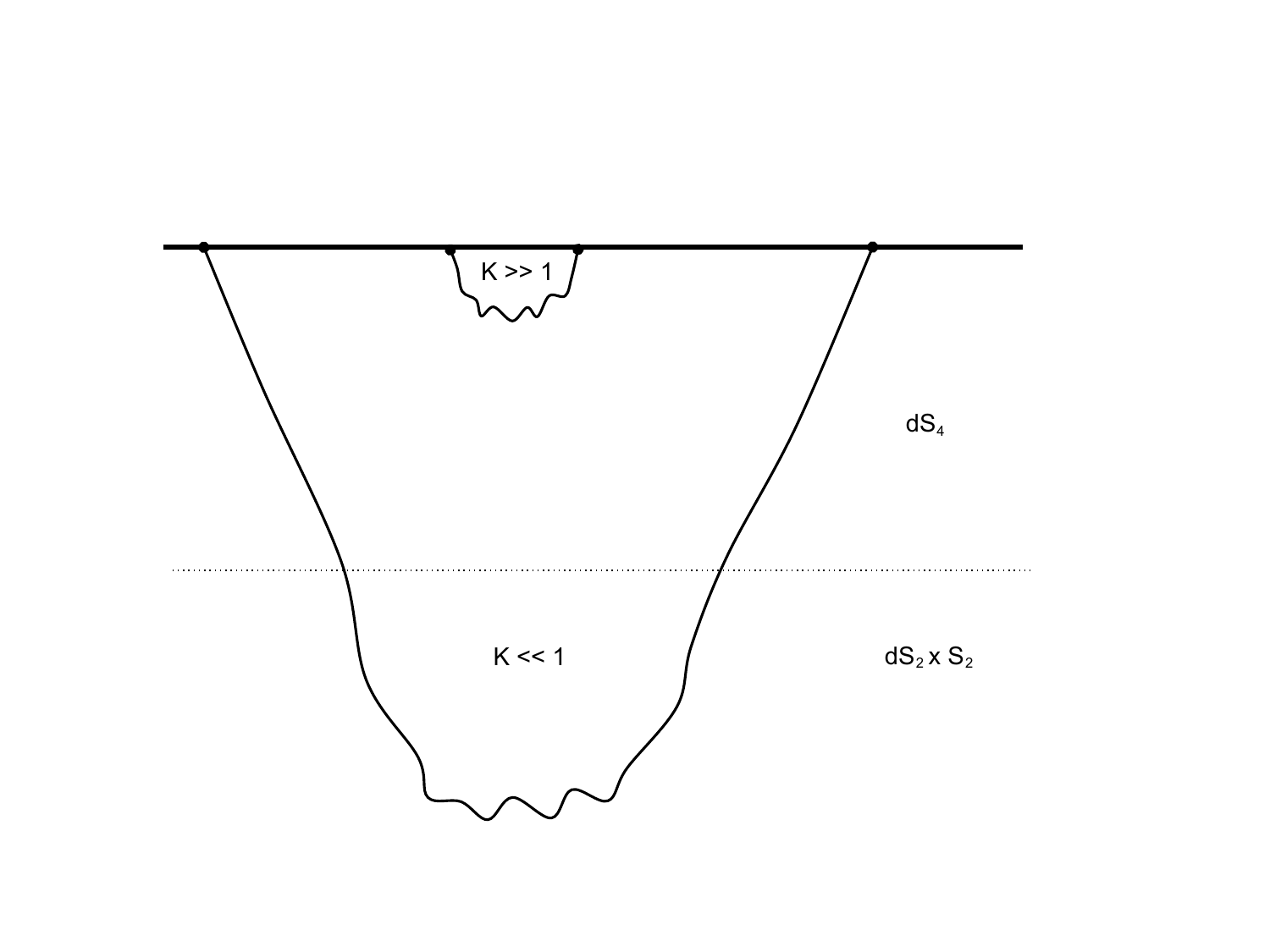}
\caption{For dimensional correlators for near extremal geometries. For short distances we have the usual four dimensional results. For long distances, distances larger than the size of the $S^2$, we have correlators dominated by the effectively two dimensional region.  }  \label{4DCorr}  
\end{center}
\end{figure}

It is useful to start first with the exactly extremal geometry \nref{ExtGeo}. 
 This asymptotes to $dS_2 \times S^2$ in the far past, for $\rho \to \rho_e$, and to $dS_4$ in the far future. 
 We have a past infinite period of $dS_2$ expansion followed by a future infinite period of $dS_4$ expansion. 
 Notice that the future boundary of $dS_4$ is a three sphere. On the other hand in \nref{ExtGeo} it looks 
 like $R \times S^2$. But $R \times S^2$ and $S^3$ are related by a conformal rescaling, which amounts to a 
 different choice of time slicing in the bulk. It is convenient for us to consider the $R\times S^2$ presentation since the correlators will only have $R\times SO(3)$ symmetry. 
 For distances smaller than the size of $S^2$ the cosmological correlators at late times will have the usual 
 four dimensional de-Sitter form. These correspond to modes that have crossed the horizon when the universe was already well approximated by four dimensional de Sitter space. 
 On the other hand, for modes at distances larger than the size of the $S^2$ (see figures \ref{CylinderCorr} and \ref{4DCorr}) the
 cosmological correlators will be essentially given by the two dimensional theory. These are modes which crossed the 
 horizon when the universe was well approximated by $dS_2 \times S^2$. 

 As a simple example, let us consider a minimally coupled 
 massless scalar field $h$. For simplicity we could expand in spherical harmonics on the sphere. If we consider the lowest harmonic and fourier transform the correlator in the $\tau_r$ direction , at late times, 
  we get 
 \bea 
 \langle h(k) h(-k) \rangle &\propto & { 1 \over |k|^3 }   ~,~~~~~~~~~|k| \gg 1  \cr
 \langle h(k) h(-k) \rangle &\propto & { 1 \over |k| } ~,~~~~~~~~~|k| \ll 1     
 \eea
 where the zero subindex indicates zero angular momentum on the $S^2$. 
 (here we assumed that the $\tau_r$ direction is non-compact.).  
 
 We expect that the exactly $dS_2$ geometry will have strong backreaction issues due to quantum gravity. 
 But we can similarly consider nearly $dS_2$ geometries by setting $\mu$ to be slightly higher or smaller than 
 $\mu_e$.  
 The case when $\mu<\mu_e$ corresponds to having a pair of black holes in de Sitter, one at the north pole and the other at the south pole of the $S^3$. Both black holes have masses very close to the extremal value. 
 Similarly, we could consider $\mu > \mu_e$. In this case, the full geometry has a past singularity, see figure 
 \ref{Penrosefd}(d). 

 In both cases, it is useful to relate the nearly $dS_2$ forms of the metric to the four dimensional metrics
 \bea
ds^2&=&{1\over 3}\left[-d\tau^2+\cosh^2\tau d\varphi^2+d\Omega_2\right];~~~~~\delta={\delta\mu^{1\over 2} 3^{1\over 4}}\sinh\tau;~~~\tau_r={\varphi\over 3^{3\over 4}\delta\mu^{1\over 2}};~~~~~~~\delta\mu>0~~~~\cr 
ds^2&=&{1\over 3}\left[-d\hat\tau^2+\sinh\hat\tau^2d\chi^2+d\Omega_2\right];~~~~~\delta={|\delta\mu|^{1\over 2}  3^{1\over 4}}\cosh\hat\tau;~~~\tau_r={\chi\over 3^{3\over 4}|\delta\mu|^{1\over 2}};~~~~\delta\mu<0~~
\cr
 & ~& \delta = { \rho \over \rho_e } - 1 ~,~~~~~~ \delta \mu = \mu - \mu_e ~,~~~~~~~ |\delta \mu | \ll 1 \la{TwoD4}
\eea
These changes of coordinates allow us to read the form of the two dimensional dilaton, which is related to 
$\rho -\rho_e$. 
Since the four dimensional geometry takes over when $\rho$ becomes different from $\rho_e$ by an order one amount, we can read off the amount of two dimensional e-foldings from \nref{TwoD4} as the value of $\tau$ where $\delta$ would 
be of order one. This gives 
\be \la{Effnum}
{ \cal N}_{\rm efolds } \sim \Delta \tau \sim  - \half \log |\mu - \mu_e | \sim - \log\left( { \phi_{g,m} \over \phi_0} 
\right) ~,~~~~~~~~ {\rm for} ~~|\mu - \mu_e | \ll 1
\ee
We also see that the number of e-folds is proportional to the log of the 
 coefficient of the dilaton in \nref{Global} or \nref{Milne}, see \nref{ParamJT}. 
For $\delta \mu > 0$ this is the number of e-folds of the expanding two dimensional phase. There is a similar number 
of e-folds for a preceding contracting two dimensional phase that comes out of a four dimensional region close to the singularity. See figure \ref{Penrosefd}(d). In the case that we compactified the direction $\tau_r$ to  $\tau_r \sim \tau_r + \lambda$, we imagined  that the universe results from a no boundary proposal. In that case we were considering solutions where $\mu -\mu_e$ was adjusted so that the circle had proper length $2\pi \rho_e$ at $  \tau =0$ so that we can join it to the euclidean sphere. In this case, we do not view the region with $\tau < 0$ as physical.   
 For $\delta \mu < 0$  \nref{Effnum}  is the number of e-folds from the cosmological horizon of the two dimensional static patch to the region that the four dimensional expansion takes over. 
  In both of these cases the two dimensional matter correlators and its quantum corrections, which we will compute in  
  sections  \ref{PureMatter}, \ref{sec:corrgb}, can be directly related to the four dimensional ones.

 Note that,  for $\mu< \mu_e$,  shifts of $\varphi$ corresponds to time evolution in the static patch, and also to a spacelike evolution in the far future along the $R$ direction of $R \times S^2$. In general, features that occur at different times along the static patch are imprinted at different points in space along the $R$ direction.

\section{Matter correlators in $dS_2$} 
\la{PureMatter}

So far we have neglected the presence of matter in de-Sitter. If we have additional matter, the wavefunctional will depend on the boundary conditions for matter. This gives an answer which is very similar to the one we have for the $AdS_2$ case. 
These matter correlators depend on the initial conditions for the matter field. We will pick a Bunch-Davies-like vacuum and discuss the various cases below. This is the vacuum that arises naturally when we consider the no boundary proposal for the wavefunction.

\subsection{Poincare patch} 

First we consider $dS_2$ in the Poincare coordinates. In order to compute in-in correlators we will begin by computing the wavefunction in the Buch-Davies vacuum for the matter field. We will focus on a massive scalar field $\psi$ of mass $m$ with action $S =\half  \int (\partial \psi)^2 - m^2 \psi^2$. 

The wavefunction at some time $\eta_0$ is given by the path integral of the matter field between the past $\eta \to -\infty$ and the present at $\eta_0$, with fixed boundary conditions at $\eta_0$ that become the argument of the wavefunction. For this path integral to be well-defined we also need to impose boundary conditions in the past $\eta \to -\infty$, which corresponds to choosing a particular quantum state. The Bunch-Davies state  corresponds to deforming the 
integration contour to \be\la{BDac}
\eta  \sim  -\alpha e^{- i \epsilon }  ~,~~~~~~~ \alpha, \epsilon > 0 
\ee
and demanding that the fields decay at early times. 
This is the analog of the no boundary proposal for these coordinates \cite{BD}. 

 We will consider light fields, meaning $0<m\leq 1/2$ in units of the dS$_2$ radius. At late times, the classical solutions   decay as $\psi \sim (-\eta)^{\Delta_- }\psi_-+ (-\eta)^{\Delta_+}\psi_+$, where we defined
\be
\Delta_\pm = \frac{1}{2} \pm \nu,~~~\nu = \sqrt{\frac{1}{4}-m^2}.
\ee
We can compute semiclassically the wavefunction for this field in the Bunch-Davies vacuum by fixing boundary conditions $\psi(\eta_0,x)=\psi_0(u=x/\eta_0)$ at a late time $\eta_0\sim 0$, together with the requirement that the field goes to zero at early times after the analytic continuation   \eqref{BDac}. We get a wavefunctional for this field of the form \cite{BD}
\begin{eqnarray}\la{poinwf}
\Psi[ \psi_0 (u)] &=& Z_{D}e^{i S_{local}} \exp\left(- i^{2 \Delta_+} c_\nu \int du\int du'{\psi_0(u) \psi_0(u')\over |u -u'|^{ 2 \Delta_+ }  }\right),\\
& =&Z_{D}e^{i S_{local}} \exp\left(-  i^{-2 \Delta_-} \tilde{c}_\nu\int \frac{dk}{2\pi}  \psi_0(k) \psi_0(-k) |k|^{  2 \Delta_+-1 }  \right),
\end{eqnarray}
where we have expressed the result both in position and momentum space \footnote{We normalize the Fourier transform of a function or correlator as $f(x)=\int \frac{dk}{2\pi} e^{ik x}f(k)$.}. We also defined the local phase 
\be \la{LocalS}
S_{local} = \frac{~\Delta_-}{2} \int du~\psi_0^2
\ee
This term diverges when written in terms of $x$ as $\sim \eta_0^{-1}$. 
$Z_{D}$ is the bulk partition function with Dirichlet boundary condition. In the expression above we defined the following real coefficients 
\be\la{coef2p}
c_\nu =\frac{\nu\Gamma(\frac{1}{2}+\nu)}{\sqrt{\pi}\Gamma(\nu)} ,~~~~~~~~\tilde{c}_\nu = \frac{\Gamma(1-\nu)}{4^\nu \Gamma(\nu)}.
\ee
The divergent piece in the wavefunctional makes the field classical at late times and only operators that are close to be diagonal in $\psi_0$ can have non-zero expectation value.

Ignoring gravitational backreaction for now, we can compute the equal-time in-in two-point function 
\be\la{inindefp}
\langle \psi(u_1)\psi(u_2)\rangle = \int \mathcal{D}\psi_0 ~|\Psi[\psi_0]|^2~ \psi_0(u_1) \psi_0(u_2),
\ee
where we use the bracket shorthand for an expectation value computed in the Bunch-Davies vacuum. We assume we always pick a normalization (possibly divergent) for the wavefunction when computing observables like \eqref{inindefp} such that $\int \mathcal{D} \psi_0~ |\Psi[\psi_0]|^2 =1$. 

The path integral is gaussian and gives 
\bea\label{eq:2pttree}
\langle \psi(u_1) \psi(u_2) \rangle &=&  \frac{N_\nu}{u_{12}^{2\Delta_-}},~~~~~~~~~~~~~N_\nu\equiv\frac{\Gamma(\nu)\Gamma(\frac{1}{2}-\nu)}{4\pi^{3/2}},\\
\langle \psi(k) \psi(-k) \rangle' &=&  \tilde{N}_\nu |k|^{2 \Delta_--1},~~~~~\tilde{N}_\nu\equiv\frac{4^\nu\Gamma(\nu)^2}{4\pi },
\ea
where $u_{12}=u_1 - u_2$. We defined $\langle \psi(k_1) \psi(k_2) \rangle = 2 \pi \delta( k_1+k_2) \langle \psi(k_1)\psi(-k_1)\rangle'$. Note that in order to obtain this result, the factor of $i^{2\Delta_+}$ appearing in the wavefunction \eqref{poinwf} is important to get the correct mass dependent prefactor. This result can be reproduced by solving a Green function equation in dS$_2$ with the appropriate boundary conditions \cite{Bousso:2001mw}. Since the theory is free, higher order correlation functions factorize into products of two-point function.

\subsection{Global $dS_2$} 

We can also compute the wavefunctional in the global time coordinates, as a simple reparametrization ($u\rightarrow \tan{u\pi\over \ell}$) of
the Poincare wavefunctional (\ref{poinwf}). In this notation, we denote by $u$ the proper length at asymptotic infinity, and the size of the circle by $\ell$. Therefore both quantities diverge at late times at the same rate. 
We find 
\bea\label{GlobalWavefunction}
\Psi[ \psi_0(u) ] &=&Z_De^{i S_{local}} \exp\left( -i^{2 \Delta_+} c_\nu \int du\int du'\psi_0(u ) \psi_0(u') \Big| {\pi\over \ell\sin [{ (u - u')\pi \over \ell}]} \Big|^{  2 \Delta_+ } \right),~~~~~~~~\\
&=&Z_De^{i S_{local}} \exp\left( -i^{2 \Delta_-} 2\pi\tilde c_\nu \sum_{k}\psi^0_k \psi^0_{-k} \left({2\pi\over \ell}\right)^{2\Delta_+-2} {\Gamma(\Delta_++k)\over \Gamma(\Delta_-+k)} \right), 
\eea
where $u \sim u + \ell$ and  the Fourier transformation is defined as $\psi_0(u)=\sum\limits_k\psi^0_ke^{i{2\pi k\over \ell}u}$, where $k$ is an integer.

Following the Poincare patch case, by taking the amplitude square of the wavefunctional and path-integrating over its argument $\psi_0$, we can obtain the equal time in-in two-point function
\beq\la{2pttreglobal}
\langle \psi(u_1) \psi(u_2) \rangle = N_\nu ~\Big|{\pi\over \ell\sin { u_{12}\pi\over \ell}} \Big|^{  2 \Delta_- };~~~~~~~\langle\psi(k)\psi(q)\rangle=\left({2\pi\over\ell}\right)^{2\Delta_-}{\tilde N_{\nu}\over 2\pi}{\Gamma(\Delta_-+k)\over \Gamma(\Delta_++k)}~\delta_{k,-q},
\eeq
where $\delta_{k,-q}$ is the Kroenecker delta imposing momentum conservation. We can see this result is proportional to the Euclidean AdS two-point function between operators of dimension $\Delta_-$. 

It is also easy to calculate the partition function, which should be the same as the sphere partition function.   That is because the $dS_2$ region is purely Lorentzian, and it cancels between the two sides of the 
Schwinger-Keldysh contour.  
We have the following path integral:
\bea
Z&=& \int d\psi_0 | \Psi[\psi_0]|^2 \nonumber\\
&=&\int d\psi_0 |Z_{D}|^2 e^{-\frac{1}{2}\sum\limits_{k}\psi^0_k \psi^0_{-k} G_{\Delta}(k)}=|Z_D|^2e^{-{1\over 2}\mathrm Tr\log {G_{\Delta}\over \ell} }
\ea
where $G_{\Delta}(k)=8\pi \cos(\pi \Delta_-) \tilde{c}_{\nu} ({2\pi\over \ell})^{2\Delta_+-2}  {\Gamma(\Delta_++k)\over \Gamma(\Delta_-+k)}$ and the additional ${1\over \ell}$ piece in the log is coming from the normalization of the mode $\psi^0_k$.  To evaluate the trace, we can use zeta function regularization.  This means the constant piece vanishes and we only have summation of the form\footnote{$G_{\Delta}(k)= G_{\Delta}(-k)$ for $k$ integer.}:
\begin{equation}
 -\half Tr \log { G_\Delta \over \ell } =	-{1\over 2}\sum_{k=-\infty}^{\infty}\log{\Gamma(\Delta_++|k|)\over \Gamma(\Delta_-+|k|)}
\end{equation}
One might think that if we regularize the summation with a hard cutoff $\sim \ell$ then there will be a possible $\ell\log\ell$ piece, but this will cancel with factors of $\ell$ appearing in the normalization of $G_\Delta/\ell$. 
One can also understand this determinant as the ratio between the partition function with Neumann Boundary condition and Dirichlet Boundary condition.  Because integrating over $\psi_0$ effectively sets the canonical momentum of $\psi_0$ to be zero.
In appendix \ref{App: MatterPartition}, we show that up to constant pieces, the partition function with different boundary conditions $Z_{D/N}$ is equal to \footnote{We will only focus on the mass dependence of the partition function. So we ignore mass independent overall constants in the partition function. }:
\begin{equation}
	\log Z_{D/N} = { 1 \over 2 } \sum_{k=-\infty}^\infty \log \Gamma(\Delta_{+/-} + |k|).
\end{equation}
Adding them up we have the whole partition function is equal to
\begin{equation}
	\log Z={1\over 2}\sum\limits_{k=-\infty}^{\infty}\log\left[\Gamma(\Delta_++|k|)\Gamma(\Delta_-+|k|)\right].
\end{equation}
which can be shown to match with the sphere partition function (Appendix \ref{App: MatterPartition}).
In addition, we saw that there is no $\ell$ dependence from the matter path integral. 

Up to now, we worked in terms of proper distance $u$ and proper size $\ell$ which is divergent at late time. One can do a conformal transformation $u\rightarrow {\tilde u \ell\over 2\pi}$ to rescale $\ell$ equal to $2\pi$, due to the anomalous dimension, $\psi_0$ will also rescaled to $({2\pi\over \ell})^{\Delta_-}\tilde \psi$. We will use this convention in section \ref{sec:corrgb} to simplify notation.

\section{Gravitational corrections to matter correlators } \la{sec:corrgb}

The gravitational degree of freedom does not totally disappear in two dimensions, as we have shown in section \ref{sec:NdSschwarzian}, but it reduces to a boundary mode. In this section, we   show how to couple it to matter and how it leads to corrections to cosmological correlators. 

One might be surprised to get physical effects from a degree of freedom that lives on a spacelike surface. However, this is common in theories with gauge symmetries, such as the instantaneous coulomb potential in some gauges. In fact, in appendix \ref{sec:almheirikang} we rederive some of the results in this section in a formalism where the gravitational effects propagate inside the lightcone (inspired by \cite{Almheiri:2014cka}). 
Furthermore, the boundary degrees of freedom are important in order to ensure that the full wavefunction obeys  the 
momentum constraint of general relativity, see appendix \ref{App: MomentumConstraint}. 
%

In our description, the effects of gravity  can be incorporated in terms of the
 boundary reparameterization field $x(u)$ and its Schwarzian action \nref{actw}
 (as in \cite{Jensen:2016pah,Maldacena:2016upp,Engelsoy:2016xyb}).
This field couples to matter as follows. 
Its primary effect is to change the location of the future boundary inside a rigid $dS_2$ space. This means that the same physical boundary condition for the matter fields along the physical boundary corresponds to 
different effective boundary conditions for the matter fields moving in the rigid 
$dS_2$ spacetime. These different boundary conditions are obtained by a conformal symmetry parametrized by the field $x(u)$. 
Finally,  
doing functional integral over the Schwarzian field then gives us the final Wheeler--deWitt wavefunctional.
We now proceed to show with formulas what we have said here in words.

\subsection{Including gravitational corrections to the wavefunction}

When matter and gravity are coupled, the Hartle-Hawking wavefunction discussed in section \ref{sec:HHWavefunction} now becomes (we use the rescaled fields and coordinates in this section)
\be\label{HHWaveBackreacted}
\Psi_s[\phi_b,\tilde\psi(\tilde u)] = \int \frac{\mathcal{D}\varphi}{SL(2)}~ e^{iS_{\rm Sch}[\varphi(\tilde u)]} ~\Psi_{\rm mat}[\tilde\psi(\varphi(\tilde u))], 
\ee
where the Schwarzian action is given by (see equation \eqref{actw})
\be \la{Phir}
iS_{\rm Sch}[\varphi(\tilde u)]=i2\phi_b  \int \{ \tan \frac{\varphi(u)}{2},u\}=i\phi_r \int \{\tan{\varphi(\tilde u)\over 2},\tilde u\};~~~\phi_r={4\pi \phi_b\over \ell}.
\ee  
The Schwarzian action has $SL(2)$ symmetry inherits from the isometry of the $dS_2$ metric. This is a gauge symmetry in our system since an overall 
$SL(2)$ transformation does not  change   the  position  of the matter relative to the boundary.    This is why we divide it out in   (\ref{HHWaveBackreacted}).
The subindex $s$ in \nref{HHWaveBackreacted} stands for ``$Schr\ddot{o}dinger$'', and it is the overall factor of the full Wheeler de Witt wavefunction, that also contains
a highly oscillatory piece responsible for enforcing the expansion of the universe. 
The derivative in the Klein Gordon inner product acts on this oscillatory piece to 
produce the standard norm, $|\Psi_s|^2$ for the $Schr\ddot{o}dinger$ piece\footnote{ This is analogous to how the second order Klein Gordon wave equation becomes a first order $Schr\ddot{o}dinger$ equation in the non-relativistic limit. }.

Note that we evaluate the wavefunction at a surface with constant $\phi_b$. $\phi_b$ acts as a clock in the future, expanding region which we are considering. 
The effects we discuss have a simple scaling dependence on time or $\phi_b$. 
In particular $\phi_r$ in \nref{Phir} becomes time independent at late times.

The factor $\Psi_{\rm mat}[\tilde\psi(\varphi(\tilde u))]$ is a conformal transformation of the matter wavefunctional in global coordinates (\ref{GlobalWavefunction}).
More explicitly, after we turn on gravity, the coupling with matter is given by 
\bea
\Psi_{\rm mat}[\tilde\psi(\varphi(\tilde u))] &=&e^{i S_{local}}\exp\left[ - i^{2 \Delta_+}c_\nu  \int \tilde\psi(\tilde u) \tilde\psi(\tilde u') G^\varphi_{\Delta_+}(\tilde u,\tilde u')\right],\\
G^\varphi_{\Delta_+}(\tilde u,\tilde u')&=&\Bigg( \frac{\varphi'(\tilde u)\varphi'(\tilde u')}{ 4 \sin^2 \frac{\varphi(\tilde u)-\varphi(\tilde u')}{2} }\Bigg)^{ \Delta_+ }.
\ea
The pure phase $S_{local}$, \nref{LocalS}, is independent of the Schwarzian mode $\varphi(\tilde u)$ and will cancel out when we compute $|\Psi_s|^2$. 

In order to compute an equal time in-in n-point correlator, we need to compute its expectation value using the backreacted wavefunction
\beq\la{inindef}
\langle \psi(\tilde u_1) \ldots \psi(\tilde u_n) \rangle = \int \mathcal{D}\tilde\psi ~|\Psi_s[\phi_b,\tilde\psi]|^2~ \tilde\psi(\tilde u_1)\ldots  \tilde\psi(\tilde u_n),
\eeq
where hidden inside of the amplitude square of the wavefunction we have a path integral over two copies of the Schwarzian mode (one from the bra and one from the ket). We assume we always pick a normalization (possibly divergent) for the wavefunction when computing observables like \eqref{inindef} such that $\int \mathcal{D} \tilde\psi~ |\Psi_s[\tilde\psi]|^2 =1$. 

\subsection{Tree Level}\label{treelevel}
We will compute correlators perturbatively in large $\phi_r = {\phi_b\over \ell}$, for which the Schwarzian mode is weakly coupled. First of all, we will spell out the path integrals involved in equation \eqref{inindef}, namely 
\bea
&&\langle \psi(\tilde u_1) \ldots \psi(\tilde u_n) \rangle = \int \mathcal{D}\tilde\psi \frac{\mathcal{D}\varphi}{SL(2)} \frac{\mathcal{D}\bar{\varphi}}{SL(2)} ~\tilde\psi(\tilde u_1)\ldots  \tilde\psi(\tilde u_n)~ \exp{(iS_{\rm Sch}[\varphi]-iS_{\rm Sch}[\bar{\varphi}])}\nonumber\\
&&~~~\exp{\left( - c_\nu\int~\tilde\psi(\tilde u) \tilde\psi(\tilde u') [i^{2\Delta_+} G^\varphi_{\Delta_+}(u,u') +i^{-2\Delta_+} G^{\bar{\varphi}}_{\Delta_+}(\tilde u,\tilde u') ]\right)},
\ea
When we compute the absolute value of the wavefunction, the mode $\varphi$ comes from the wavefunction $\Psi_s$ (the ket) while the other Schwarzian mode $\bar{\varphi}$ comes from its complex conjugate $\Psi_s^*$ (the bra). After complex conjugation, the action for the $\bar{\varphi}$ mode changes sign, as shown in the equation above.  

Semiclassically, the Schwarzian path integral has a saddle point at $\varphi(\tilde u)=\tilde u$. To organize the expansion around this solution we will write 
\beq
\varphi(\tilde u) \to \tilde u+ \varepsilon(\tilde u), 
\eeq
where $\varepsilon(\tilde u)$ is periodic in $\tilde u$ and for large $\phi_r$ it is a small correction. In this approximation we can expand the Schwarzian action as $iS_{\rm Sch}[\varepsilon]= -i \frac{ \phi_r}{2} \int [(\varepsilon '')^2-(\varepsilon')^2]$ and its coupling with matter as 
\beq
G^{\varphi=u+\varepsilon}_{\Delta_+}(\tilde u,\tilde u')=G^0_{\Delta_+}(\tilde u,\tilde u') + \delta_\varepsilon G_{\Delta_+}(\tilde u,\tilde u') + \delta_\varepsilon^2 G_{\Delta_+}(\tilde u,\tilde u') + \ldots,
\eeq
where it will be convenient to define
\beq
G^0_{\Delta_+}(\tilde u,\tilde u')\equiv\left( 4 \sin^2 \frac{\tilde u-\tilde u'}{2}\right)^{-\Delta_+}
\eeq
 and $\delta_\varepsilon^n G_{\Delta_+}(\tilde u,\tilde u')$ indicates the order $n$ term in the expansion of $G^{\varphi=u+\varepsilon}_{\Delta_+}(\tilde u,\tilde u')$ for small $\varepsilon(\tilde u)$ and this induces an interaction between two matter fields and $n$ reparametrization modes. The first orders in this expansion are given by 
\bea
\label{dG1}\delta_\varepsilon G_{\Delta}(\tilde u_1,\tilde u_2)&=&\frac{\Delta}{|2 \sin \frac{\tilde u_{12}}{2}|^{2\Delta}}\left[\varepsilon'_1+\varepsilon'_2-\frac{\varepsilon_1-\varepsilon_2}{\tan \frac{\tilde u_{12}}{2}} \right],\\
\label{dG2}\delta_\varepsilon^2 G_{\Delta}(\tilde u_1,\tilde u_2)&=& \frac{1}{|2 \sin \frac{\tilde u_{12}}{2}|^{2\Delta}} \left[\Delta\left( \frac{(\varepsilon_1-\varepsilon_2)^2}{4\sin^2 \frac{\tilde u_{12}}{2}}-\frac{\varepsilon_1'^2+\varepsilon_2'^2}{2}\right)+{\Delta^2\over 2}\left(\varepsilon'_1+\varepsilon'_2-\frac{\varepsilon_1-\varepsilon_2}{\tan \frac{\tilde u_{12}}{2}}\right)^2\right].~~~~~~~~~
\ea 
The Schwarzian tree-level propagator is given by 
\beq
\langle \varepsilon(\tilde u) \varepsilon(0)\rangle_0 = \frac{1}{2\pi~i\phi_r}\left[-\frac{(|\tilde u|-\pi)^2}{2} + (|\tilde u|-\pi)\sin|\tilde u| + a + b \cos \tilde u \right],
\eeq
with $a$, $b$ arbitrary coefficients that disappear for $SL(2)$ invariant observables \cite{Maldacena:2016upp}. From higher order expansion of the Schwarzian action we also need to include self interactions between the reparametrization mode, although they will not be important for the observables studied below.

A similar analysis can be done for $\bar{\varphi}(\tilde u)=\tilde u+\bar{\varepsilon}(\tilde u)$ and $G^{\bar{\varphi}}_{\Delta_+}(\tilde u,\tilde u')$. The expansion of the matter coupling for small $\varepsilon$ is equivalent to equations \eqref{dG1} and \eqref{dG2} after replacing $\varepsilon \to \bar{\varepsilon}$. 
Since the action for $\bar{\varepsilon}$ is the complex conjugate of $\varepsilon$, the propagator satisfy $\langle \bar{\varepsilon}(\tilde u) \bar{\varepsilon}(0)\rangle_0=  -\langle \varepsilon(\tilde u) \varepsilon(0)\rangle_0$. 
More importantly, there is no propagator between two schwarzian fields: $\langle \varepsilon(\tilde u) \bar{\varepsilon}(0) \rangle =0$.  One can understand this as the fact that there is no gravitational operators that can be inserted at the future time slice. 
 The tree level propagator for the matter field is given by $\langle \psi(\tilde u_1)\psi(\tilde u_2) \rangle_0= N_\nu G^0_{\Delta_-}(\tilde{u}_1,\tilde{u}_2)$, where $N_\nu$ is a mass dependent prefactor defined in equation \eqref{2pttreglobal}.

With this information we can start doing perturbative calculations. We effectively have three fields, $\tilde\psi(\tilde u)$, $\varepsilon(\tilde u)$ and $\bar{\varepsilon}(\tilde u)$. The matter field $\tilde\psi$ interacts with $\varepsilon$ and $\bar{\varepsilon}$ through bilocal couplings $\delta_\varepsilon^n G_{\Delta_+}(\tilde u,\tilde u')$ and $\delta_{\bar{\varepsilon}}^n G_{\Delta_+}(\tilde u,\tilde u')$. 

\begin{figure}
\begin{center}
\includegraphics[scale=0.6]{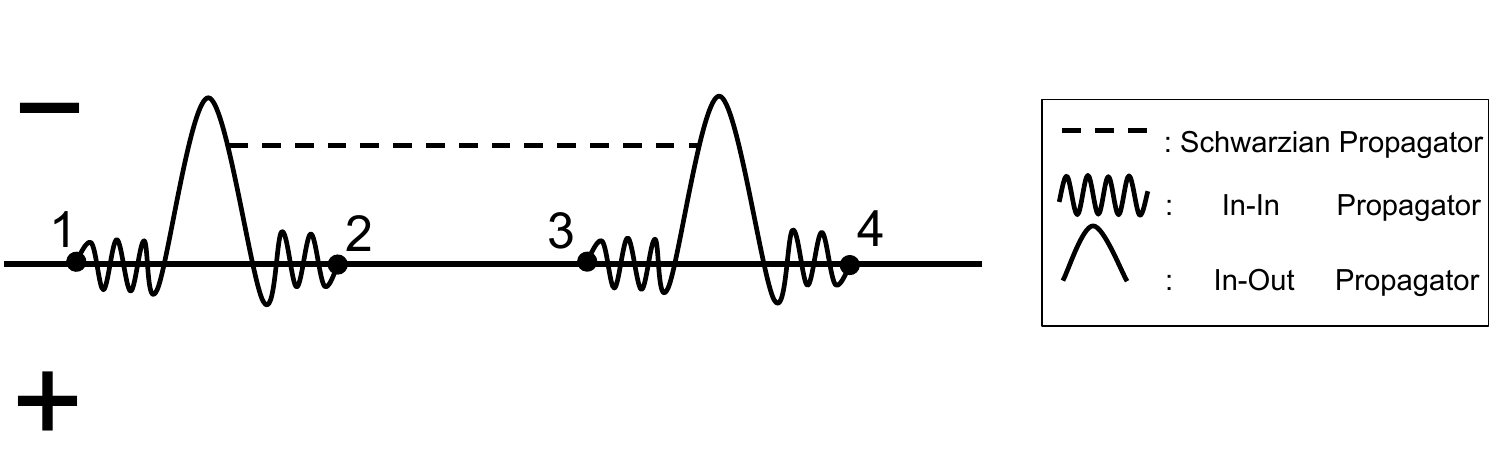}
\end{center}
\caption{Tree level diagram for in-in four point function. The horizontal black line indicates the asymptotic future time slice where we insert operators. We draw the space above and below (denoted by $\mp$) the future time slice as bra ($-$) and ket ($+$) state. Various propagators are dictated in the right box.  Here we only draw one channel with backreaction from the Schwarzian mode in bra state, there are two other channels coming from permutation with external legs. Adding the other Schwarzian contribution (coming from ket state) corresponding to add its complex conjugate.}
\label{fig:4ptdiagram}
\end{figure}

A general feature of this expansion is that the two Schwarzian modes never interact with each other\footnote{In figure \ref{fig:4ptdiagram}, there are no gravitational interactions generated between the matter at the upper half and lower half plane. This happens because $\varepsilon$ and $\bar{\varepsilon}$ are independent. We will give a different argument of this in appendix \ref{sec:almheirikang}.}. To leading order in ${1\over \phi_r}$, this allows us to compute corrections to correlators coming from a single Schwarzian mode, then we can account for the second Schwarzian by adding the complex conjugate. 

We will start by computing the connected in-in four-point function to leading order. This is given by a sum of diagrams like the one shown in figure \ref{fig:4ptdiagram}, given by
\bea
&&\hspace{-0.6cm}\langle \psi(\tilde u_1)\psi(\tilde u_2) \psi(\tilde u_3) \psi(\tilde u_4) \rangle_{\rm conn} =8 \frac{ i^{4 \Delta_+}c_\nu^2 N_\nu^4 }{2}\sum_{\rm channels} \int d\tilde u_a d\tilde u_b d\tilde{u_c} d\tilde{u_d}~G^0_{\Delta_-}(\tilde{u}_1,\tilde{u}_a) \nonumber\\
&& G^0_{\Delta_-}(\tilde{u}_2,\tilde{u}_b)G^0_{\Delta_-}(\tilde{u}_3,\tilde{u}_c)G^0_{\Delta_-}(\tilde{u}_4,\tilde{u}_d)\langle \delta_\varepsilon G_{\Delta_+}(\tilde u_a,\tilde u_b)\delta_\varepsilon G_{\Delta_+}(\tilde{u}_c,\tilde{u}_d)\rangle + {\rm c.c}.~~~~
\ea
The sum is over the three different non-equivalent Wick contractions (bulk channels). In figure \ref{fig:4ptdiagram} we show one of them. This means the other terms in the sum consist of different non-equivalent ways of pairing up $\{\tilde u_1,\tilde u_2,\tilde u_3,\tilde u_4\}$ with $\{\tilde u_a,\tilde u_b,\tilde{u}_c,\tilde{u}_d\}$. The brackets in the RHS means to take expectation value with respect to Schwarzian mode and the explicit result can be found in section 4 of \cite{Maldacena:2016upp}. The term $+{\rm c.c.}$ sums the complex conjugate coming from the other Schwarzian mode. The overall factor of $8$ comes from counting equivalent Wick contractions of the matter fields. Also, we ignore terms in the Taylor expansion coming from disconnected diagrams since they account for the correction in the normalization of the wavefunction. 

In appendix \ref{app:identity} we show the following identity 
\bea\la{4ptidentity}
&&\hspace{-0.8cm}\int d\tilde u_a d\tilde u_b d\tilde u_c d\tilde u_d \Big| 2^4 \sin \frac{\tilde u_{1a}}{2} \sin \frac{\tilde u_{2b}}{2} \sin \frac{\tilde u_{3c}}{2} \sin \frac{\tilde u_{4d}}{2}\Big|^{-2\Delta_-}  \delta_\varepsilon G_{\Delta_+}(\tilde u_a,\tilde u_b)\delta_\varepsilon G_{\Delta_+}(\tilde u_c,\tilde u_d)= \nonumber\\
&&\hspace{-0.2cm} ~= \Big(\frac{2\pi \tan \pi \Delta_-}{2\Delta_--1}\Big)^2 ~\delta_\varepsilon G_{\Delta_-}(\tilde u_1,\tilde u_2) \delta_\varepsilon G_{\Delta_-}(\tilde u_3,\tilde u_4),
\ea
which is valid even before taking the expectation value over Schwarzian mode. Using this relation, adding up the contribution and simplifying the prefactor, the final answer for the four-point function is given by 
\bea\la{4pttreeee}
&&\hspace{-1cm}\langle \psi(\tilde u_1)\psi(\tilde u_2) \psi(\tilde u_3) \psi(\tilde u_4) \rangle = N_\nu^2 \tan \pi \Delta_-  ~i^{-1}\Big[\langle \delta_\varepsilon G_{\Delta_-}(\tilde u_1,\tilde u_2) \delta_\varepsilon G_{\Delta_-}(\tilde u_3,\tilde u_4)\rangle \nonumber\\
&&+\langle \delta_\varepsilon G_{\Delta_-}(\tilde u_1,\tilde u_3) \delta_\varepsilon G_{\Delta_-}(\tilde u_2,\tilde u_4)\rangle + \langle \delta_\varepsilon G_{\Delta_-}(\tilde u_1,\tilde u_4) \delta_\varepsilon G_{\Delta_-}(\tilde u_2,\tilde u_3)\rangle \Big],
\ea
where $N_\nu$, defined in equation \eqref{eq:2pttree}, gives basically the two-point function normalization. Since the Schwarzian propagator is purely imaginary the right hand side of \eqref{4pttreeee} is real. The term in brackets in the right hand side is the Euclidean $AdS_2$ four-point function between identical operators of dimension $\Delta_-$ after replacing $C\to i \phi_r$. Therefore its explicit dependence with the four coordinates can be obtained directly from \cite{Maldacena:2016upp}. 
Up to the factor of $\tan{\pi \Delta_-}$, this result can be understood as coming from reparametrization of the in-in four point function and taking the imaginary part.

To give an explicit final answer it is more convenient to work in Fourier space (with respect to $\tilde{u}$). The first step is to find the Fourier transform of the Schwarzian four-point function. This can be done using equation (3.126) of \cite{Maldacena:2016hyu}. Combining this with the identity derived above, the dS in-in correlator is given by 
\beq
\langle \psi_{n_1}\psi_{n_2}\psi_{n_3}\psi_{n_4} \rangle = - \frac{ \tilde{N}_\nu^2 \tan \pi \Delta_-}{ \phi_r} [K_{global}(1,2;3,4)+K_{global}(1,3;2,4)+K_{global}(1,4;2,3)],
\eeq
where $\tilde{N}_\nu^2$ is a prefactor appearing in the momentum space tree level two-point function. The single channel Fourier transform is given by 
\beq\la{4glob4ptf}
K_{global}(1,2;3,4)=-\frac{B_{\Delta_-}(n_1,n_2) B_{\Delta_-}(n_3,n_4)}{(n_1+n_2)^2((n_1+n_2)^2-1)}    ~{\delta_{n_1+n_2+n_3+n_4,0}\over 2\pi},
\eeq
where we defined the function
\beq
B_{\Delta_-}(n_1,n_2)=\frac{(\Delta_-(n_1+n_2)-n_1)\Gamma(\Delta_-+n_1)}{\Gamma(1-\Delta_-+n_1)}+\frac{(\Delta_-(n_1+n_2)-n_2)\Gamma(\Delta_-+n_2)}{\Gamma(1-\Delta_-+n_2)}.
\eeq

Due to the $SL(2)$ symmetry, the momentum space four-point function above is defined for $|n_1+n_2|\geq 2$ and vanishes otherwise. With this explicit expression we can verify the identity \eqref{4ptidentity} derived in position space. The Fourier transform of the nearly-$EAdS_2$ Schwarzian four-point function for operators of dimension $\Delta$ is given by the same expression, in terms of $B_\Delta(n_1,n_2)$. 


It is interesting also to consider these correlators in the limit that 
$|u_i - u_j|  \ll \ell$ (or $|\varphi_i -\varphi_j| \ll 1$), 
  where they reduce to the Poincare patch correlators. 
Then the above results reduce to

\beq
\langle \psi(k_1) \psi(k_2)\psi(k_3) \psi(k_4)\rangle =- \tilde{N}_\nu^2 \frac{\tan \pi \Delta_-}{\phi_r}[K_P(1,2;3,4)+K_P(1,3;2,4)+K_P(1,4;2,3)].
\eeq 
 The single channel Fourier transform of the correlators is given by
\bea\la{4ptFourier}
K_P(1,2;3,4)&=&-2\pi \delta\Big(\sum_{i=1}^4 k_i\Big) ~\frac{1}{|k_{1+2}|^4}\left[\frac{\Delta_- (k_1+k_2) - k_1}{\left| k_1 \right|^{1-2 \Delta_-}}+\frac{\Delta_- (k_1+k_2) - k_2}{ \left| k_2\right| ^{1-2 \Delta_-}}\right] \nonumber\\
&&\left[\frac{\Delta_- (k_3+k_4)-k_3}{ \left| k_3\right| ^{1-2 \Delta_-}}+\frac{\Delta_- (k_3+k_4)-k_4 }{\left| k_4 \right|^{1-2 \Delta_-}}\right].
\ea
As explained in \cite{Maldacena:2016upp} the expressions in position space of this amplitudes depend on the particular order of the insertions in space.

This tree level computation is the two dimensional analog of the gravitational correction to the four point function computed in four dimensions in \cite{Seery:2008ax}.

\subsection{Loop corrections}
\begin{figure}
\centering
	  \subfigure{%
    \includegraphics[width=0.4\textwidth]{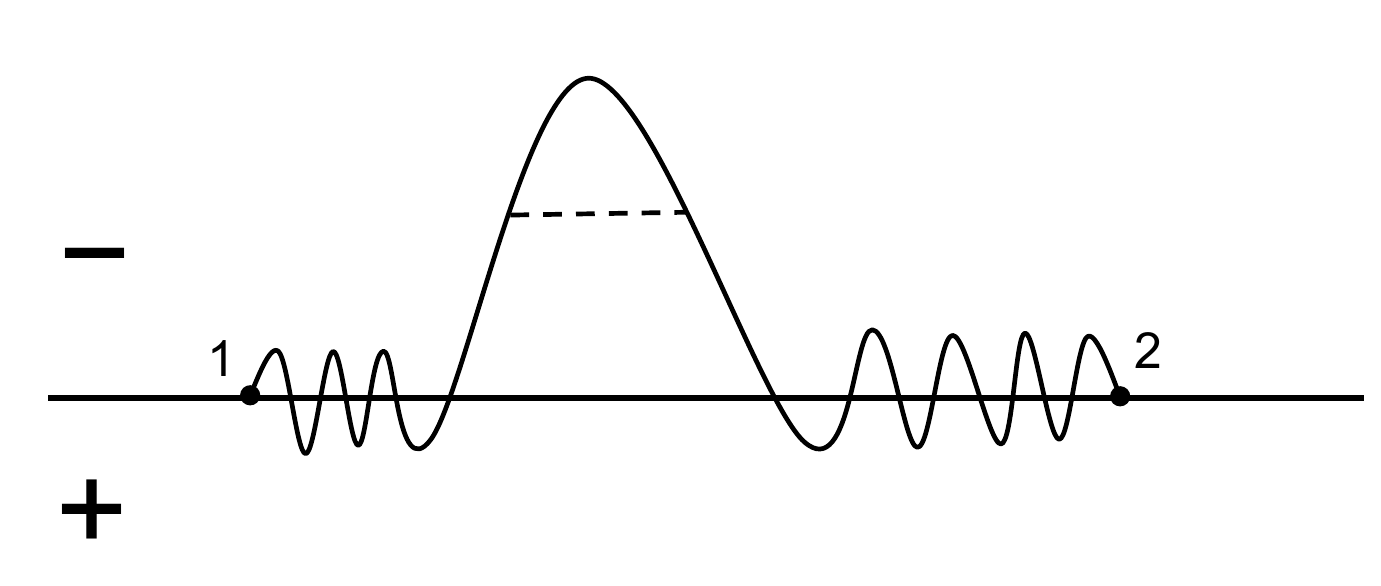}
    }\hspace{2cm}
    \subfigure{%
    \includegraphics[width=0.4\textwidth]{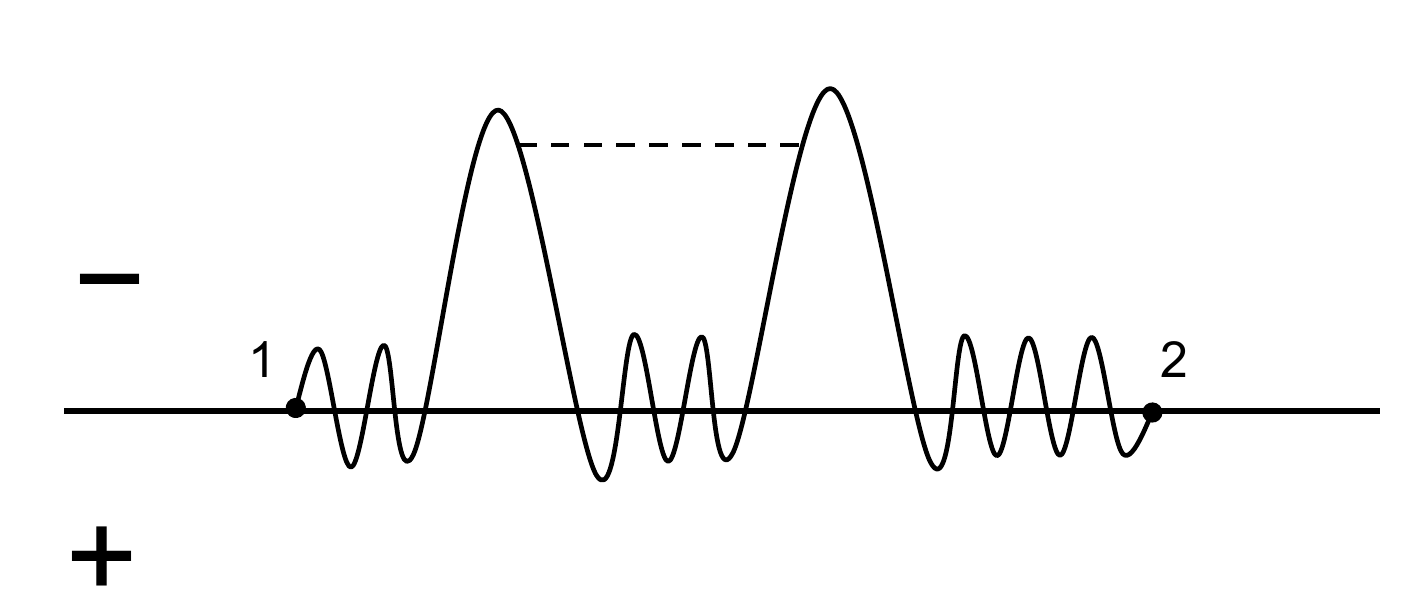}
  }%
\caption{Diagrams contributing to two-point function to one loop. The left figure shows the contribution from the one-loop correction of in-out propagator, and the right figure shows the contribution from interaction between two in-out propagators. }
\label{fig:2ptdiagram}
\end{figure}
We can also compute one-loop corrections to the two-point function. To simplify the discussion we will focus on the Poincare patch correlators. We need to sum diagrams such as the ones shown in figure \ref{fig:2ptdiagram}. The left diagram corresponds to expanding the matter coupling to second order $\langle \delta_\varepsilon^2 G_{\Delta_+}\rangle$ while the diagram to the right comes from a quadratic contribution from the linear variation $\langle \delta_\varepsilon G_{\Delta_+}\delta_\varepsilon G_{\Delta_+}\rangle $, integrated over two of its arguments\footnote{Note that there is no tadpole diagram for the Schwarzian mode since the Schwarzian propagator vanishes at zero momentum.}.

The result in Fourier space is given by  
\beq\la{2ptback}
\frac{\delta \langle \psi(k)\psi(-k)\rangle'}{ \langle \psi(k)\psi(-k)\rangle_{0}' } =-\frac{\Delta_-(1-\Delta_-)(1-2\Delta_-)}{6}\frac{1}{\phi_r|k|}+ \ldots,
\eeq
where $\langle \psi(k)\psi(-k)\rangle_{0}'$ is the Fourier transform of the correlator in \eqref{eq:2pttree} and the dots are subleading in $1/\phi_r$. 

In the case of nearly-Euclidean-$AdS_2$, the correction to a two-point function of operators of dimension $\Delta_-$ is given by $ \delta \langle \psi(k)\psi(-k)\rangle'_{EAdS, \Delta_-} =\langle \psi(k)\psi(-k)\rangle_{0}' ~\frac{\Delta_-(1-\Delta_-)(1-2\Delta_-)}{6 \tan \pi \Delta_- }\frac{1}{\phi_r|k|}$. 
Notice that as in the case of the four point function, the dS one-loop result can also be understood as coming from reparametrization of the two point function since comparing with \eqref{2ptback} we see that $\delta \langle \psi(\tilde u_1) \psi(\tilde u_2) \rangle_{dS} = -\tan{\pi \Delta_-} ~\delta \langle \psi(\tilde u_1) \psi(\tilde u_2) \rangle_{EAdS, \Delta_-}$. In momentum space, this relation is also valid for the one-loop correction to the two-point function in the global patch. This can be shown by methods similar to appendix \ref{app:identity}.

From \eqref{2ptback} we can see the one-loop correction is negative and decreases correlation at long distances. Even though we can take $|k|\ll 1$ the correction computed in this section is only valid for $|k| \phi_r\gtrsim 1$, otherwise the boundary mode becomes strongly coupled. 

It is instructive to study the correction in position space since it involves the handling of IR divergences. The momentum space one-loop two-point function diverges at small $|k|$. Since we can only trust this for $|k| \phi_r\gtrsim 1$ we will introduce a hard IR cut-off $L_{IR}^{-1}<|k|$, with large $L_{IR} \sim \phi_r $. Then the one-loop correction is given by
\beq
\langle \psi(\tilde{u}) \psi(0) \rangle \sim \frac{1}{|\tilde{u}|^{2\Delta_-}} - \frac{ \Delta_-(1-\Delta_-)}{6\cos \pi \Delta_- \Gamma(2\Delta_-)\phi_r}L_{IR}^{1-2\Delta_-  } + \frac{ \Delta_-(1-\Delta_-)}{6\cot \pi \Delta_-~\phi_r} |\tilde{u}|^{1-2\Delta_- }
\eeq
Since we strictly focus on $|\tilde{u}|\ll L_{IR}$ this implies that the leading correction is negative in position space and gravitational corrections tend to decrease correlations at separate points. To see this we can compare a difference of two-point functions such that the cut-off dependent piece drops out. Moreover, since $\Delta_- < 1/2$, we can see that the gravitational correction decreases as we increase $|\tilde{u}|$, but it decreases slower than the tree level correlator.

\begin{figure*}
	\centering
	\includegraphics[scale=0.7]{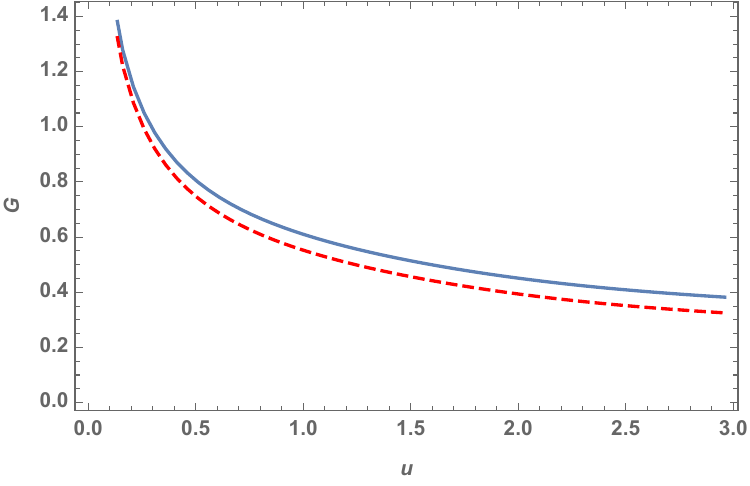}
	\caption{Here we show the two point function in position space.  The top blue (solid) curve is the tree level correlator and the bottom red (dashed) curve is the one-loop corrected correlator. We see that the gravitational effects suppress the correlation at long distances.  For this plot, we chose $\Delta_-={1\over 2}-{\sqrt{3}\over 6}$ and $\phi_r=20$.}
	\label{fig:2ptPositionSpace}
\end{figure*}

Another way to regulate the IR divergences of the one-loop correction is to sum over one-particle-irreducible diagrams.  This gives 
\beq
 \langle \psi(k)\psi(-k)\rangle' \sim \frac{|k|^{2\Delta_- -1}}{1 +\frac{\Delta_-(1-\Delta_-)(1-2\Delta_-)}{6}\frac{1}{\phi_r|k|}+ \ldots},
\eeq
We can do the Fourier transformation of this integral and show that for large distances $|\tilde u|\lesssim \phi_r$, the two-point function decreases due to gravitational corrections (see Figure \ref{fig:2ptPositionSpace}). Even if this expression is convergent, the answer can only be trusted for $|\tilde{u}|\ll \phi_r$. Still, this regulator might be preferred since it comes from resuming a subset of the diagrams.

\subsection{Further comments on corrections to the wavefunction} 

Finally, we would like to point out that we can formally compute the gravitationally dressed wavefunction of the matter field. The exact wavefunction is generically non-gaussian. We can compute it term by term in an expansion for small matter fluctuations $\tilde\psi$ as
\bea
&&\hspace{-1cm}\Psi_s[\phi_b,\tilde\psi] =\Psi_{\rm Schwarzian}[\phi_b,\ell]~ e^{i S_{ct}}\exp\Big[ - i^{2 \Delta_+} c_\nu \int \tilde\psi(\tilde u_1) \tilde\psi(\tilde u_2) \langle G_{\Delta_+}(\tilde u_1,\tilde u_2)\rangle \nonumber\\
&&+ \frac{i^{4 \Delta_+} c_\nu^2}{2}\int \tilde\psi(\tilde u_1)\tilde\psi(\tilde u_2)\tilde\psi(\tilde u_3)\tilde\psi(\tilde u_4) \langle G_{\Delta_+}(\tilde u_1,\tilde u_2)  G_{\Delta_+}(\tilde u_3,\tilde u_4)\rangle_{\rm conn}  +\ldots \Big],
\ea
where in each kernel $\langle G_{\Delta_+}\ldots G_{\Delta_+} \rangle$ one can replace the exact Schwarzian connected correlators computed in \cite{Bagrets:2016cdf, Mertens:2017mtv, Lam:2018pvp, Kitaev:2018wpr, Yang:2018gdb}, identifying $C$ in those references as $C\to i \phi_r$. The dots denote terms that are higher order in $\tilde\psi$. But each of the shown terms also contains 
an expansion in terms of $1/\phi_r$. 
 These terms have the interpretation of self-interactions of the scalar field mediated by gravity. The factor $\Psi_{\rm Schwarzian}[\phi_b,\ell]$ is given in equation \eqref{WdWtot}. 

Let's focus now on the Poincare case. As a simple application of this result, we will see the IR limit of the quadratic kernel in a regime where the gravitational mode is strongly coupled. The Schwarzian correlator behaves as $\langle G_{\Delta_+}(\tilde u,0)\rangle \sim {\tilde u}^{-3/2}$ for $\phi_r\rightarrow 0$. Then the kernel in the quadratic term of the wavefunction behaves as 
\beq \la{QuadWf}
\Psi_s[ \phi_b, \tilde\psi ] \sim \exp\left(- i^{2\Delta_-}\int \frac{dk}{2\pi}  \tilde\psi(k) \tilde\psi(-k) |k|^{1/2} + \ldots   \right),
\eeq
where we rescaled the field $\tilde\psi$ to eliminate the momentum-independent prefactor. This should be contrasted with the leading semiclassical conformal kernel given by $|k|^{2\Delta_+-1}$. Instead, the strongly coupled IR behavior is universal and lies between the massless ($m=0$, $\sim |k|^1$) and threshold case ($m=1/2$, $\sim |k|^0$). We see that strongly coupled gravitational effects in the IR break the conformal invariance of the problem at distances bigger than $\phi_r$. A similar analysis could be done for the higher order kernels in the wavefunction. 

Note, however, that \nref{QuadWf} represents only the 
expansion of the wavefunction around $\tilde \psi =0$ and we expect that the higher order terms are equally important to determine the probabilities. In other words, we expect the wavefunction to be strongly non-gaussian. It would be very nice to  determine its properties, at least in the very long distance regime...

 \section{Conclusions and discussion}

  In this paper we have considered simple two dimensional theories of gravity in nearly-$dS_2$, and discussed 
  some four dimensional situations where they arise as an approximation. We will divide the discussion 
  into a few different topics.

\subsection{Pure JT gravity} 
    
We have discussed pure JT gravity with positive curvature\footnote{For related work see \cite{Anninos:2018svg} and \cite{CJM}.}.
 The main observation is that many features are similar to the more extensively studied case of negative curvature. 
In particular, the computation of the no-boundary wavefunction of the universe is essentially identical to the computation of the partition function for the euclidean $AdS_2$ case. The only difference is the 
change ${ \beta \over \phi_b}  \to - i { \ell \over \phi_b}$. It is perhaps not surprising that this holds for the disk contribution and its perturbative gravity corrections, which include a functional integral over a boundary reparametrization mode. As in the $AdS_2$ case, this mode can be viewed as the result of a spontaneous and explicit breaking of the asymptotic symmetries of $dS_2$.  

 More interestingly,   
 one can also include other topologies in the computation of the wavefunction of the universe. Here we are merely reinterpreting the results of \cite{SSS}. The key point that makes this reinterpretation possible is the observation that in the no boundary computation  of the wavefunction of the universe one considers analytically continued geometries. Among these geometries one can consider geometries which are given by the 
same ones contributing to the asymptotically $AdS_2$ problem, but with  {\it minus} the metric. This makes them 
valid no-boundary configurations. 

As usual in string theory, the sum over topologies is divergent. In \cite{SSS} a resummation was proposed in terms of  hermitian random matrices. We can do exactly the same here. We have the same hermitian matrix model but
we simply consider a different observable, we change $Tr[e^{-\hat \beta H} ] \to Tr[ e^{ i \hat \ell H } ]$.  
Where $\hat \beta$ and $\hat \ell $ are the renormalized lengths of the boundary (i.e. the proper lengths divided
by the value of the dilaton, for example).  
  
Now, we can wonder what the interpretation of this Hamiltonian is. In the $AdS_2$ case it is supposed to be
 the Hamiltonian of the dual quantum system, or the full microscopic Hamiltonian of the theory, which 
 describes the unitary evolution of black hole microstates.  
In the $dS_2$ case, it is associated to translations in space, rather than translations in time. 
Nevertheless, 
we know that in general de Sitter dynamics, processes that occur   in the static patch of de Sitter leave their imprint on spatial correlators at infinity. And this evolution in space is the same generator that produces evolution along the static patch. Therefore it is tempting to think that this Hamiltonian and its states are related to the states in the static patch. Normally the evolution of fields in the static patch give rise to quasinormal modes, particles can fall into the horizon. But these results suggest that perhaps there is a dual description is fully unitary. Now, before reaching such a strong conclusion we should remember that the pure JT theory has no matter, so the only modes we have are the microstates themselves and we cannot probe them with simple fields. 
We should also mention the prefactor in the density of states in 
e.g. \nref{Dens}, which in $AdS_2$ was the extremal entropy of the black hole, here it is equal to half of the usual  $dS_2$ entropy of the static patch. It looks as if it is only the contribution of one of the two horizons of the static patch.

Given that we have been talking about the energies of the Hamiltonian $H$ appearing above, it is worth asking for
a classical picture of what states with different ``energies'' look like. 
These correspond to spacetimes in the Milne coordinates, as in \nref{Milne}, with various values of $\phi_m$. 
These can be viewed as black holes in de-Sitter, and we have that $E \propto \phi_m^2 $. So the extremal value 
corresponds to $E\to 0$.  When we connect this picture to the four dimensional picture involving a four dimensional black hole, then $E \propto - (\mu - \mu_e)$. Such black holes have a maximum mass and the 
extremal value corresponds to $E=0$ and larger values of $E$ correspond to  masses smaller 
 than the extremal value.

    Here we have discussed geometries with a single boundary. When we compute expectation values we will get geometries with two boundaries, one for the bra and one for the ket of the wavefunction. A new aspect of those configurations is that we will have geometries connecting the two boundaries. The matrix model certainly will lead to such geometries when we compute $\langle Tr[e^{ i \hat \ell  H } ] Tr[ e^{ - i \hat 
\ell H } ] \rangle$. We will leave a more detailed analysis of those geometries to the future. 

However, we cannot resist making a comment. The average over Hamiltonians that looks a bit funny in the 
asymptotically $AdS_2$ problem seems more natural from the $dS_2$ point of view. 
The idea is that, perhaps, some version of $dS/CFT$ is true. But, as in $AdS$, in order to have a completely well defined quantum system we need to go all the way to the boundary. On the other hand, when we do observations 
at finite times in de-Sitter, then we do not wait forever, we consider the bra and the ket and we integrate over what will happen in the future. It seems plausible that after doing this integration we do not have a definite Hamiltonian any longer, but we have an average over quantum systems.

\subsection{Gravitational corrections in two dimensions} 

 If we consider matter in $AdS_2$, we can simply compute the gravitational corrections to their propagators 
 by including the quantum mechanics of the boundary reparametrization modes. 
 In the regime where the quantum gravity corrections are small, we can include the effects of the reparametrization modes also for the $dS_2$ observables. These can be the wavefunction of the universe, or
 cosmological correlation functions. These are computed on a slice with constant value of $\phi$ and as a function of the proper time along the slice. The computation involves {\it two} reparametrization modes, one for the ``bra'' and one for the ``ket''. We have discussed two types of corrections, first a
  tree level correction to the four point function. This is a ``graviton exchange" diagram. 
  And then a loop correction to the two point function. (In four dimensions they were computed in \cite{Seery:2008ax} and \cite{Giddings:2010nc} respectively).
     The main point of these computations is that the simplicity of two dimensional gravity enables us to do them explicitly and in a relatively simple manner. Loop corrections in inflation are notoriously difficult and this offers a simple context where they are calculable. As far as we know the loop correction due to gravity has not been evaluated in standard single field inflation.  
 Perhaps these simple computations could be useful for researchers trying to understand the behavior of IR fluctuations in cosmology. 
 
 In the $AdS_2$ case, one can sum the whole gravitational perturbation theory for some observables, such as the two or higher point functions, where exact expressions can be obtained \cite{Mertens:2017mtv, Lam:2018pvp, Kitaev:2018wpr, Yang:2018gdb}. 
   In $dS_2$ this has some implications for the first derivatives of the wavefunctional around zero values of the field. However, observables in the 
   $dS_2$ are more complicated because we are interested in the wavefunctional for more general values of the boundary fields. This involves computing higher and higher order correlators and we could not see how to 
   sum exactly the effects of the boundary mode. 
   For example, one could wonder whether for the de-Sitter two point expectation values we can also integrate out the reparametrization modes exactly \footnote{We thank L. Iliesiu for discussions on this point.}.

\subsection{Relation to SYK} 

The formula for the wavefunction of the universe is the same as the one we would have obtained if we 
considered the low energy Euclidean partition function for the SYK model \cite{Sachdev:1992fk, KitaevTalks}, and then continued the inverse temperature as $\beta \to - i \ell$. The net effect is that we are computing 
$Tr[ e^{ i \ell H_{SYK} } ]$. But this looks like a Lorentzian SYK computation, were we evolve the  system backwards in time, over a time $\ell$ and compactify the time direction, restricting to the low energy sector. The Lorentzian computation would also get a contribution from the very high energy sector of the SYK spectrum, since the spectrum of SYK is statistically symmetric under flipping the sign of the energy.
 Maybe this should be viewed as another oscillatory branch of the wave function. An equivalent continuation would 
 be to take the SYK model in Euclidean space, on a circle of length $\ell$, and then take $J \to - i J$. 
 
 The matter correlators that we considered for light scalar fields, are also the same as the ones we would have obtained for the time ordered correlators in Lorentzian signature, at least for bosonic fields. 
 If we consider light fermionic fields, then the results seem a bit different since the dimension of the fermion is related to the mass by $\Delta = \half  \pm m$ in Euclidean signature or in $AdS$ and 
 $\Delta = \half \pm i m $ in de Sitter. So standard massive fermions in de-Sitter would lead boundary operators of complex dimensions. This is not what we have in the standard SYK model\footnote{An exception is the case of massless bulk fermions and $q=2$ SYK, which might be seen as a lower dimensional version of \cite{Anninos:2011ui}. We thank D. Anninos for discussion on this.}, but perhaps there are some variations of the model that produce this structure... 
 
 Another interesting point is that when we compute expectation values we are coupling the ``bra'' and the 
 ``ket''. This looks like the coupling of two models (as in e.g. \cite{Maldacena:2018lmt}).

\vspace{0.75cm}
{\bf Acknowledgments } 

We thank  L. Iliesiu, R. Mahajan for initial collaboration on this project.
We thank A. Almheiri, D. Anninos, V. Gorbenko, J. Hartle, G. Horowitz, D. Marolf, S. Shenker, E. Silverstein, D. Stanford and  A. Vilenkin for discussions. J.M. is supported in part by U.S. Department of Energy grant
de-sc0009988 and by the Simons Foundation grant 385600. G.J.T is supported by a Fundamental
Physics Fellowship. G.J.T. thanks the IAS for hospitality while this work was being completed.
Z.Y is supported by Charlotte Elizabeth Procter Fellowship from Princeton University.

We benefited from the workshop ``Chaos and Order: From
strongly correlated systems to black holes" at KITP, supported in part by the National
Science Foundation under Grant No. NSF PHY-1748958 to the KITP.

\appendix

\section{Partition functions and Nearly-$AdS_2$ WdW wavefunctions }
\la{AdSPartition}

 In this appendix we explain the connection between solutions of the Wheeler de Witt equation and 
 the partition function of the dual theory in the nearly-$AdS_2$ case. 
 
The idea is that we should view the partition function as an inner product of two solutions of the 
Wheeler de Witt equation. One of the solutions $\Psi_{IR}$ is determined by the sum over smooth geometries,
 or the Euclidean analog of the no boundary proposal. 
The other, $\Psi_{UV}$, is determined by the fact that we fix some of the metric parameters at the 
boundary of $AdS_2$. In particular we fix the temperature, or the renormalized length of the circle. 

Writing the metric as $ds^2 = e^{ 2 \hat \rho} ( dz^2 + d\varphi^2)$, with $\varphi \sim \varphi + 2 \pi$, 
we find that for Euclidean nearly-$AdS_2$, the Wheeler de Witt equation takes the form 
\be  
( - \partial_\phi \partial_{\hat \rho} + 16 \pi^2 \phi e^{ 2 \hat \rho} )  \Psi =0 ~, ~ 
~ ~~~{\rm or} ~~~ [ 4 \partial_u \partial_v + 4] \Psi=0 ~,~~~~ u \equiv \phi^2 ~,~~~~~v\equiv - (2\pi)^2 e^{2\hat \rho} 
\ee
This can be compared with \nref{WdWE}. We have defined $u$ and $v$ so that we get exactly the same wave 
equation for  a massive field as in \nref{WdWE}. But now we are looking at that equation in the Rindler wedge 
\bea
ds^2 &=& - du dv = d R^2 - R^2 d\sigma^2 ~,
\\
 & ~& ~ {\rm with} ~~~~~~~R=\sqrt{-uv} = \phi e^{\hat \rho} = \phi \beta ~,~~~~~~e^\sigma = \sqrt{ -v \over u} = { \beta \over \phi } 
 \eea
 where $\beta = 2\pi e^{\hat \rho}$ is the proper length of the circle. 
For large $R$ the solutions go like  
\be \la{GSoE}
\Psi_\pm   = { 1 \over \sqrt{R} } e^{ \pm  2 R } f_\pm(\sigma )   
\ee
where $f_\pm$  are arbitrary functions. 
The two solutions we are interested in are 
\bea \la{PsiIR}
\Psi_{IR} &=&  { 1 \over \phi_b } \left( {\phi_b \over \beta } \right)^{3\over 2 }  \exp\left[   2 \phi_b \beta  +  S_{0,AdS}  +   { 4 \pi^2  \phi_b\over \beta} \right]   ~,~~~~ f_+(\sigma) = 
 e^{-\sigma}   
 e^{ 4 \pi^2 e^{-\sigma }}
\\ \la{PsiUV}
\Psi_{UV} &=&   \delta\left( { \beta  } - { \beta_0 }  \right)  \exp\left[ { - 2 \phi_b \beta } \right] ~,~~~~~~~f_-(\sigma ) = e^{ - {\sigma \over 2}}  \delta ( \sigma - \sigma_0) 
\eea
Both of these functions have the general form \nref{GSoE} with  for different choices of function $f_\pm$.
\nref{PsiIR} is the  growing solution  and \nref{PsiUV} is the decaying solution.
 For $\Psi_{IR}$ the function $f_+$ was computed by summing over smooth geometries,  
 or the no boundary idea applied to the 
Euclidean $AdS$ problem. For $\Psi_{UV}$ we  chose a function $f_-$ that fixes the length of the circle. This 
is just a choice because we want to compute the answer as a function of $\beta_0/\phi_b$. Notice
that this ratio is what is normally called the ``renormalized temperature'' since 
both $\beta$ and $\phi$  are proportional to $1/z$ when  $z \to 0$. In other words, on a solution of the bulk gravity 
equations, $\beta/\phi$ is 
independent of $z$ near $z=0$.   The factor of 
 $e^{ - {\sigma \over 2} }$ in $\Psi_{UV}$  was chosen so that we get the expected answer as function of     $\sigma_0$ after we take the inner product below, \nref{PartAdS}. 
 However, it is likely that it could be fixed from first principles. In fact, if we look at the  
 first expression for $\Psi_{UV}$, we note that we get a pure delta function for $\sigma$. 

Note that the good classical solutions obey $\phi = \phi_h \cosh \tau$ and $\beta = 2\pi \sinh \tau$, so that
${ u \over \phi_h^2} + { v \over ( 2 \pi)^2 } =1$. This implies that for $u=0$, all solutions go through 
the point $v = (2\pi)^2$, which is exactly 
 the same as the one we had found in the de-Sitter discussion, see figure \ref{SuperSpaceFig}(b). This point is
outside the Rindler wedge we are considering, which is the left wedge   in figure \ref{SuperSpaceFig}(b). 

Then the Klein Gordon inner product gives us the partition function, 
or the object that we view as the partition function of the dual boundary theory
 \be
  Z\left({  \beta_0 \over \phi_b } \right)= \langle \Psi_{UV} , \Psi_{IR} 
  \rangle = \left( {\phi_b \over \beta_0 } \right)^{3\over 2 }  \exp\left[    S_{0,AdS}  +   { 4 \pi^2  \phi_b\over \beta_0} \right]  \la{PartAdS}
\ee
The inner product, computed on a surfaced of fixed $\phi=\phi_b$, 
 involves a $\beta$ derivative and a $\beta$ integral, as in \nref{KGProd}. The derivative only acts on the 
leading exponential, bringing down a factor of $2 \phi_b$. The integral is done by using the $\delta$ function in \nref{PsiUV}.  

Note that in contrast with the situation  in \cite{Heemskerk:2010hk,Lee:2009ij}, here $\Psi_{UV}$ 
{\it does } solve the Wheeler de Witt equation in the asymptotic region. 
The fact that the inner product does not depend on the choice of slice, seems related to the invariance
of the partition function under the choice of UV regulator. 
Notice that the exponential dependence on $R$  in \nref{PsiIR} and \nref{PsiUV} is such that it cancels in 
the computation of the partition function \nref{PartAdS}. We can view this as the subtraction of the 
``counterterms'' that are UV divergent. The fact that the Klein Gordon product is independent of the 
slice where it is evaluated can be interpreted as the statement of the renormalization group flow. 

  It seems very likely that a similar analysis could be done also in the higher dimensional case, but we leave this to the future.

\section{Details on Matter Correlators}

\subsection{Matter Partition Function}\label{App: MatterPartition}

\subsubsection{Partition Function on H$_2$}
As argued in section \ref{sec:HHWavefunction}, path integral over the global dS$_2$ geometry can be achieved by analytic continuing the AdS$_2$ result with negative metric.
This allows us to relate the matter partition function on dS$_2$ with the partition function on H$_2$, and the finite piece is unchanged under analytic continuation. 

Considering a scalar field $\psi$ with mass $m$ in dS$_2$, this corresponding to a scalar field on H$_2$ with imaginary mass $im$ (coming from the flip of sign of the metric). That is the Klein-Gordon equation is $(\Box+m^2)\psi=0$.
After we expand in angular momentum modes on the circle, indexed by $k$, we have the radial equation ($ds^2 = dr^2 + \sinh^2 r d\varphi^2 $):
\be
{ 1 \over \sinh r } \partial_r ( \sinh r \partial_r \psi) - { k^2 \over \sinh^2 r }\psi  + m^2 \psi =0 
 \ee
 Choosing the solution that are regular and normalized at $r=0$ we 
   get 
   \bea
 \psi & =&    { ( \tanh r )^{|k|} \over (\cosh  r)^{\Delta_+}  }  ~_2F_1( { 1 \over 4 } + { |k| \over 2} + {\nu\over 2} ,  { 3 \over 4 } + { |k| \over 2} + {\nu\over 2} ,1+|k|, \tanh^2 r ) 
 \cr
 & & \Delta_{\pm} = \half \pm \sqrt{ { 1 \over 4 } - m^2 } ~,~~~~~ \nu =\sqrt{ { 1 \over 4 } - m^2 } 
  \eea
  Now, to compute the determinant we will use the Coleman formula \cite{coleman_1985} which expresses it in terms of the value of these functions at the boundary  $r=r_c \gg 1$. 
  \be
  \psi\sim {\Gamma(1+|k|) 2^{|k|}\over \sqrt{\pi}}\left[{1\over (2\cosh r_c)^{\Delta_-}}{\Gamma(\Delta_+-{1\over 2})\over ~\Gamma(\Delta_++|k|)}+{1\over (2\cosh r_c)^{\Delta_+}}{\Gamma(\Delta_--{1\over 2})\over ~\Gamma(\Delta_-+|k|)}\right]
  \ee
   The Coleman formula then tells us that the partition function with Dirichlet (or Neumann) boundary condition is the logarithm sum of the coefficient of the leading (subleading) vanishing piece.
   We will only focus on the finite piece that depends on the mass parameter, which is the following:
   \be\label{HyperbolicPartition}
   \log Z_{D/N} = { 1 \over 2 } \sum_{k=-\infty}^\infty \log \Gamma(\Delta_{+/-} + |k|).
\ee
 
 \subsubsection{Partition function on S$^2$}
 
 The partiton function on S$^2$ is given by considering the euclidean theory and summing over all the modes of the laplacian 
 \be
 - \nabla^2 + m^2 = l ( l+1) + m^2 = \lambda_l 
 \ee
 We get 
 \be\label{SpherePartition}
 \log Z_{S^2} = - \half \sum_{l=0}^\infty (2l+1) \log \lambda_l =-\half \sum_{l=0}^{\infty}(2l+1)\log(l(l+1)+m^2)
\ee

\subsubsection{S$^2$ partition function and Z$_D$Z$_N$}
We will show that the sphere partition function (\ref{SpherePartition}) is the product of partition function on H$_2$ with Dirichlet and Neumann boundary condition (\ref{HyperbolicPartition}).
The general argument is the following, when we do path integral over S$_2$, we can do Lorentizian evolution at its equator to global dS$_2$ forwards and backwards. Such Lorentizian evolution does not affect the path integral since they cancel, but meanwhile one can write the full path integral on this geometry as:
\be
\int \mathcal{D} \psi_0 |\Psi_m|^2\sim Z_D^2\int \mathcal{D}\psi_0 \exp\left[-C\int \int{\psi_0(u)\psi_0(u')\over (2\sin {u-u'\over 2})^{2\Delta} }\right]\sim Z_D^2 {\rm Det}\left[{1\over (2\sin {u-u'\over 2})^{2\Delta}}\right]^{-\half}
\ee
where we are ignoring local pieces during the derivation.
Meanwhile, $Z_N$ is related to $Z_D$ by Legendre of $\psi_0$ at the boundary, the difference of which is precisely the determinant:
\be
Z_N=\int \mathcal{D}\psi_0 e^{iS(\psi)}\sim Z_D\int \mathcal{D}\psi_0 \exp\left[-C'\int \int{\psi_0(u)\psi_0(u')\over (2\sin {u-u'\over 2})^{2\Delta} }\right]\sim Z_D {\rm Det}\left[{1\over (2\sin {u-u'\over 2})^{2\Delta}}\right]^{-\half}
\ee
In consequence, the sphere partition function is equal to $Z_DZ_N$.

We can check this relation explicitly with (\ref{SpherePartition}) and (\ref{HyperbolicPartition}).  Taking derivative of $\log Z_{S^2}$ with respect to $m^2$, we have \footnote{We use the relation for digamma function $\psi(z)=-\sum\limits_{k=0}^{\infty}{1\over z+k}-\gamma+\sum\limits_{k=1}^{\infty}{1\over k}$. In the derivation we only keep the mass dependence.}:
\bea
\partial_{m^2}\log Z_{S^2}=-{1\over 2}\sum_{l=0}^{\infty} {1\over l+\half +\nu}+{1\over l+\half -\nu}=\half\left[\psi(\half+\nu)+\psi(\half -\nu)\right]
\eea

Taking derivative of $\log Z_{D/N}$ with respect to $\Delta_{\pm}$, we have:
\bea
\partial_{\Delta}\log{Z}&=&-{1\over 2}\psi(\Delta)+\sum_{k=0}^{\infty}\psi(\Delta+k)=-\half \psi(\Delta)-\sum_k\sum_l {1\over \Delta+k+l} \nonumber \\
&=&-\half \psi(\Delta)+(\Delta-1)\sum_{m}{1\over \Delta+m}=({1\over 2}-\Delta)\psi(\Delta).
\eea
Therefore we have
\be
\partial_{m^2}\log (Z_DZ_N)={1\over 2\nu}(\partial_{\Delta_-}\log Z_N-\partial_{\Delta_+}\log Z_D)={1\over 2}\left[\psi({1\over 2}+\nu)+\psi({1\over 2}-\nu)\right]=\partial_{m^2}\log Z_{S^2}.
\ee

\subsection{Backreaction: Perturbation theory}\la{App:FeynmanDiagrams}

In this appendix we will give details on how to use perturbation theory to compute gravitational backreaction on matter correlators. For simplicity we will focus on the Poincare patch.

Instead of doing perturbation theory with the matter field and the two copies of the Schwarzian mode, here we will use a different method to illustrate another possible approach. We will begin by integrating out the Schwarzian mode in the wavefunction, which give an expansion in gravity mediated self-interactions. The wavefunction is 
\bea
&&\Psi[\tilde{\psi}] = \exp\Big[ - i^{2 \Delta_+} c_\nu\int \tilde{\psi}(u_1) \tilde{\psi}(u_2) \langle G^\varphi_{\Delta_+}(u_1,u_2)\rangle \nonumber\\
&&+ i^{4 \Delta_+}\frac{c_\nu^2}{2}\int \tilde{\psi}(u_1)\tilde{\psi}(u_2)\tilde{\psi}(u_3)\tilde{\psi}(u_4) \langle G^\varphi_{\Delta_+}(u_1,u_2)  G^\varphi_{\Delta_+}(u_3,u_4)\rangle_{\rm conn}  +\ldots \Big],
\ea
We will work to leading order $1/\phi_r$ and therefore we can take only the leading correction to the $\tilde{\psi}^2$ and $\tilde{\psi}^4$ kernels. Using the perturbative Schwarzian correlators from \cite{Maldacena:2016upp} we can expand to this order 
\begin{eqnarray}
\Psi[\tilde{\psi}] &=& \exp\Big[- i^{-2 \Delta_-}  \tilde c_\nu \int \frac{dk}{2\pi} \tilde{\psi}_k \tilde{\psi}_{-k} |k|^{2\Delta_+-1} \nonumber\\
&&- i^{2 \Delta_+ -1} \tilde c_\nu \frac{\Delta_+(\Delta_+-1)(1-2\Delta_+)}{6\phi_r \tan \pi \Delta_+}  \int \frac{dk}{2\pi} \tilde{\psi}_k \tilde{\psi}_{-k}  |k|^{2\Delta_+ -2}\nonumber\\
&&+\frac{i^{4 \Delta_+-1}}{2\phi_r}\tilde c_\nu^2 \int \tilde{\psi}_{k_1} \ldots \tilde{\psi}_{k_4} K(k_1\ldots k_4) \Big],
\end{eqnarray}
where the term in the second line is the leading correction to $\langle G^\varphi_{\Delta_+}(u_1,u_2)\rangle$ while the term in the third line is the fourier transform of the connected four point function for the Schwarzian in Euclidean signature $\langle G^\varphi_{\Delta_+}(u_1,u_2)G^\varphi_{\Delta_+}(u_3,u_4)\rangle_{\rm conn}$. This Fourier transform can be easily computed starting from equation (3.126) of \cite{Maldacena:2016hyu}, to leading order in large $C$, and gives the result quoted in equation \eqref{4ptFourier}. One can do the same with the global correlators but, as we saw in the main text, the expressions are more complicated.
 
Taking the absolute square and redefining $\tilde{\psi} \to \tilde{N}_\nu^{1/2} \tilde{\psi}$ to simplify the tree level propagator term, we get a probability distribution given by
\bea\la{WFSQ}
|\Psi[\tilde{\psi}]|^2 &=& \exp\{- \frac{1}{2}  \int \frac{dk}{2\pi} \tilde{\psi}_k \tilde{\psi}_{-k} |k|^{2\Delta-1} -\frac{1}{2}\frac{\Delta_+(\Delta_+-1)(2\Delta_+-1)}{6\phi_r}  \int \frac{dk}{2\pi} \tilde{\psi}_k \tilde{\psi}_{-k}  |k|^{2\Delta_+-2}\nonumber\\
&&+\frac{\tan \pi \Delta_+}{8\phi_r} \int \tilde{\psi}_{k_1} \ldots \tilde{\psi}_{k_4} K(k_1\ldots k_4) \}
\ea 
We can write Feynman diagrams if we interpret $\int [d\tilde{\psi}] |\Psi[\tilde{\psi}]|^2$ as an in-out path integral with action $  - \log |\Psi[\tilde{\psi}]|^2$. We denote interactions by blue dots, given by the terms written in the equation above. 

Finally, when computing in-in correlators we should normalize by the norm of the state, which also might have gravitational corrections. As well known in QFT, this is taken care of by simply ignoring disconnected graphs. Therefore, for the two point function, we have a sum of only these three terms
\beq\la{2ptfuncfeyndiag}
\langle \psi_{k} \psi_{-k} \rangle_{\rm in-in}=  \ 
 \raisebox{-1mm}{ \begin{tikzpicture}[scale=0.6, baseline={([yshift=0cm]current bounding box.center)}]
\draw[thick] (-2,0) -- (2,0);
\draw[fill,black] (-2,0) circle (0.1);
\draw[fill,black] (2,0) circle (0.1);
\draw (-0,-.4) node {\small $k$};
\end{tikzpicture}}
~~+  \ 
 \raisebox{0.5mm}{ \begin{tikzpicture}[scale=0.6, baseline={([yshift=0cm]current bounding box.center)}]
\draw[thick] (-2,0) -- (2,0);
\draw[fill,black] (-2,0) circle (0.1);
\draw[fill,black] (2,0) circle (0.1);
\draw (-0,.4) node {\small };
\draw (-1.1,-0.4) node {\small $k$};
\draw (1.1,-0.4) node {\small $k$};
\draw[fill, blue] (0,0) circle (0.12);
\end{tikzpicture}}
~~+  \ 
 \raisebox{2.2mm}{ \begin{tikzpicture}[scale=0.6, baseline={([yshift=0cm]current bounding box.center)}]
\draw[thick] (0,0.6) circle (0.6);
\draw[thick] (-2,0) -- (2,0);
\draw[fill,black] (-2,0) circle (0.1);
\draw[fill,black] (2,0) circle (0.1);
\draw (-1.1,-0.4) node {\small $k$};
\draw (1.1,-0.4) node {\small $k$};
\draw (0,0.7) node {\small $p$};
\draw[fill, blue] (0,0) circle (0.12);
\end{tikzpicture}}
\eeq

The first term in the right hand side is given by the tree level contribution $G(k)=|k|^{2\Delta-_-1}$. The second term is the leading quadratic interaction in the exponent of \eqref{WFSQ}, which gives 
\beq
   \ 
 \raisebox{0.5mm}{ \begin{tikzpicture}[scale=0.6, baseline={([yshift=0cm]current bounding box.center)}]
\draw[thick] (-2,0) -- (2,0);
\draw[fill,black] (-2,0) circle (0.1);
\draw[fill,black] (2,0) circle (0.1);
\draw (-0,.4) node {\small };
\draw (-1.1,-0.4) node {\small $k$};
\draw (1.1,-0.4) node {\small $k$};
\draw[fill, blue] (0,0) circle (0.12);
\end{tikzpicture}} ~~=-|k|^{1-2\Delta_+} \frac{\Delta_+(\Delta_+-1)(2\Delta_+-1)}{6} \frac{1}{\phi_r|k|}
\eeq

The third term can be computed using \eqref{4ptFourier} and gives 
\bea
 \ 
 \raisebox{2.2mm}{ \begin{tikzpicture}[scale=0.6, baseline={([yshift=0cm]current bounding box.center)}]
\draw[thick] (0,0.6) circle (0.6);
\draw[thick] (-2,0) -- (2,0);
\draw[fill,black] (-2,0) circle (0.1);
\draw[fill,black] (2,0) circle (0.1);
\draw (-1.1,-0.4) node {\small $k$};
\draw (1.1,-0.4) node {\small $k$};
\draw (0,0.7) node {\small $p$};
\draw[fill, blue] (0,0) circle (0.12);
\end{tikzpicture}} ~~&=&\frac{\tan \pi \Delta_+}{\phi_r} ~G^2(k) \int \frac{dp}{2\pi}~ G(p)K(k,p;-k,-p)\nonumber\\
&=& |k|^{1-2\Delta_+} \frac{\Delta_+(\Delta_+-1)(2\Delta_+-1)}{3} \frac{1}{\phi_r |k|}
\ea
Contractions involving $K(k,-k;p,-p)=0$ automatically vanish. Notice that the final answer for these two diagrams are proportional to each other. This can be shown in general and is also true for global correlators. The proof is more complicated and can be done by using the method presented in the next subsection. 

Also, to each order in $\phi_r$ we need to renormalize the wavefunction. This is standard in QFT and its accounted for by neglecting disconnected graphs. Putting these diagrams together we recover the result presented in equation \eqref{2ptback}. 

Using the Feynman rules derived above, we can compute the connected in-in four-point function in Fourier space. It is given only by the diagram 
\beq
  \ 
 \raisebox{-1mm}{ \begin{tikzpicture}[scale=0.7, baseline={([yshift=-0.15cm]current bounding box.center)}]
\draw[thick] (-1,-1) -- (1,1);
\draw[thick] (-1,1) -- (1,-1);
\draw[fill,black] (-1,-1) circle (0.1);
\draw[fill,black] (-1,1) circle (0.1); 
\draw[fill,black] (1,1) circle (0.1); 
\draw[fill,black] (1,-1) circle (0.1); 
\draw[fill, blue] (0,0) circle (0.14);
\end{tikzpicture}}~~~= G(k_1)G(k_2)G(k_3)G(k_4) (K(k_1,k_2;k_3,k_4)+K(k_1,k_3;k_2,k_4)+\ldots )
\eeq
In the formula above we need to sum over $K$'s with different pairings. This is equivalent to a sum over different channels in the bulk. 

Using the explicit expression in \eqref{4ptFourier} we can easily show that 
\beq
G_\Delta(k_1)G_\Delta(k_2)G_\Delta(k_3)G_\Delta(k_4) K_\Delta(k_1,k_2;k_3,k_4) = K_{1-\Delta}(k_1,k_2;k_3,k_4),
\eeq
where $G_\Delta(k)=|k|^{2\Delta-1}$. Therefore, in our calculation of the four-point function, the effect of the convolution with the two-point function is simply to replace $\Delta_+ \to \Delta_-$. We will give a general proof of this identity in the next subsection that will also be valid for global correlators. The final result for the 4pt function is therefore proportional to the Euclidean AdS correlator as anticipated in section \ref{treelevel}
\beq
  \ 
 \raisebox{-1mm}{ \begin{tikzpicture}[scale=0.7, baseline={([yshift=-0.15cm]current bounding box.center)}]
\draw[thick] (-1,-1) -- (1,1);
\draw[thick] (-1,1) -- (1,-1);
\draw[fill,black] (-1,-1) circle (0.1);
\draw[fill,black] (-1,1) circle (0.1); 
\draw[fill,black] (1,1) circle (0.1); 
\draw[fill,black] (1,-1) circle (0.1); 
\draw[fill, blue] (0,0) circle (0.14);
\end{tikzpicture}}~~~= K_{1-\Delta}(k_1,k_2;k_3,k_4)+K_{1-\Delta}(k_1,k_3;k_2,k_4)+K_{1-\Delta}(k_1,k_4;k_2,k_3)
\eeq
where we sum over bulk ``s", ``t" and ``u" channels. 

As a final comment, we could also use these diagrams to compute corrections to the norm of the HH state. This would be given by two diagrams where we  close the lines in the two point functions in \eqref{2ptfuncfeyndiag}. When we correct for permutation factors, these two diagrams cancel. Therefore, the leading correction, if any, to the norm appears at order at least $1/\phi_r^2$. 

\subsection{Four-point function identity} \la{app:identity}
To show the identity \eqref{4ptidentity} we will need the following result. If we start from the reparametrized two point function $G_{\Delta_+}^\varphi(u_1,u_2)=\Big( \frac{\varphi'(u_1)\varphi'(u_2)}{4 \sin^2 \frac{\varphi(u_1)-\varphi(u_2)}{2}}\Big)^{\Delta_+}$ then we can show that 
\beq\la{Ginverse}
\int du' G^{\varphi}_{\Delta_+}(u_1,u')G^{\varphi}_{\Delta_-}(u',u_2) =\frac{2 \pi \tan \pi \Delta_+}{2\Delta_+ -1} \delta(u_1-u_2)
\eeq
which is valid for an arbitrary $\varphi(u)$. This follows from the reparametrization symmetry of this integral which allows us to undo $\varphi(u)$. This is the same argument that allows to solve the low energy Schwinger-Dyson equation in SYK \cite{KitaevTalks}.

Using this result, we can write the following trivial identity
\beq\la{GGGid}
G^{\varphi}_{\Delta_-}(u_1,u_2) = \frac{2\Delta_+ -1}{2 \pi \tan \pi \Delta_+}\int du_a du_b ~G^{\varphi}_{\Delta_-}(u_a,u_1) G^{\varphi}_{\Delta_+}(u_a,u_b) G^{\varphi}_{\Delta_-}(u_b,u_2)
\eeq
valid for any $\varphi (u)$. When we expand $\varphi(u) = u + \varepsilon(u)$, this identity will give us non-trivial relations order by order. To linear order, the right hand side will produce a sum of 3 terms for each propagator. Two of them involve a reparametrization of the $G_{\Delta_-}$ propagators. Using the identity \eqref{Ginverse}, both terms give the same answer
\beq
 \frac{2\Delta_+ -1}{2 \pi \tan \pi \Delta_+} \int du_a du_b~ \delta_\varepsilon G_{\Delta_-}(u_a,u_1) G^0_{\Delta_+}(u_a,u_b) G^0_{\Delta_-}(u_b,u_2) =\delta_\varepsilon G_{\Delta_-}(u_1,u_2)
\eeq
where to simplify notation we denote $G^0_{\Delta_+}(u_1,u_2)=( 4 \sin^2 \frac{u_1-u_2}{2})^{-\Delta_+}$. The other term proportional to $\delta_\varepsilon G_{\Delta_-}$ can be similarly obtained from the formula above exchanging $u_a \leftrightarrow u_b$ and give the same final result. Moving these two terms to the left hand side and including the term proportional to $\delta_\varepsilon G_{\Delta_+}$ we can deduce that to linear order \eqref{GGGid} is equivalent to  
\beq
\delta_\varepsilon G_{\Delta_-}(u_1,u_2) =- \frac{2\Delta_+-1}{2 \pi \tan \pi \Delta_+} \int du_a du_b ~G^0_{\Delta_-}(u_a,u_1)  G^0_{\Delta_-}(u_b,u_2)~ \delta_\varepsilon G_{\Delta_+}(u_a,u_b)
\eeq
Multiplying this identity with a similar expression involving $(u_1,u_2) \to (u_3,u_4)$ we obtain the equation \eqref{4ptidentity}.

So far we looked at the identity derived from expanding \eqref{GGGid} to linear order in $\varepsilon$. We can do the same to second order but we will not give the details here. This will give us the identity needed to show that the one loop correction to the two-point function in the global patch satisfies $\delta \langle \psi(\tilde u_1) \psi(\tilde u_2) \rangle_{dS} = -\tan{\pi \Delta_-} ~\delta \langle \psi(\tilde u_1) \psi(\tilde u_2) \rangle_{EAdS, \Delta_-}$.

\subsection{Linearized Graviton Calculation}\la{sec:almheirikang}

In section (\ref{App:FeynmanDiagrams}), we solve the gravitational backreaction using the Schwarzian techniques where we fix the metric and leave the dilaton fluctuates.  In this appendix, we use a different method to study the gravitational backreaction.  
This method corresponds to choose a gauge to fix the dilaton field and then study the metric fluctuation.  Such a linearized graviton method was studied in $AdS_2$ by Almheiri and Kang \cite{Almheiri:2016fws}.

Using the time and spatial diffeomorphism transformations, we can fix the metric and dilaton to the following form:
\begin{equation}
ds^2={1\over \eta^2}(-e^{h-g}d\eta^2+e^{h+g}dx^2);~~~~~~\phi={1\over\eta}
\end{equation}
where $h$ and $g$ describes the metric fluctuation. By a direct evaluation, one can find the JT action for $h, g$ to quadratic order: 
\begin{equation}
iS=i\int \phi(R-2)=i\int d\eta dx {1\over \eta^2}g\partial_{\eta}h-{h^2\over \eta^3}.
\end{equation}
Under a change of variable $h\rightarrow h-{\eta^3\over 2}\partial_{\eta}({g\over \eta^2})=h+g-{\eta\over 2}\partial_{\eta}g$, we now have:
\begin{equation}
iS=-i\int d\eta dx {1\over 4}{(\partial_{\eta} g)^2\over \eta}+{h^2\over \eta^3}.
\end{equation}
The new $h$ field does not contain a kinetic term and is not propagating, so we can fix it to its classical value 0. The action for $g$ has only time derivative, and this means $g$ only propagates in time direction but not in spatial direction.  This phenomenon is consistent with the fact that two-dimensional dilaton gravity only contains a single quantum mechanical degree of freedom. 
Including a massive scalar field $\psi$, the full action (to quadratic order) becomes\footnote{This analysis is only valid at tree level, at one-loop level one need also consider the effect of ghost modes coming from gauge fixing.}:
\begin{equation}\label{LinearizedAction}
\begin{split}
iS&=-{i\over 2}\int d\eta dx {1\over 2}{(\partial_{\eta} g)^2\over \eta}-(\partial_{\eta}\psi)^2+(\partial_x\psi)^2+{m^2\over \eta^2}\psi^2-g[(\partial_{\eta}\psi)^2+(\partial_x\psi)^2-{m^2\psi^2\over 2\eta^2}-{m^2\over 2\eta}\partial_{\eta}(\psi^2)]\\
\end{split}
\end{equation}
We see that the coupling constant for $g$ is decreasing at time grows. Therefore there is a competition between the diminishing of gravitational coupling and the matter proliferate during the history of $dS_2$ inflation. 
The propagator for $g$ satisfies differential equation:
\begin{equation}
{i\over 2}\partial_{\eta}\eta^{-1}\partial_{\eta}G^g=-\delta(\eta-\eta')\delta(x-x').
\end{equation}
The boundary condition we are imposing is that it should not grow at the horizon ($\eta\rightarrow -\infty$).
Since we are calculating the in-in propagator, there are four types of propagators depends on the location at the Schwinger-Keldysh contour:
\begin{equation}
\begin{split}
&G^g_{++}(x,\eta;x',\eta')=-i\delta(x-x')(\eta^2\theta(\eta-\eta')+\eta'^2\theta(\eta'-\eta)-\eta_c^2);\\
&G^g_{--}(x,\eta;x',\eta')=i\delta(x-x')(\eta^2\theta(\eta-\eta')+\eta'^2\theta(\eta'-\eta)-\eta_c^2);\\
&G^g_{+-}(x,\eta;x',\eta')=G^g_{-+}(x,\eta;x',\eta')=0.~~~~~~
\end{split}
\end{equation}
$\eta_c$ is the cutoff, and by taking $\eta_c\sim 0$ we can drop the constant piece. 
The vanishing of $G^g_{+-}$ and $G^g_{-+}$ is a manifestation of the fact that there is no local propagating degree of freedom between two contours.
We can also read that the excitation of $g$ goes to zero at a late time as a direct consequence of the running of coupling constant.
That means the spatial metric does not change at future time slice. Hence we do not need to do any final rescaling of the spatial coordinate.
The interaction between matter and graviton is also simple:
\be
iS_{\rm int}=i{g\over 2}\int (\partial_{\eta}\psi)^2+(\partial_x\psi)^2-{m^2\psi^2\over 2\eta^2}-{m^2\over 2\eta}\partial_{\eta}(\psi^2).
\ee

Now let's look at how this reproduces the Schwarzian tree level answers.
In section \ref{treelevel}, we showed that the four point function receives a correction from gravitational backreaction as a reparameterization of the in-in correlator.
We can reproduce this result from this linearized graviton analysis. To make things clear, let's
focus on only one channel (this can be achieved for instance by introducing flavor indices to the matter).
Then the tree level gravitational backreaction of the four point function $\langle \psi_1\psi_2\psi_3\psi_4\rangle$ is:
\begin{equation}\label{BulkFourPointFunctionCorrection}
-\int_{+} {d\eta_i dx_i d\eta_j dx_j\over \eta_i^2\eta_j^2}D_{i}[G^{++}_{1,i}G^{++}_{2,i}]G^g_{++}(i,j)D_{j}[G^{++}_{j,3}G^{++}_{j,4}]+(c.c).
\end{equation}
This simple factorization is due to lacking of gravitatonal propagator between $+-$ contours. Here $G^g$ is the graviton propagator and $G^{++}_{ij}$ is the matter propagator.  $D_{i}$ is the differential operator from the coupling between graviton and matter:
\begin{equation}
D_{i}(G^{++}_{1,i}G^{++}_{2,i})=\partial_{\eta_i}G^{++}_{1,i}\partial_{\eta_i}G^{++}_{2,i}+\partial_{x_i}G^{++}_{1,i}\partial_{x_i}G^{++}_{2,i}-{m^2\over 2}({1\over \eta_i^2}+{1\over\eta_i}\partial_{\eta_i})G^{++}_{1,i}G^{++}_{2,i}
\end{equation}
The integral (\ref{BulkFourPointFunctionCorrection}) after analytic continuation of $\eta$ to $iz$ is essentially the same as the EAdS four point function correction done in \cite{Almheiri:2016fws}, and 
there they match it to the Schwarzian result \footnote{There is an overall minus $i$ sign coming from the change of coupling constant of the Schwarzian action.}.
Here the only difference is that there is a change of conformal dimension from $\Delta_+$ to $\Delta_-$ due to the change from AdS bulk to boundary correlator $K(x_1;z_2,x_2)\sim{z_2^{\Delta_+}\over (z_2^2+x_{12}^2)^{\Delta_+}}$ to dS bulk to boundary correlator $G^{\pm\pm}(x_1;z_2,x_2)\sim{(\pm i z_2)^{\Delta_-}\over (z_2^2+x_{12}^2)^{\Delta_-}}$. 
This explains the simple relation of four point function backreaction between $dS_2$ and $AdS_2$.

\section{ADM decomposition}\label{App: MomentumConstraint}
In this appendix we will show that the Schwarzian equation of motion is the same as Momentum Constraint.
We will illustrate this simple relation by coupling JT gravity with massless scalar field $\psi$.
Using the ADM decomposition of the metric:
\begin{equation}
ds^2=-N^2dt^2+\omega^2(dz+N^zdt)^2,
\end{equation}
one can show that the action is equal to:
\begin{equation}
iS=-2i\int N(\omega^{-1}\phi''-\omega^{-2}\omega'\phi'+ \omega\phi+{1\over 4}\omega^{-1}\psi'^2)+N^{-1}[(\dot\phi-N^z\phi')(\dot\omega-(N^z\omega)')-{\omega \over 4}(\dot\psi-N^z\psi')^2].
\end{equation}
The lapse $N$ and shift $N^z$ are Lagrangian Multipliers, whose equations of motion are the Hamiltonian and Momentum Constraints:
\bea
{\rm Hamiltonian ~Constraint:}~~~~~~~~~~\partial_u^2\phi+\phi -K \partial_n \phi +\half T^{\rm m}_{nn}=0& \label{HamiltonianConstraint}\\
{\rm Momentum ~Constraint:}~~~~~~~~~~~~~~ K \partial_u\phi +\partial_u\partial_n\phi+\half T^{\rm m}_{n u}=0& \label{MomentumConstraint}
\eea
where we use the following expression for the extrinsic curvature $K$ and $\partial_n \phi$ :
\begin{equation}
K=-{\Pi_{\phi}\over2 \omega}={\dot\omega-(N^z\omega)'\over \omega N};~~~~~~~\partial_n\phi=-{\Pi_{\omega}\over 2}={\dot \phi-N^z\phi'\over N};~~~~~~~du=\omega dz;
\end{equation}
$u$ is the proper length along the future time slice, and the normal direction ($\vec{n}$) is pointing to the future.
$\Pi_{\omega /\phi}$ are momentum densities for $\omega$ and $\phi$.
Equation (\ref{MomentumConstraint}) is the Momentum Constraint of the system and it relates with change of the normal derivative of dilaton along the time slice with the energy flux from matter, after we choose the gauge $\partial_u \phi=0$.  
By taking variation with respect to the boundary metric $\sigma$, one can show that up to a constant piece, the normal derivative of $\phi$ is equal to $\phi_b\{x(u),u\}$ \cite{Maldacena:2016upp}.
In addition, it is easy to check that the Schwarzian equation of motion is:
\be \la{MoCon}
2\phi_b\partial_u \{ x(u),u \}+ T^{m}_{nu}=0
\ee 
This means that integrating over the Schwarzian field automatically implement the momentum constraint on the Wheeler de-Witt wavefunction.  In this sense, the boundary mode describes the dynamics of the constraint.
If the spatial direction is a circle, then integrating \nref{MoCon} over the full circle we get that the total momentum of matter should be zero. By multiplying (\ref{MoCon}) with ${k(u)\over x'(u)}$, where $k(u)=\lbrace 1,x(u),x(u)^2 \rbrace$, one can get conservation law for three $SL(2)$ charges\footnote{Restricting to the disk topology.}.  Again, when the spatial direction is a circle, all three $SL(2)$ matter charges are zero, so 
that we have a fully de-Sitter invariant matter state. 

The Hamiltonian constraint gives us the WdW equation (\ref{WdWE}). To see that, lets work in the gauge where $\partial_u\phi=0$ and $N^z=0$ and turn off the matter field.
As we mention before, the momentum constraint will force $\Pi_{\omega}$ to be a constant with respect to $u$. 
This means that the WdW wavefunctional $\Psi$ only depends on constant mode of $\phi$ and $\Pi_{\omega}$ and therefore is only a simple function.
Redefining $\omega=e^{\rho}$ and specifying the period of $z$ to be $2\pi$, the Hamiltonian constraint \ref{HamiltonianConstraint} becomes:
\be
[4\omega \phi-\Pi_{\omega}\Pi_{\phi}]\Psi=[\partial_{\phi}\partial_{\rho}+16\pi^2\phi e^{2\rho}]\Psi=0
\ee
where we use the relation between momentum density and momentum: $\Pi_{\phi/\omega}={1\over 2\pi}\pi_{\phi/\omega}$.

\section{Nearly $dS_2$ from $D$ dimensional gravity } 
\la{GeneralD}
 
 Here we briefly discuss the generalization of section \ref{4dSec} to arbitrary dimensions. We will point out the main formulas required to carry a similar analysis as done for four dimensions. The metric for the Schwarzschild de-Sitter black holes is given by
 \be
 ds^2 = -f d\tau^2 + \frac{d\rho^2}{f}+\rho^2 d\Omega_{D-2}^2,~~~~~~~f=1 - \rho^2  -\frac{\mu}{\rho^{D-3}}.
 \ee
The temperature, defined as the periodicity of $\tau$ at $\rho_+$, the first zero of $f(\rho)$, is given by 
\be\la{TempBHD}
\beta = \frac{4\pi \rho_+}{ (D-3) -(D-1) \rho_+^2 },~,~~~~~ f(\rho_+)=0
\ee
 If we compactify $\tau \sim \tau + \lambda$, we
 get the same equation as in \nref{RhoPE} but with 
 \be\la{Drel}
 \lambda_c = { 2 \pi \over \sqrt{(D-1)(D-3)}} ~,~~~~~\rho_e = \sqrt{D-3 \over D-1} ~,~~~~~
 \mu_e = { 2 \over (D-3) } \rho_e^{ D-1},
 \ee
which corresponds to the D-dimensional critical length, Nariai radius and Nariai mass respectively. Therefore, for any dimension $D$ the path followed by the solutions in the $\rho_+$ plane as one varies $y$ is identical to figure \ref{Saddles}. The off shell classical action is 
 \be
 iS =   { R_{dS}^{D-2} \Omega_{D-2}  
 \over 16 \pi G_N} \left[  - i { \lambda  ( D-2) }  ( \rho_+^{D-3} - \rho_+^{D-1} ) + 4 \pi \rho_+^{D-2} \right].
 \ee
 If we expand in large $\lambda$ and focus on the same saddle as explained in \ref{4dSec} then the wavefunction has the same form as \eqref{PsiExpan} after replacing 
\bea
M_e &=& {R_{dS}^2 \Omega_{D-2} \over 16 \pi G_N } { (D-2) \mu_e  } ~,~~~~~~~~~~~S_e ={ R_{dS}^2 \Omega_{D-2} \over 4 G_N}  \rho_e^{D-2}, \\
  C_1 &=&  {R_{dS}^2 \Omega_{D-2} \over 8 \pi G_N} \frac{2\pi^2(D-2)}{D-3}\rho_e^{D-1} ~,~~~ C_2 =  { R_{dS}^2 \Omega_{D-2} \over 8 \pi G_N}  { 4 \pi^3(D-2)^2  \over 3(D-3)^2 }  \rho_e^{D}.
\eea

\section{Computation of the  action for   geometries with Lorentzian conical singularities.  } 
\la{Cones}

In this section we will evaluate the gravitational action  for geometries given by 
\nref{SchdS4} with $\tau_r \sim \tau_r + \lambda$. These have a lorentzian conical singularity at $\rho_+$, where
$f(\rho_+)=0$.   We will define an off shell  regularization of the lorentzian conical singularity and compute its
contribution to the action.  

\subsection{Evaluating the action away from the singularity } 
 
First we evaluate the bulk part of the action, without including any contribution from 
 the singularity at $\rho = \rho_+$. 
 We write 
 \be
 iS  = { 1 \over 16 \pi G_N} \left[ \int( R -6)  - 2 \int K \right] 
 \ee
 We now use that
 \be
 \varphi \sim \varphi + \lambda' ~,~~~~~ \lambda = \lambda' { \sqrt{ - f(\rho_c) } \over \rho_c} ~,~~~~~~R = 12 ~,~~~~
 \sqrt{h} K = \partial_n \sqrt{h} = \sqrt{ - f} \partial_\rho { \sqrt{ - f } \rho^2 } 
 \ee
 Note that $\lambda$ is the actual ratio of the proper length of the $S^1$ over the radius of the $S^2$ at the 
 cutoff surface $\rho_c$ \cite{Hawking:1982dh}. 
Pulling out a volume of the sphere we get 
\bea
i S &=&  { \lambda' \over 4 G_N} \left[ 6 \int_{\rho_+}^{\rho_c} d\rho \rho^2 - 2  \sqrt{ - f} \partial_\rho { \sqrt{ - f } \rho^2 } \right] 
\cr
i S  & = & { \lambda \over 4 G_N } \left[ 2 { \rho_c \over \sqrt{ f(\rho_c) } } ( \rho_c^3 - \rho_+^3 ) - 2  \partial_\rho { \sqrt{ - f } \rho^2 }|_{\rho = \rho_c} \right] 
\eea
After expanding $
\sqrt{ - f } =  \rho_c ( 1 - \half 1/\rho_c^2 + \half
 \mu/\rho_c^3)$ we get 
 \bea
 iS  & =& i S_{\rm Large} + i S_{\rm Finite} ~,~~~~~~~~iS_{\rm Large } = i { \lambda \over 4 G_N } \left[ - 4    \rho_c^3 +  2 \rho_c\right]  \la{ActVol}
 \cr
 iS_{\rm Finite} & =&  i  { \lambda\over 4 G_N } \left[   - \mu - 2 \rho_+^3      \right] 
 \eea

\subsection{Evaluating the regularized cone action} 

Near $\rho = \rho_+$ we have a metric of the form 
\be
ds^2 = - d\tau^2 + \gamma^2 \tau^2 d\psi^2 ~,~~~~~~\psi \sim \psi + 2 \pi 
\ee
This is a conical Lorentzian space. The idea is that we fill it in using a slightly complex no boundary manifold, which is not a solution of the Einstein equations by deforming the tip a bit. 
An example is 
\be \la{regepl}
ds^2 = - d\tau^2 + \tau^2 { - \epsilon^2 + \gamma^2 \tau^2 \over \tau^2 + \epsilon^2  } d\psi^2 
\ee
with $\epsilon$ very small. Where $\tau \propto \rho - \rho_+$. 
Our goal here is to evaluate 
the integral of the Einstein action over a small region in $\tau $. Since it is a small region, we naively
expect a small contribution, but because the curvature is high, we can have a finite contribution.
However, the curvature is high only in the two dimensional space parametrized by $\tau$ and $\psi$, 
but not the directions of the transverse sphere. So the problem reduces to evaluating the two dimensional curvature, which we write as\footnote{This is multiplied by the area of the sphere and $1/(16 \pi G_N )$.} 
\be \la{TopAc}
i \int \sqrt{g} R =i  \left[ \int \sqrt{g} R - 2 \int K \right]  + i  2 \int K  ~,~~~~~ i 2 \int K = { 4 \pi  \lambda }
\ee
The first term is a topological invariant and turns out to be equal to $4 \pi$. This is slightly subtle since naively we are getting a real answer in the end. So, we can do it very explicitly for a general metric 
$ ds^2 = - d\tau^2 + h^2 d\psi^2 $ we get 
\be \la{Eto}
i \int \sqrt{g} R - 2 i \int K = -  i 4 \pi  h'(0) 
\ee
Where we used that $\sqrt{g} R  = 2 h''$  is a total derivative. The extrinsic curvature term cancels the
boundary term at $\tau \gg \epsilon$ and we are left with $h'(0)$. 
When we apply this to \nref{regepl} we get 
\be
\sqrt{ h} = \tau \sqrt{ - \epsilon^2 + \tau^2 \gamma^2  \over \tau^2 + \epsilon^2 } 
\ee
In order to evaluate $h(0) $ we will need $\sqrt{-1} =+i$. The fact that we get $+i$ comes from the 
fact that we are supposed to continue towards the upper half plane in the complex time direction. 
This is part of the no boundary prescription over which metrics we are integrating over. After 
we get this  in \nref{Eto} we  get 
\be
  i [  \int \sqrt{g} R - 2 K ] = 4 \pi 
  \ee
  
  Finally, in our  problem this two dimensional action is multiplied by the volume of the $S^2$, 
  proportional to $\rho_+^2$ and the appropriate power of $G_N$. In addition,  in our problem 
  \be
  \gamma = { \lambda (-f') \over  4\pi} 
  \ee
So that in the end of the day we get a total contribution from the singularity equal to 
\be
 iS_{\rm Sing} = { \pi \over G_N } \rho_+^2 + i { \lambda   \over 4 G_N} ( 2\rho_+^3 - \mu ) 
 \ee
 \be
 i S^{\rm total}_{\rm Finite} =   { \pi \over G_N } \rho_+^2 -  i { \lambda \over 2 G_N} \mu = 
  { \pi \over G_N } \rho_+^2 -  i { \lambda\over 2 G_N} ( \rho_+ - \rho_+^3 ) = S- i \lambda M  \la{ActionFin}
 \ee
 The saddle point equation for $\rho_+$ is given by 
 \be
 i \lambda = { 4 \pi \rho_+ \over 1 - 3 \rho_+^2 } 
 \ee
 which is the right one. 
 As we mentioned, we have assumed that $f'(\rho_+) < 0$, or $\rho_+ > \rho_e$. 

Redoing this in the case $0 < \rho_+ < \rho_e$ we find the same final action. 
 
\subsection{Euclidean Actions in general two dimensional gravity theories}

\begin{figure}
\centering
\subfigure[AdS Type Prepotential]{%
    \includegraphics[width=0.4\textwidth]{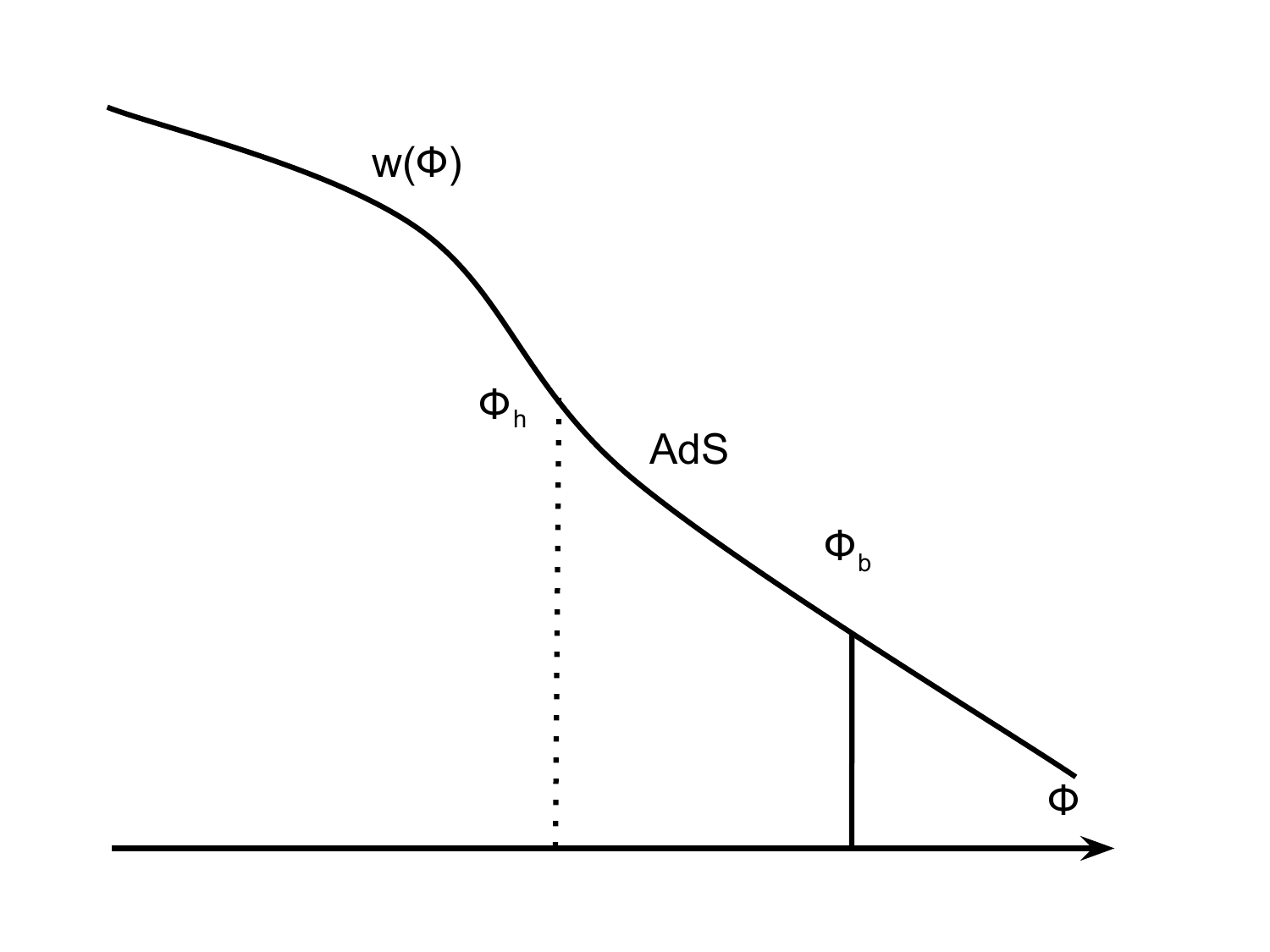}
    \label{fig:SupAdS}%
    }\hspace{2cm}
    \subfigure[dS Type Prepotential]{%
    \includegraphics[width=0.4\textwidth]{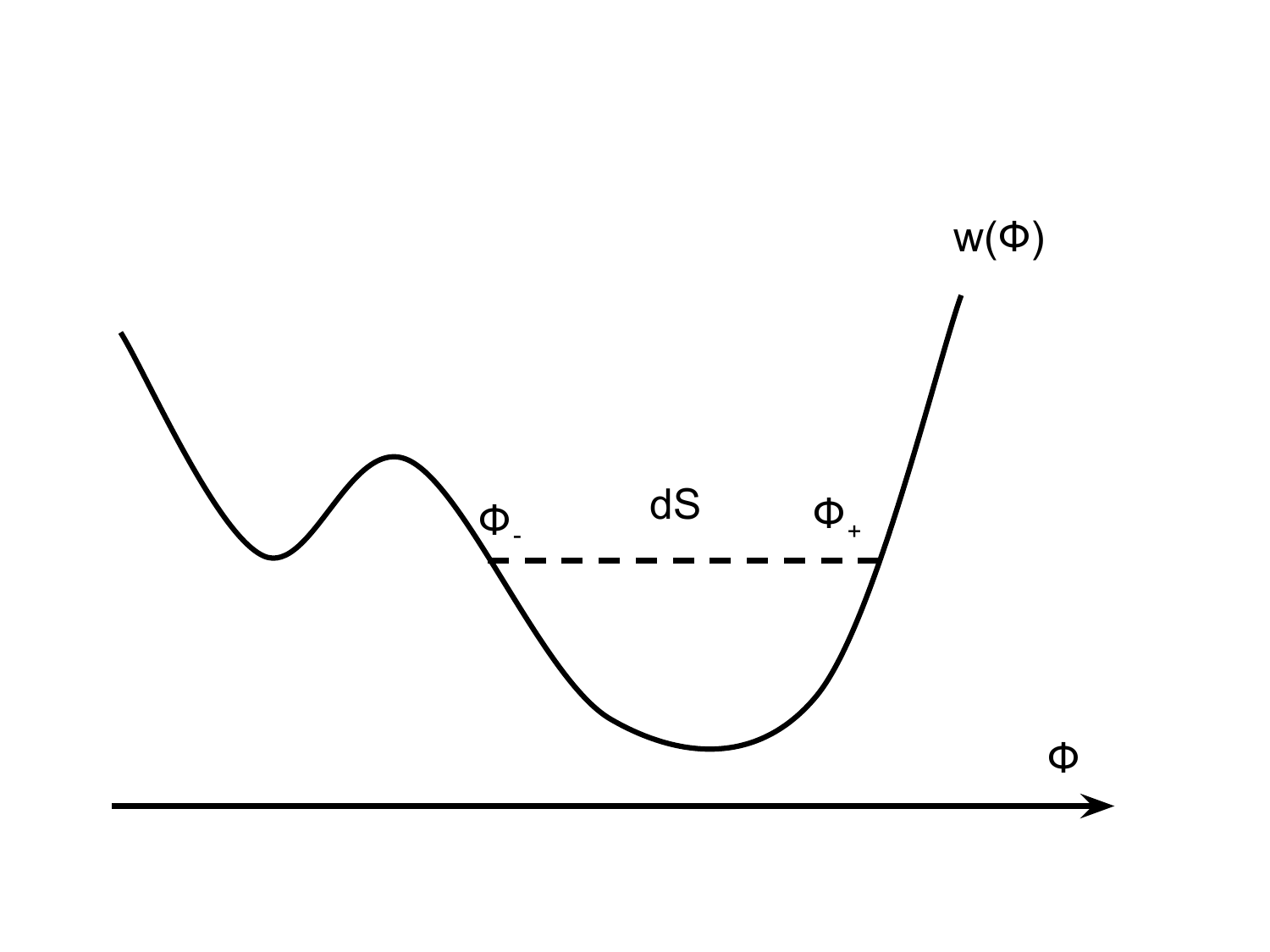}%
    \label{fig:SupdS}%
  }\hspace{2cm}%
\caption{The type of prepotentials that we consider. (a) for the case of one horizon. (b) for the case of two horizons. }
\label{Superpotential}
\end{figure}

We can also do the analysis for general gravitational system under the assumption of spherical symmetry so that
we reduce the problem to a two dimensional gravity problem. 
Such systems can be reduced (after a field redefinition)  to a two dimensional action
  \be
-S_E=\int \phi R-U(\phi).
\ee
with a general $U(\phi)$ that depends on the system we are considering. 
Such actions always have a killing vector \cite{Mann:1992yv}. 
$\zeta^\mu = \epsilon^{\mu \nu } \partial_\nu \phi$. 
 We can  choose the coordinates so that $\zeta^\mu \propto  \partial_t$. Then we can write
\begin{equation}
ds^2=-e^{2\Omega( z)}dx^+ dx^-;~~~~~\phi=\phi(z );~~~~z={x^+-x^-\over 2};~~t={x^++x^-\over 2}.
\end{equation}
The classical equations of motion are \cite{Almheiri:2014cka}:
\begin{equation}
-2\partial_z\Omega\partial_z\phi+\partial_z^2\phi=0 ~,~~~~~~~
e^{2\Omega}U+\partial_z^2\phi=0
\end{equation}
The first equation relates the growth of dilaton with the scale factor:
\begin{equation}
\partial_z\phi=- e^{2\Omega} \la{DerPhiOm}
\end{equation}
where we have fixed an undetermined constant by a rescaling of $z$. Now when $z$ increases $\phi$ decreases.
Using this, we can solve the second equation and it gives:
\begin{equation}
\partial_z\phi={ W(\phi)-  W(\phi_h) }~,~~~~~~e^{2\Omega}={ W(\phi_h) - W(\phi)   }~,~~~~W(\phi) - 
W(\phi_h) =\int_{\phi_h}^\phi  d\phi' U(\phi')
\end{equation}
where we defined $\phi_h$ to be the value of $\phi$ where $\partial_z\phi=0$, which corresponds to a horizon. 
We call $W(\phi)$ a ``prepotential''.  
Near the horizon, we can expand the equation and gives:
\begin{equation}
\partial_z\phi= {U(\phi_h) }(\phi-\phi_h);~~~\phi=\phi_h+  e^{ z U(\phi_h) }.
\end{equation}
where we  absorbed an integration constant by a shift of $z$. 
We are assuming that the combination $z U(h) \to -\infty $ as we approach the horizon. 
The metric near the horizon is 
\begin{equation}
ds^2=-e^{2\Omega}dx^+dx^-=-{U(\phi_h)} e^{z U(\phi_h)}(-dt^2+dz^2)
\end{equation}
If the horizon is smooth, then we will have a relation between Euclidean time period $ t \sim t + \beta$ 
 and the magnitude of $|U(\phi_h)|$:
\begin{equation}
\beta = {4\pi  \over |U(\phi_h)|} 
\la{TempU}
\end{equation}
We will consider two general situations, one corresponding to black hole in a space with an asymptotically "cold" region, so that we have only one horizon. The second is black hole in dS where we have a region with two horizons.  
As shown in figure \ref{Superpotential}(a), in the case of AdS, one usually encounter the situation that the dilaton keeps growing away from its horizon value towards the asymptotic region. 
We can put a cutoff and specify  its boundary value $\phi_b$. 
We define $z$ such that the horizon is located at $z\rightarrow+\infty$, which is consistent with the sign we picked in \nref{DerPhiOm},  and the boundary is at finite $z$ location $z_b$. 
One can calculate the classical Euclidean action
\bea
-S_E&=&\int(\phi R- U)-2\phi_b\int K=\beta \int_{z_b}^{\infty} dz(-2\partial_z^2\Omega\phi-e^{2\Omega}U)-2\beta \phi_b \partial_z\Omega(z_b) \nonumber\\
&=&-2\beta\partial_z \Omega \phi_h +2\beta \int_{z_b}^{\infty} dz\partial_z^2\phi={4\pi  }\phi_h -{2\beta [W(\phi_b)-  W(\phi_h)] }. \la{BulkAc}
\eea
where we used \nref{TempU} (with $U(\phi_h) <0$). 
At the boundary, we define the proper UV temperature, given by the proper length of the circle,  as 
\be
\la{RenBeta}
\hat \beta ={~e^{\Omega(z_b)}\beta\over \sqrt{-W(\phi_b)}}=\sqrt{1-{W(\phi_h)\over W(\phi_b)}}{\beta }
\ee
This gives  
\bea
-S_E&=&{4\pi  }\phi_h + 2 \hat \beta \sqrt{-W(\phi_b) }  \sqrt{ W(\phi_h) - W(\phi_b) }  \la{OffFB}
\\
-S_E&\sim &  - 2 \hat \beta W(\phi_b)  +{4\pi  }\phi_h   + \hat \beta   W(\phi_h) \la{OffSh2}
\eea
where in the second equation we took the limit where $\phi_b\gg \phi_h$. The first term is infinite in that limit, but proportional to $\hat \beta$ (a vacuum energy contribution). The last two terms can be viewed as the entropy and energy contributions. And we saw that the energy has a simple relation with the prepotential:
\be
E=-W(\phi_h).
\ee
Since the gravitational entropy is equal to $4\pi \phi_h$, this simple relation is actually the thermodynamics relation for general Black Holes.

Here we have obtained \nref{OffSh2} as an on shell equation, where $\phi_h$ and $\beta$ are linked as in \nref{TempU}.
 However, it is possible to obtain it also as an off shell expression where $\phi_h$ is an independent variable, which becomes fixed by using a saddle point equation, which fixes it to the value given by \nref{TempU}. We proceed to explain this in more detail. 
 
 If we call the identification of the circle $\beta$, but we do not link them as in \nref{TempU}, then the bulk action computation 
 described in \nref{BulkAc} would give us 
\be \la{BulkOff}
-S_E= -{ \beta U(\phi_h)  }\phi_h +\hat \beta W(\phi_h) -2\hat \beta W(\phi_b).
\ee
However, when the geometry is not smooth at the horizon we should 
add an extra contribution from the conical singularity 
\be
-S_{\rm sing}=4\pi \phi_h \left[1-{\beta { (-U(\phi_h) )\over 4 \pi } }\right]
\ee
When added to \nref{BulkOff} we get 
\be
-\tilde S_E^{\rm Off-Shell}  =4\pi \phi_h +\hat \beta  W(\phi_h)
\ee
Here we can think of the first term as the entropy and the second 
as the energy. As promised, extremizing over $\phi_h$ gives us 
\nref{TempU}. Here we derived it for the case that we take the limit $\phi_b \gg \phi_h$, but we get the right answer even if we do not take this limit. Namely, we get \nref{OffFB}, and 
 with \nref{TempU}, and remembering the relation \nref{RenBeta}. This type of off shell configurations was discussed before, see e.g. \cite{Carlip:1993sa}.

We now turn to the case  where we have two horizons. An example is the static patch of de-Sitter space.  Now the typical prepotential 
 is shown in   figure \ref{Superpotential}(b).  We are interested in the region between the two horizons. 
 In this case, we can also go to Euclidean space and consider off shell geometries with conical singularities at one or both horizons. We could divide this region in two half, and each half would look like the computation we did above. When we put the two halves together all terms involving $\phi_b$ cancel out, since such terms were arising after integration by parts.    So get that the final off shell action is 
  \be
  -S_E^{\rm Off-Shell} =  4\pi \phi_- + 4\pi\phi_+ 
  \ee
  where $\phi_\pm$ are the values of $\phi$ at the two horizons\footnote{Similar calculation was done in \cite{Bousso:1998na}.}.
  Notice that it is independent of the period that we have chosen for the Euclidean time direction. In this action we included a bulk piece and also the contributions for the curvature at the tips of the cones. It is interesting that we only have the contributions from the entropies of both horizons.
  If we vary with respect to $\phi_+$ and $\phi_-$ we get the condition that the two temperatures are equal, which is not possible if we solve the equations between the two horizons. In some special circumstances we have solutions where they are equal. An example is the black hole in $dS_4$, where the two temperatures can be equal only for the Nariai solution, 
  $dS_2 \times S^2$. 
 
\section{Integration contours and contributing saddles } 
\la{Contours}

We have heuristically argued in appendix \ref{Cones} 
that we could reduce the whole path integral to an integral over one 
  parameter. The argument is heuristic because we have not honestly done the path integral, we have just used a saddle point approximation for most variables except for one. 
The final one dimensional integral has the form 
\be \la{dSConeq}
\int_\mathcal{C} d \rho ~\exp\left\{  { 1 \over \hbar }E(\rho)  \right\} ~,~~~~~E(\rho) =  - i y ( \rho - { 1 \over 3 } \rho^3 ) +  \rho^2  
\ee
We are not paying attention to the measure. 
The variables $\rho$ and $y$ correspond to the variables $\rho_+/\rho_e$ and $\lambda/\lambda_c$ in the main text, while $\hbar \sim G_N/R_{dS}^2$. $\mathcal{C}$ denotes the contour that defines the integral. We will start by analyzing the case for which we integrate over $\rho\in(0,+\infty)$. In the end, we will compare with other choices of contours. 

\begin{figure}[h!]
\begin{center}
\begin{tikzpicture}[scale=1]
\node[inner sep=0pt] (russell) at (0,0)
    {\includegraphics[width=.35\textwidth]{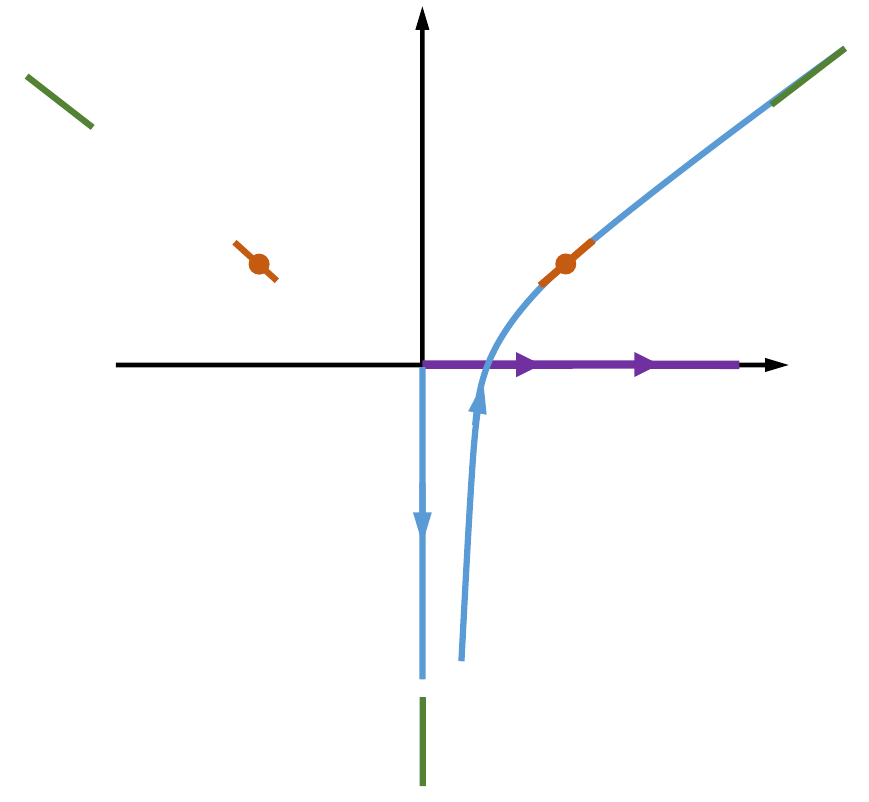}};
\draw (1.3,0.8) node {\small $\rho_1$};
\draw (-1.5,0.8) node {\small $\rho_2$};
\draw (2.5,-0.05) node {\small $\rho$};
\draw (0,-3.2) node {\small (a) $y>1$};
    \end{tikzpicture}~\hspace{2cm}
     \begin{tikzpicture}[scale=1]
\node[inner sep=0pt] (russell) at (0,0)
    {\includegraphics[width=.35\textwidth]{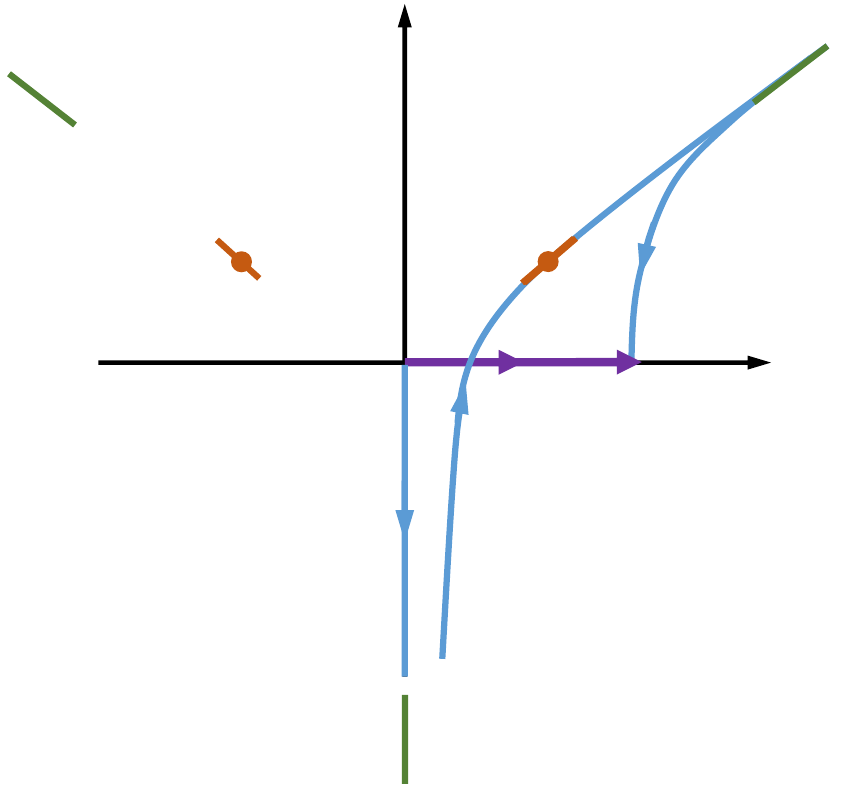}};
\draw (1.1,0.8) node {\small $\rho_1$};
\draw (-1.5,0.8) node {\small $\rho_2$};
\draw (0,-3.15) node {\small (b) $y>1$};
\draw (1.3,-0.06) node {\small $\sqrt{3}$};
\draw (2.5,-0.05) node {\small $\rho$};
    \end{tikzpicture} ~\\
    \vspace{0.2cm}
    \begin{tikzpicture}[scale=1]
\node[inner sep=0pt] (russell) at (0,0)
    {\includegraphics[width=.35\textwidth]{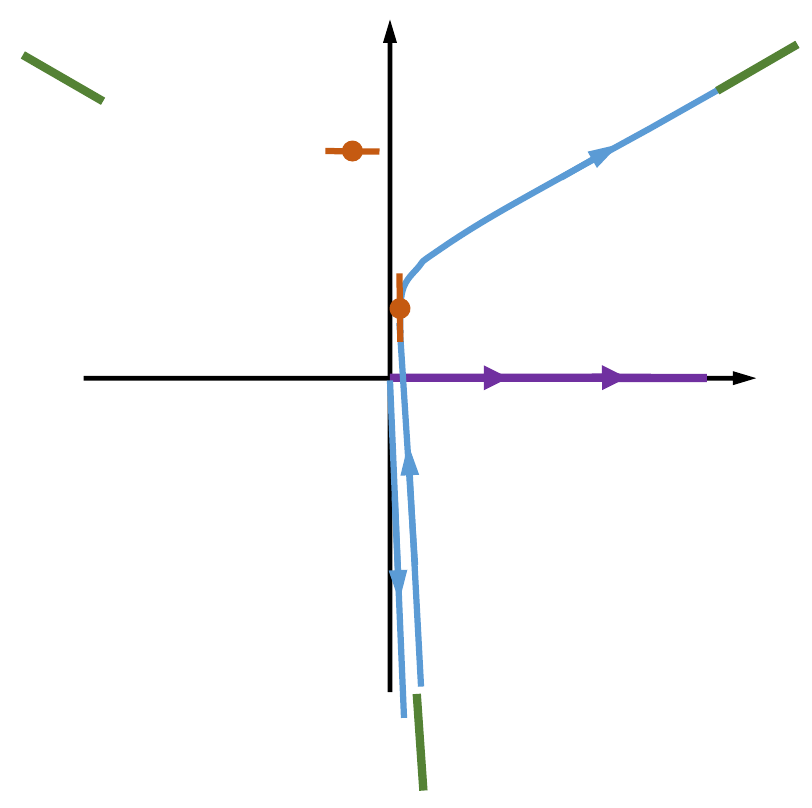}};
\draw (0.3,0.8) node {\small $\rho_1$};
\draw (-0.9,1.8) node {\small $\rho_2$};
\draw (2.5,-0.05) node {\small $\rho$};
\draw (0,-3.1) node {\small (c) $y<1$};
    \end{tikzpicture}~\hspace{2cm}
    \begin{tikzpicture}[scale=1]
\node[inner sep=0pt] (russell) at (0,0)
    {\includegraphics[width=.35\textwidth]{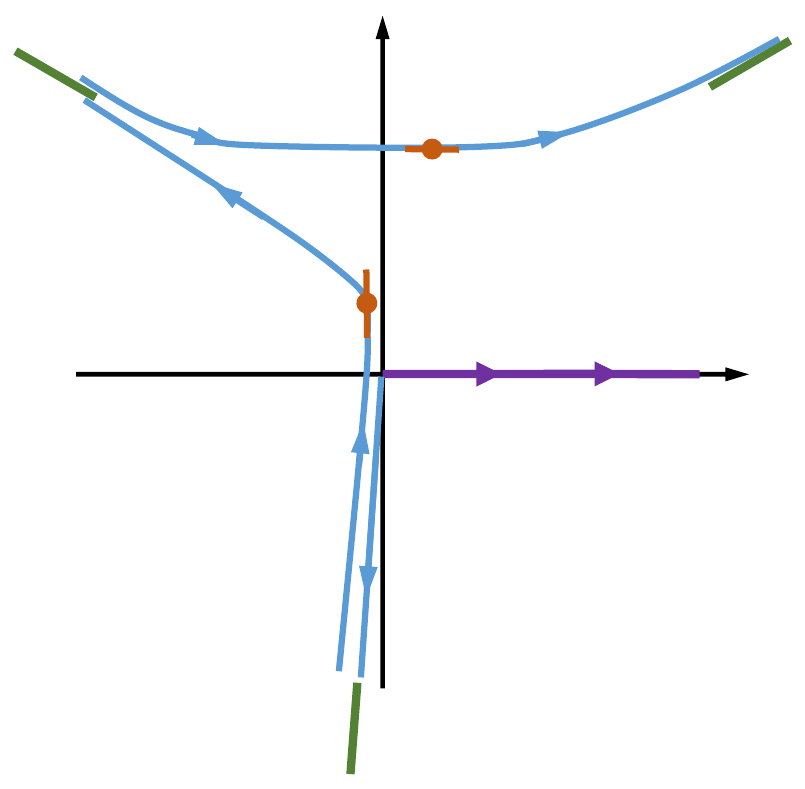}};
\draw (-0.5,0.5) node {\small $\rho_1$};
\draw (0.5,1.4) node {\small $\rho_2$};
\draw (2.5,-0.2) node {\small $\rho$};
\draw (0,-3.2) node {\small (d) $y<1$};
    \end{tikzpicture}
\caption{We show original $\rho$ contour in purple and asymptotic steepest descent directions in green.
 In red, we show saddle points and local direction of steepest descent.  In blue, we show steepest descent contours homologous to the original one.
  (a) For the case $y>1$ only $\rho_1$ contributes when we integrate along the $(0,\infty)$ contour.
  (b) Also $y>1$ but integrating along the $(0,\sqrt{3}$) contour. In 
  (c) and (d) we consider the $y<1$ case along the $(0,\infty)$ contour for two different regularizations. 
     In  (c) we 
  only pick up  $\rho_1$. In (d) we pick up both, but $\rho_1$ dominates and both choices are consistent, and
  we are sitting at at Stokes line. 
  }
\label{dScont}
\end{center}
\end{figure}
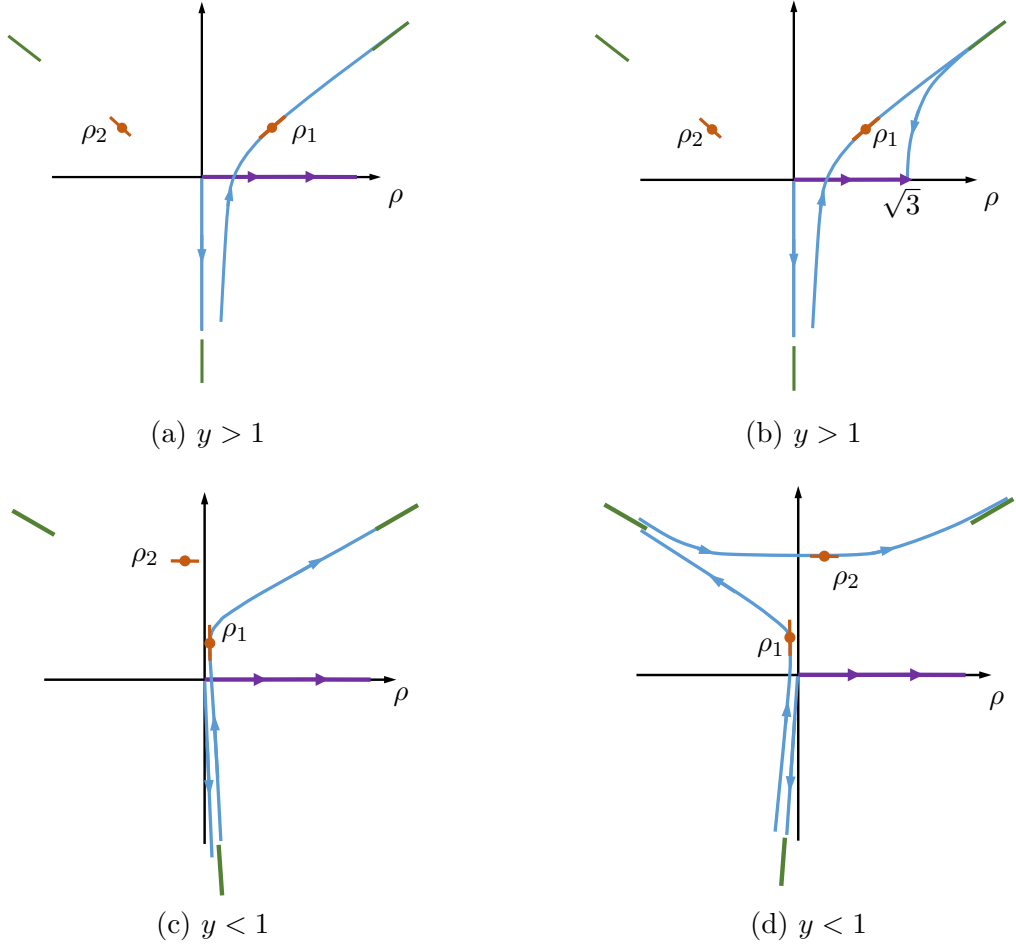

The first step is to identify the directions in the complex plane where the exponent $E(\rho)$ decreases for 
large $|\rho|$. The steepest descent directions at large $|\rho|$ are 
 determined by the  $i \rho^3$ term in the exponent and correspond to the directions with 
arg$(\rho)= \pi/6 , 5\pi/6, -\pi/2$. See green lines in figure \ref{dScont} (while the original contour is in purple). Also near the endpoint at $\rho=0$ the 
exponent goes as $E\sim - i y \rho $  (for $\rho \to 0$) 
 and so the steepest descent line is along the negative imaginary direction. 
 
In order to evaluate the integral with the steepest descent method we need to deform the contour 
so that it starts at $\rho=0$ and follows a steepest descent line at large $|\rho|$, this implies that 
it should go along the direction with arg$(\rho)= + \pi/6$. We also choose a contour where the imaginary part is constant along the contour. 
In principle we can get contributions from saddles and also from the endpoint at $\rho=0$.  The 
endpoint contribution has zero action and will always be present. 
The saddle contributions might or might not be present. Let us first analyze the case with 
$y>1$. Then the two saddles and their directions of steepest descent are given by 
\bea
\rho_1 &=& { i \over y } +  \sqrt{ 1 - { 1 \over y^2 } } ~,~~~~~ {\rm arg}(\delta \rho) = { \pi \over 4 }  
\cr
 \rho_2 &=& { i \over y } -  \sqrt{ 1 - { 1 \over y^2 } } ~,~~~~~ {\rm arg}(\delta \rho) = -{ \pi \over 4 }  
\eea
where the argument of $\delta \rho $ tells us the direction of steepest descent (of course if we add $\pi$ to the argument of $\delta \rho$ we also get a line of steepest descent). As we see in figure \ref{dScont} we can consider a contour with constant imaginary part that passes through the first root coming from the direction $ - i \infty $ and goes to infinity along the arg$(\rho) = \pi/6$ line. This is combined with a contour that starts at $\rho=0$ and goes to $-i\infty$ along a steepest descent line. This gives us a full contour homologous to the original one. This picks up the saddle $\rho_1$ but not the one with $\rho_2$. We also have an endpoint contribution at $\rho=0$. 

The exponent $E$  at $\rho_1$ is 
\be
{\rm Re }\left[ E(\rho_1) \right] = 1 - { 2 \over 3 } { 1 \over y^2 } 
\ee
We see that this is always bigger than zero in for $y>1$, so that this saddle dominates over the 
endpoint contribution, which has $E=0$. 

We now analyze the case $y< 1$. In this case,    the two saddles are along the line 
$\rho = i r $ with $r> 0$. The two roots and their directions of steepest descent are given by 
\bea
\rho_1 &=&  i \left[ { 1 \over y } - \sqrt{ { 1 \over y^2} -1} \right] ~,~~~~~~ \rm{arg}( \delta \rho) ={ \pi/2}
\cr
\rho_2 & =&  i \left[ { 1 \over y } + \sqrt{ { 1 \over y^2} -1} \right] ~,~~~~~~ \rm{arg}( \delta \rho) = 
0 
\eea
In order to analyze this configuration it is convenient to add a small imaginary part to $y$ so that these
two roots are displaced away from the real axis. It does not matter much which way we displace them. 
Let us imagine we displace the first one towards the region of positive real part, as shown in figure 
\ref{dScont}(c). 
Then we can run a contour that passes through $\rho_1$ and then asymptotes to arg$(\rho) = \pi/6$. 
Then only $\rho_1$ would contribute. If we had given the imaginary part in the other direction 
(figure \ref{dScont}d) we would have a slightly more complicated contour that would also pick up $\rho_2$. So we see that 
we are sitting precisely at a Stokes line where $\rho_2$ is appearing or disappearing. 
Of course, the contribution of $\rho_2$ is subdominant. And the contribution of $\rho_1$ 
has an exponent $E(\rho_1) > 0$ which makes it dominate over the endpoint contribution. 

Note that for $y<1$ the geometries look similar to Euclidean AdS black holes. But the saddle we pick is 
$\rho_1$, which corresponds to the ``small'' black hole, in contrast with the one picked in 
\cite{Banerjee:2013mca}. We discuss in more detail the $AdS$ problem in appendix \nref{AdSContour}.  
 
\begin{table}[]
\begin{center}
    \begin{tikzpicture}[scale=1.2]
\draw (0,0) node {\small Contours};
\draw (2.6,0) node {\small Saddles};
\draw (5.3,0) node {\small Endpoints};
\draw (-1.5,-0.3) -- (7.5,-0.3);
\draw (1.6,0.3) -- (1.6,-2.5);
\draw (4,0.3) -- (4,-2.5);
\draw (-0.1,-0.7) node {\small $\rho \in (0,+\infty) $};
\draw (5.2,-0.7) node {\small $0 $};
\draw (2.7,-0.7) node {\small \color{red} $\hspace{0.25cm}\rho_1$};
\draw (0,-0.7*3) node {\small ~~$\rho \in (-\infty,+\infty) $};
\draw (2.7,-0.7*3) node {\small \color{red}~~\hspace{-0.1cm}$\rho_1$ and $\rho_2$};
\draw (-0.17,-0.7*2) node {\small $\rho \in (0,\sqrt{3}) $  };
\draw (2.8,-0.7*2) node {\small $\rho_1$    };
\draw (5.45,-0.7*2) node {\small $0 $ and  { \color{red} $\sqrt{3}$ } };
   \end{tikzpicture}
   \vspace{-0.3cm}
\end{center}
\caption{Possible contours (described by their endpoints) together with
 their contributing  saddles and endpoints for $y>1$.
The ones  in red dominate.}
\label{tablecont}
\end{table}

Finally, we can also study which saddles are picked depending on the choice of contour $\mathcal{C}$. Since it is the case most relevant for the discussion in this paper, we will focus on the $y>1$ case (a similar analysis can be done for $y<1$). We summarize the results in table \ref{tablecont}.
 
Another option is to integrate over $\rho \in (0,\sqrt{3})$, or $\rho_+ \in (0,1)$.
 Along this contour the mass goes from $\mu=0$ to $\mu_e$ at $\rho=1$ and then goes back to $\mu=0$ at $\rho=\sqrt{3}$. We had argued in section \ref{sec:ContChoices} that the contour is expected not to stop at $\rho=1$, or $\rho_+ =\rho_e$. For this reason, the only natural place to stop seems to be at $\rho_+=1$. 
 The attractive feature of this contour is that we are summing only over regular black hole solutions in de-Sitter, 
 with masses $0\leq \mu \leq \mu_e$. 
 For this contour,  
 the only modification is the addition of an endpoint contribution at $\rho =\sqrt{3}$, for which  
 the steepest
 descent direction (for large $y$) is essentially along the positive imaginary direction. 
 This endpoint contribution looks like empty $dS_4$ in Milne-like coordinates, given by \nref{SchdS4} with 
 $\tau$ identified. We can still deform the contour so that $\rho_1$ still contributes, see figure \ref{dScont}(b).  
  The endpoint with $\rho=\sqrt{3}$ dominates since $E(\rho=\sqrt{3})=3$. This corresponds to the entropy of
 $dS_4$.  But the $\rho_1$ saddle is still present, though as a subleading contribution. 
  
Finally, we could integrate over $\rho\in(-\infty, \infty)$ then we can choose a steepest descent contour that goes from ${\rm arg}(\rho)=5\pi/6$ to $-i \infty$ passing through $\rho_2$ and then goes from $-i\infty$ to ${\rm arg}(\rho)=\pi/6$ passing through $\rho_1$ and picking both saddles.

 \subsection{Euclidean AdS blackhole saddle point analysis } 
 \la{AdSContour}
 
 \begin{figure}[h!]
\begin{center}
\begin{tikzpicture}[scale=1]
\node[inner sep=0pt] (russell) at (0,0)
    {\includegraphics[width=.35\textwidth]{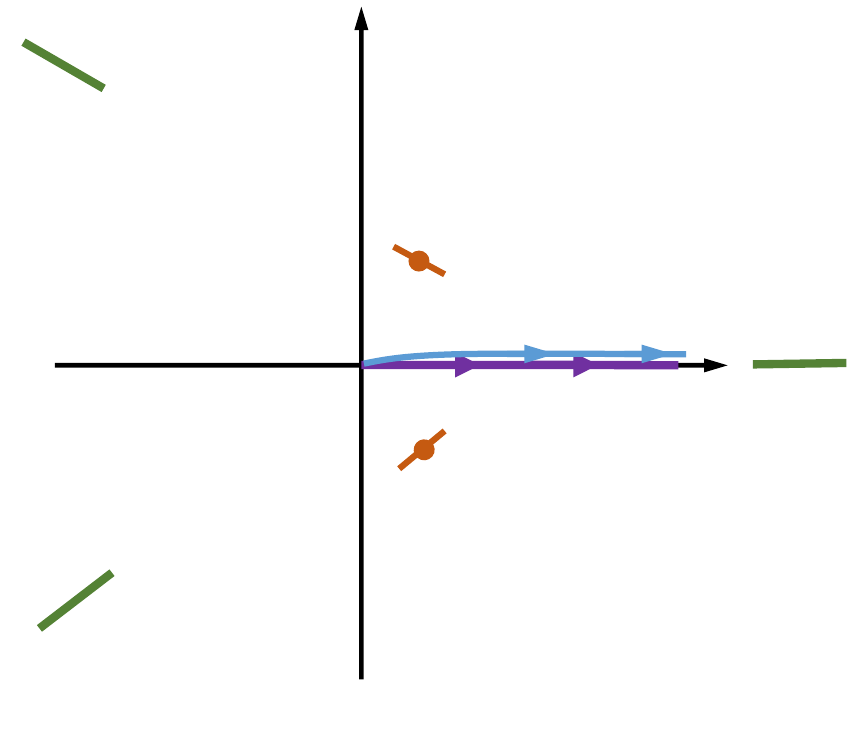}};
\draw (0.2,-1.1) node {\small $r_1$};
\draw (0.2,0.8) node {\small $r_2$};
\draw (0,-3) node {\small (a) $y>1$};
\draw (2.1,-0.35) node {\small $r$};
    \end{tikzpicture}~\\
    \vspace{0.2cm}
    \begin{tikzpicture}[scale=1]
\node[inner sep=0pt] (russell) at (0,0)
    {\includegraphics[width=.35\textwidth]{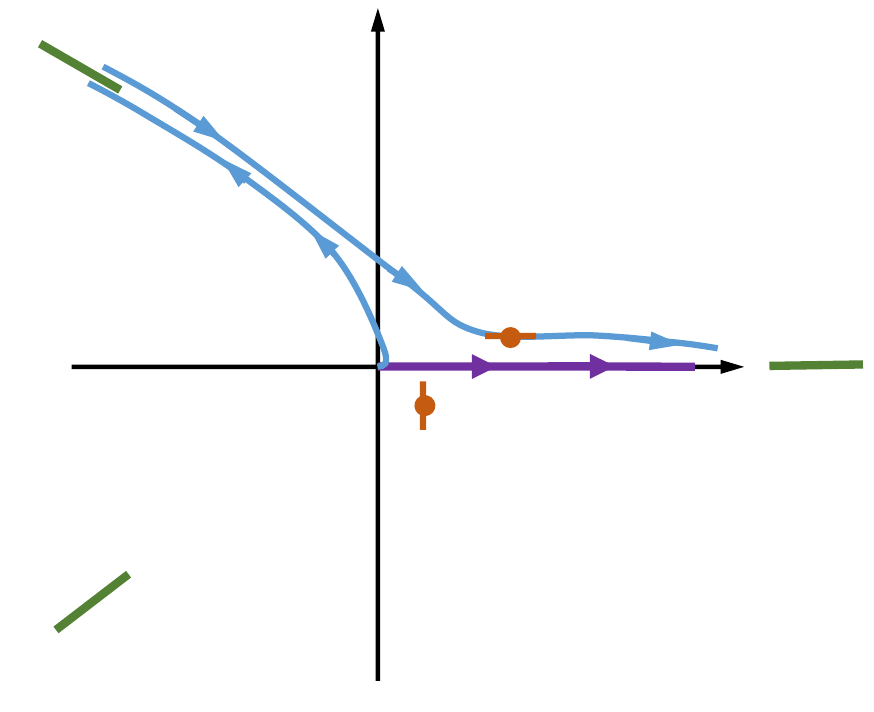}};
\draw (0.4,0.5) node {\small $r_2$};
\draw (0.1,-0.8) node {\small $r_1$};
\draw (0,-3.1) node {\small (b) $y<1$};
\draw (2.1,-0.35) node {\small $r$};
    \end{tikzpicture}~\hspace{2cm}
    \begin{tikzpicture}[scale=1]
\node[inner sep=0pt] (russell) at (0,0)
    {\includegraphics[width=.35\textwidth]{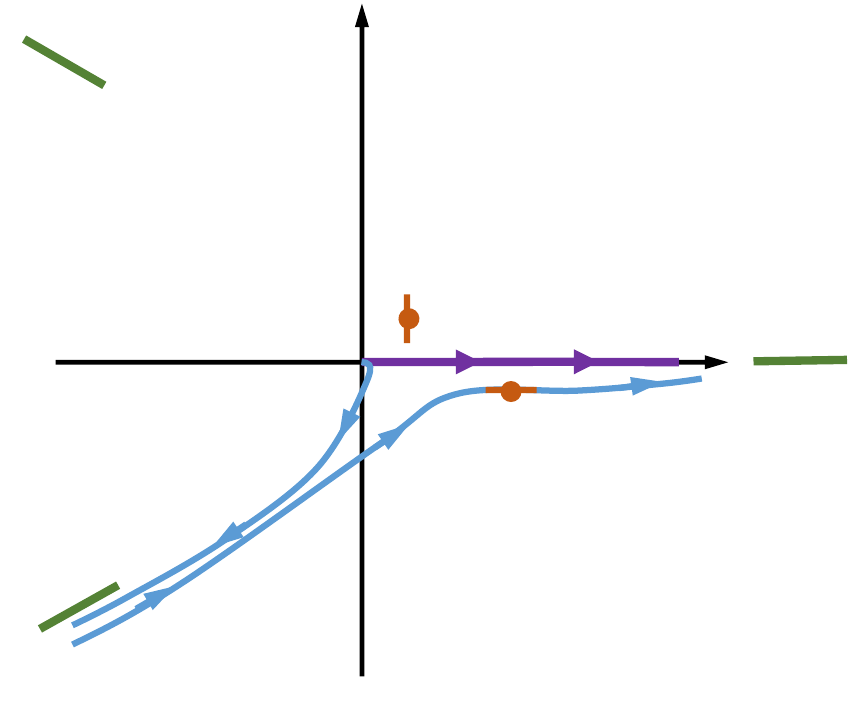}};
\draw (0.1,0.45) node {\small $r_1$};
\draw (0.6,-0.6) node {\small $r_2$};
\draw (0,-3.2) node {\small (c) $y<1$};
\draw (2.1,-0.35) node {\small $r$};
    \end{tikzpicture}
\caption{We show original $r$ contour (purple) and asymptotic steepest descent directions (green). In blue we show steepest descent contours homologous to the original one. In red we show saddle points and local direction of steepest descent. (a) For the case $y>1$ only end-point contributes. Case $y<1$ needs regulation: Both curves (b) and (c) pick $r_2$ and the end-point.  
}
\label{AdScont}
\end{center}
\end{figure}
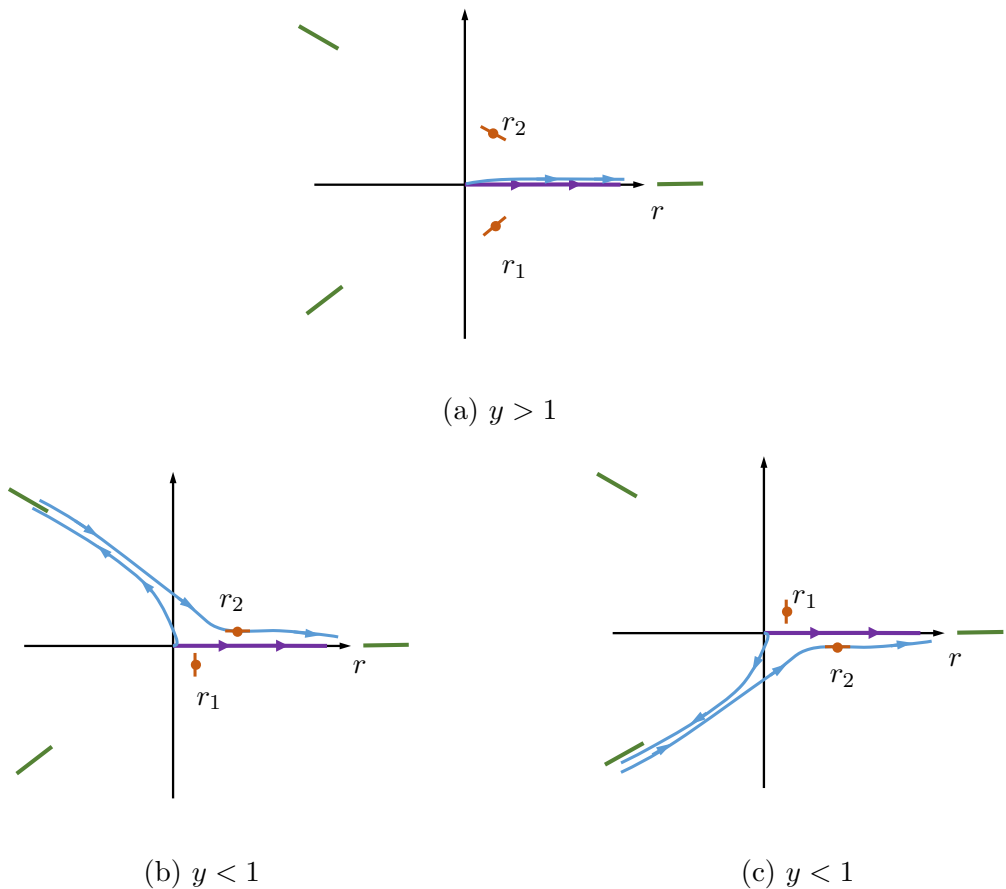

 In the $AdS$ case,  we encounter a very similar problem.  
 We have various saddle points and we want to know which ones 
 contribute. 
 Again we can imagine solving the equations holding the area or mass of the black holes fixed. This leads us to 
 spacetimes with a conical singularity. Smoothing out the cone slightly gives us an action which is essentially 
 $-I = S - \beta M$ where $S$ is the entropy and $M$ the mass of the black hole. Then the integral over the last
 variable is 
 \be
 \int_\mathcal{C} d r ~\exp\left\{  { 1 \over \hbar }\tilde E(r)  \right\} ~,~~~~~
 \tilde E(r) =   - y ( r + { 1 \over 3 } r^3 ) +  r^2  ~,~~~~~ r =   \sqrt{3} r_+ ~,~~~~y = { \beta \over \beta_c} 
 \ee
 with $\beta_c = \lambda_c$. 
 This has a form similar to \nref{dSConeq} with $\rho =i r$ and an overall change in sign, $\tilde E(r) = -E(ir)$. 
    Relative to the problem analyzed above 
 there are two changes, first we change the sign of the exponent. Second, the   natural integration 
 contour is along real $r$ or positive imaginary $\rho$. The saddle points continue to be the same, but
 the lines of steepest descent are rotated by $\pi/2$ due to the change in action. In addition, the 
 lines of steepest descent at large $\rho$ are rotated so that now the line with positive imaginary $\rho$
 (or real $r$) is a line of steepest descent. We show these directions in figure \ref{AdScont} in green. In the AdS case,   the natural contour  is between $r\in(0,+\infty)$, the purple line in figure \ref{AdScont}.
 This can be viewed as the integral over energies when we go from the microcanonical to the canonical ensemble. 
 
 In figure \ref{AdScont}(b) and \ref{AdScont}(c) we analyze the steepest descent curves homologous to the original contour of integration. For $y<1$ we are forced to regulate the saddle points by moving them in the imaginary direction (this can be done by a shift $y \to y+ i \epsilon$). The steepest descent curves are shown in blue and the saddle $\rho_2$ always contributes. This corresponds to the large black hole in AdS. Whether it dominates or not relative to the endpoint depends on the temperature, this defines the Hawking-Page transition. It is interesting to note that the small black hole, besides having subleading action, is not even part of the steepest descent contour.
 
 When $y>1$, there is a steepest descent curve that goes along the original contour of integration (blue curve in figure \ref{AdScont}a). Therefore the endpoint at $\rho=0$ always dominates, this is thermal $AdS$. There are no real black hole solutions in $AdS$, and from this analysis we see the two complex solutions should not be included, not even as subleading contributions.

\section{Note added in v4, which corrects the identification of two wavefunctions } \label{App:corr}

In this appendix, we will correct the identification between the wavefunction that appears in the gravitational path integral for de Sitter JT gravity \nref{FinWf},  $\Psi_{\rm path}$,  with the wavefunction that appears in the Wheeler de Witt equation, $\Psi_{WdW}$,  \nref{AsFor}. In the first versions of this paper, we had identified these two objects. However, another indentification was pointed out in the paper \cite{Cotler:2024xzz}, see appendix B in \cite{Cotler:2024xzz}. 

One way to motivate the new identification is to match the inner products computed using the two methods. 

It is clear that the path integral that leads to \nref{FinWf} involves fixing the boundary metric.
If we now want to take an inner product we need to take two such path integrals, identify the boundaries,  and sum over all boundary metrics up to a diffeomorphisms.  More precisely, we have set $\phi = \phi_b$   as a gauge condition and then integrated over the boundary metrics. Here we are assuming that the $\phi = \phi_b$ gauge condition does not lead to any $\ell $ dependent factor, which seems reasonable since this condition is completely local and $\ell$ involves some non-local information, the size of the circle. 

The boundary metric involves an einbein 
$ds = e(\gamma )d\gamma$. We can fix the gauge by the condition $e(\gamma) =1$ so that $\gamma$ measures proper distance. There is a residual gauge transformation that is not fixed by this condition, given by $\gamma \to \gamma +c$. Therefore we  need to divide by the volume of this transformation. The volume is proportional to $\ell$. 
Therefore the path integral that computes the inner product is of the form 
\be 
\int { d \ell } { 1 \over \ell } |\Psi_{\rm path} |^2
\ee 
Note that this argument has derived the measure of the path integral over $\ell $, up to numerical factors and factors involving $\phi_b$. This argument is the same as the argument for the path integral of a relativistic particle discussed around equation (9.57) of \cite{Polyakov:1987hqn}. 

On the other hand, the inner product defined in \nref{KGProd} has the form 
\be 
\int d\ell |\Psi_{WdW} |^2 
\ee
where we used that the leading term comes when the derivative in \nref{KGProd} acts on the first term of the form $e^{ - i 2\phi_b \ell }$ in the exponential, and we are ignoring overall $\phi_b$ dependent factors. (This inner product is derived for the specific operator ordering in \eqref{WdWE}, but one could also consider other choices.)

We see that there is a factor of $\ell$ difference between these two functions \be 
\Psi_{WdW} \propto { 1 \over \sqrt{ \ell } } \Psi_{\rm path } 
\ee 
as derived in appendix B of \cite{Cotler:2024xzz}. 

This changes some results later in the paper. In particular \nref{ConvExp} should be changed by replacing ${1 \over \lambda^3 } \to { 1 \over \lambda^4 }$, but the qualitative conclusions discussed there remain the same.
Similarly, in \nref{WdWtot} we should include an extra factor of $1/\sqrt{\ell}$  
in the right hand side if we interpret the left hand side as $\Psi_{WdW}$.

\mciteSetMidEndSepPunct{}{\ifmciteBstWouldAddEndPunct.\else\fi}{\relax}
\bibliographystyle{utphys}
\bibliography{TwoDimensionaldeSitter.bib}{}

\end{document}